\documentclass[%
 prx,
 amsmath,amssymb,
reprint,%
]{revtex4-1}

\usepackage{graphicx}
\usepackage{dcolumn}
\usepackage{bm}
\usepackage{braket}
\usepackage{dsfont}

\usepackage[usenames,dvipsnames]{color}

\usepackage{xr}
\usepackage{amsmath}
\usepackage[boxruled]{algorithm2e}

\newcommand{\XP}{\mathbf{x}}       
\newcommand{\PP}{\mathcal{P}}      
\newcommand{\DD}{\mathcal{D}}

\newcommand{\fref}[1]{Fig.~\ref{fig:#1}}

\newcommand{\flabel}[1]{\label{fig:#1}}

\newcommand{\beq}{\begin{equation}}
\newcommand{\eeq}{\end{equation}}

\begin{document}

 


\title{ Force field optimization by  imposing kinetic constraints   
 with path reweighting}

\author{P.G.~Bolhuis} 
\affiliation{ van 't Hoff Institute for Molecular Sciences, University of Amsterdam,  PO Box 94157, 1090 GD Amsterdam, The Netherlands}
\author{Z.F.~Brotzakis}
\affiliation{Department of Chemistry, University of Cambridge, Cambridge CB2 1EW,UK}
\author{B.G.~Keller} 
\affiliation{
Department of Biology, Chemistry, Pharmacy, Freie Universität Berlin, Arnimallee 22, D-14195 Berlin}

\date{\today}

\begin{abstract}
Empirical force fields employed in molecular dynamics simulations of complex systems
can be optimised to reproduce experimentally determined
structural
and thermodynamic properties. In contrast,  experimental knowledge about
the rates of interconversion between metastable states in such
systems, is hardly ever
incorporated in a force field, due to a lack of an efficient
approach. 
Here, we introduce such a framework,  based on the relationship between dynamical observables such as rate constants, and the underlying force field parameters, using the statistical mechanics of trajectories.
Given a prior ensemble of molecular trajectories produced with  imperfect force field parameters, the approach allows the optimal adaption of these parameters, such that  the imposed constraint of equal  predicted and experimental rate constant is obeyed. 
To do so, the  method  combines  the continuum path ensemble Maximum Caliber
approach with path reweighting methods for stochastic dynamics.
When multiple solutions are found, the
method selects automatically the combination that corresponds to the
smallest perturbation of the entire path ensemble, as required by the
Maximum Entropy principle. To show the validity of  the approach we
illustrate the  method on simple test systems undergoing rare event
dynamics. Next to simple 2D potentials we explore particle models
representing molecular isomerisation reactions as well as protein-ligand
unbinding. Besides optimal interaction parameters the methodology gives physical insight into what parts of the model  are most sensitive to the kinetics.  We discuss the generality and broad implications of the methodology.
\end{abstract}
\maketitle

\section{Introduction}
Often encountered in molecular biology, chemistry,
material science and soft condensed matter physics, complex molecular systems 
can be, in principle, fully characterised by determining their
structure, thermodynamics and kinetics. Such characterisation enables  
understanding these systems'
function  and  how  their macroscopic properties arise, and eventually
allowing control over their behavior. 
Experimentally, the first step in characterising (bio)molecular systems
is usually to determine the structural and thermodynamic properties~\cite{Brunger1998}. The
next step is to
identify ``kinetic ensembles'', which besides  the structure  and
population of the different states also determine their interconversion rates~\cite{Bonomi2017,Capelli2018,Brotzakis2021}.
Theoretically, (bio)molecular systems can be modeled by molecular
dynamics (MD), which, provided with a faithful underlying interaction
potential,  yields quantitative structural, thermodynamic,
as well as kinetic predictions in  microscopic detail~\cite{Alder1957,Frenkel2001}.
The molecular dynamics community has grown tremendously since the
first simulations in the 1950's~\cite{Alder1957}, a trend greatly
boosted by the increase in computer power provided by advancements in
computer architecture~\cite{Shaw08,Stone2007}, and arguably even more by  algorithmic
improvements\cite{Ciccotti2022},
leading to e.g. powerful 
multiscale models for
complex chemical systems. Indeed, molecular dynamics has been shown  widely
applicable to systems and processes relevant for biology, physics, chemistry and
material science\cite{Hollingsworth2018}.

While the above is true in principle, in practice there are two
obstacles to obtain kinetic ensembles by MD:
First, the interconversion rates are determined by time scales often
way beyond what direct MD can access.  This is known as the  ``rare event''
problem, or the ``sampling''
problem~\cite{Eyring1935,Chandlerbook}. The long time scales are
often connected to/caused by high free  barriers between states. 
A vast spectrum
of enhanced sampling methods  can be applied
to overcome such high free energy barriers and address this ''sampling problem''\cite{Bolhuis2009,Valsson2016}. 

The second potentially more severe and open problem is that the
atomistic molecular dynamics force fields are far from perfect, and  often face challenges in reproducing the relevant experimental data. While atomistic force fields coarse-grain the quantum mechanical (QM) molecular interactions, thereby making the modelling of many body molecular systems feasible, they suffer from errors associated with the model selection and the  usually sparse experimental data sets used for their parametrization 
~\cite{Harrison2018,Piana2011},e.g. vaporization data for small fragments~\cite{Harrison2018,VanderSpoel2021}.
Modern approaches also employ Machine Learning to coarse grain the electronic
degrees of freedom\cite{Tkatchenko2009,Zhang2018,Bonati2018,Singraber2019}. 
Current
force fields can reach experimental accuracy for such
thermodynamic ensemble data \cite{Lindorff2012}, but can differ vastly in
their kinetic properties \cite{Vitalini2015,Piana2011}, because ensemble
averages are very sensitive to the (free) energy differences of the
minima in the potential energy function, but not so much to the (free)
energies of the barriers.
Thus, by parametrizing against thermodynamic ensemble averages one
is unlikely to obtain a kinetically accurate force field.

While an accurate bottom-up correction of force field parameters is
still challenging for complex molecular systems, one can take a different strategy, namely to correct for
force field parametrization inaccuracies in a top-down manner by reweighting atomistic molecular dynamics ensembles using constraints or restraints based on maximum entropy principle in order to match experimental
structural and thermodynamic data ~\cite{Cavalli2013a,Boomsma2014,Bonomi2015}. Taking this strategy further, in previous work we introduced a technique based on the Maximum Caliber approach for continuum path ensembles (CoPE-MaxCal) to
incorporate dynamical constraints into unbiased atomistic classical
molecular dynamics\cite{Brotzakis2021,Bolhuis2021}.
The CoPE-MaxCal framework ameliorates the effects of force field parametrization inaccuracies
on the kinetic properties of complex molecular systems, by
reweighting trajectories based on how far they progress, as measured along
a collective variable. Such a strategy was recently further integrated with deep reinforcement learning~\cite{Tsai2022} to enrich molecular simulation ensembles while making it agree with experimental rate constants.
While very powerful and generally applicable, the CoPE-MaxCal  method  does not
alter the dynamics of the trajectories in the path
ensemble, and therefore  is not (directly) suitable for improving the underlying
force field parameters.

Systematic and efficient top-down approaches to optimize force field parameters to reproduce thermodynamics and kinetics data are still in their infancy due to sparse solution experiment datasets and lack of efficient, data- or physics-driven optimization algorithms. Even when considering matching just thermodynamics, only a few systematic force field optimisation procedures exist.
For instance, recent method developments have enabled parametrizing force fields of large molecules using solution experiment data from NMR, by systematic thermodynamic reweighting methods~\cite{Cesari2019,Tesei2021,Frohlking2020}. In addition, artificial intelligence and new open
science platforms have entered the field~\cite{Varela-Rial2022,Qiu2021}.
However,  so far, there is
no existing systematic strategy to optimize the force field parameters 
such that an
improved force field would match the target (experimental)
interconversion rates directly, although procedures for model optimization for dynamical trajectories has been recently proposed\cite{Rose2021,Das2021,Das2022}). Such force fields, capable of representing experimental kinetics, would have a large
impact in  molecular dynamics, since they would accurately report on transition state structures and populations or processes that are impossible to resolve by experiments~\cite{Brotzakis2021}, thus offering a leverage in, for instance, protein design
(e.g by mutations), or regulation (e.g by transition state small
molecule binders). 

To propose parameters for such a force field, one could perform a naive exhaustive trial and
error search, 
in order to match experimental kinetic data.
However, while this  sounds straightforward and should work in principle, in practice,
this is extremely inefficient as 1) recomputing even a single rate
constant is computationally
expensive due to the rare event problem and 2) moving randomly in the
high dimensional 
force field parameter space, if at all possible, would take many steps to converge. 

In this work we therefore explore an effective way  to infer the relationship between force field parameters and kinetic data using only prior ensembles of  reference trajectories, and
employing techniques to reweight these trajectories. 
Recently, Donati, Kieninger and Keller explored such path reweighting
techniques, which explicitly compute the change in  the path action
based on a force field perturbation\cite{Donati2017, Donati2018, Kieninger2021}.
Here, we combine this path reweighting technique with the CoPE-MaxCal
approach in
order to  impose the dynamical constraint, and at the same time 
select the best solution among multiple solutions:
multiple sets of parameters that all give the correct kinetics.
This selection thus corresponds to a  minimal perturbation, i.e. the change
in the force field cause the smallest possible  perturbation to the
entire path ensemble, while still obeying the constraint.

The computation of the reference trajectory ensemble and the rate
constant can be obtained by direct MD, but as mentioned above this is
not very efficient.
Therefore we employ path sampling methodology, in particular TPS \cite{Bolhuis2002,Dellago2002} and
its descendent single replica transition interface sampling (SRTIS) to
efficiently obtain path ensembles\cite{Du2013}.
We stress than any rare event method that can compute the reference
path ensemble (e.g. FFS\cite{Allen2005} or weighted ensemble \cite{Zuckerman2017}) is suitable to be used with our methods. 

The remainder of the  paper is organised as follows. In the next section we develop the above sketched approach. In section \ref{sec:results}  we first validate  that the path reweighting can predict rate constant changes. We then optimize the parameters for 
several model systems to illustrate the effectiveness of the methodology.
We end by giving an outlook to which challenges in molecular sciences and other fields our method could be applied.

\section{Theory}
\subsection{Maximum Caliber and path reweighting}
Consider a system consisting of $N$ atoms. $x \in  \mathbb{R}^{3N}$
denotes the configurational state of the system, where $\mathbb{R}^{3N}$ is the $3N$-dimensional position space.
We assume that the system evolves according to the overdamped Langevin dynamics \cite{Leimkuhler2015, Oksendal2003} in a force field $-\nabla V(x)$ and
 note that our method can be generalized to underdamped Langevin
dynamics \cite{Kieninger2021}, so that $x \in  \mathbb{R}^{6N}$.
We simulate the system using the Euler-Maruyama (EM) method~\cite{Platen1992} to obtain time-discretized trajectories.
A trajectory is defined as an ordered sequence of frames $\XP = \{x_0,   x_1, ...x_L\}$, where  the subscripts denote the time index.
 Subsequent frames are separated by a time interval $\Delta t$, such that the total duration
of a path is $\mathcal{T} = L \Delta t$. 
These paths $\XP$ live in a  domain $\mathcal{S}$. 
The probability for a trajectory in this  domain $\mathcal{S}$  
is defined as
\begin{align}
\PP[\XP] &= \frac{1}{\mathcal{Z}}\rho(x_0) \prod_{i=1}^L  p(x_{i-1} \rightarrow x_{i}),
\label{eq:priordist}
\end{align}
where $\rho(x_0)$ denotes the probability density of the initial condition, usually the Boltzmann
distribution $\rho(x) \sim \exp(- \beta V(x))$, with $V(x)$ the potential
energy of configuration $x$, $\beta=1/k_BT$ the reciprocal
temperature, $T$ the temperature and $k_B$ Boltzmann's constant.
$p(x_i \rightarrow x_{i+1})$ is a short-time Markovian probability
representing the dynamical evolution, as given by the  integration algorithm
and thus depends on $V(x)$.(See eq.~13 in Ref.~\onlinecite{Kieninger2021} for $p(x_i \rightarrow x_{i+1})$ in the EM algorithm).
$\mathcal Z$ is a  normalization constant such that  $\PP[\XP]$ is normalised with respect to integration over the path ensemble $\int_{\mathcal{S}} \DD \XP \mathcal{P}[\mathbf{x}]=1$.
($\DD \XP$ indicates a path integral over all trajectories $\XP \in
\mathcal{S}$, 
 in the domain $\mathcal{S}$, see the Appendix for a
discussion on the definition of $\mathcal{S}$ in relation to path integrals).

The (relative) path entropy, $S$ or caliber, for any path distribution
$\PP [\XP]$ is given by the Kullback-Leibler divergence $D_{KL}$ 
\begin{align} D_{KL} =- S = \int_{\mathcal{S}} \DD \XP \PP [\XP] \ln
  \frac{\PP[\XP]}{\PP^0[\XP]}.
  \label{eq:DKL}
\end{align}
Here, $\PP^0[\XP]$ denotes the probability of trajectory $\XP$ in the
reference path ensemble. 
  The  maximum caliber principle \cite{Hazoglou2015} states that the optimal path
  probability distribution $\PP^{MC}[{\XP}]$ follows from maximising the
  caliber while satisfying an external constraint
$s^{\mathrm{exp}}$.
  \begin{align}
  \PP^{MC}[{\XP}]=&\operatorname*{argmax}_{\PP[\XP]}  S[\PP||\PP^0]  \label{eq:MaxCal}\\    
&\textrm{subject to: }\begin{cases}
\int_{\mathcal{S}} \DD{\XP }  \PP[{\XP }] s [{\XP }] = \langle s [{\XP}] \rangle =s^{\mathrm{exp}} \\
  \int_{\mathcal{S}} \DD{\XP } \PP[{\XP }]=1.
\end{cases}\notag
\end{align}
That is, $\PP^{MC}[{\XP}]$  maximizes the path entropy or caliber,
while obeying the constraints given by external constraint
$s^{\mathrm{exp}}$ and keeping the probability normalized.
Even though the specification of the domain $\mathcal{S}$ that is associated to the external constraints is important, we will nonetheless drop $\mathcal{S}$ from the following equations to keep the notation manageable.

Solving eq.~\ref{eq:MaxCal} can be addressed using the method of Lagrange multipliers. 
The path Lagrange function  is
\begin{align}
  \mathcal{L} &=  D_{KL} 
				- \mu \left( \int
                \mathcal{D}\mathbf{x}\, \mathcal{P}[\mathbf{x}]
                s(\mathbf{x})  - s^{\mathrm{exp}}\right) \notag\\
				&- \nu \left( \int \mathcal{D}\mathbf{x}\, \mathcal{P}[\mathbf{x}] - 1\right),			
\label{eq:pathLagrangian}					
\end{align}
where the second term imposes the experimental constraint, $\mu$ and $\nu$ stand for Lagrange multipliers, and the final
constraint enforces normalisation.
$\mathcal{L}$ depends on the potential energy function $V(x)$ via Eq.~\ref{eq:priordist}.
The task is now to find the stationary points of the Lagrange function,
which constitutes setting to zero the derivatives of $\mathcal L$  with respect to the adjustable parameters of the potential energy function.

We make the following ansatz: the adjusted potential energy function $\tilde V(x; \mathbf{a})$ differs from the current/prior $V(x)$ by a perturbation $U(x; \mathbf{a})$
\begin{eqnarray}
	\tilde V(x; \mathbf{a}) &= V(x) + U(x; \mathbf{a}) \, ,
\end{eqnarray}
%
where the change from the current to the new potential energy function can be expressed in terms of $m$ parameters
%
$	\mathbf{a} = (a_1, a_2 \dots a_m)$.
%

The path probability of the new force field $\mathcal{P}[\mathbf{x}]$
and the path probability of the prior force field
$\mathcal{P}^0[\mathbf{x}]$ are related by  the relative path probability\cite{Kieninger2021} $\frac{\mathcal{P}[\mathbf{x}]}{\mathcal{P}^0[\mathbf{x}]}= \frac{\mathcal{Z}^0}{\mathcal{Z}(\mathbf{a})} W(\XP; \mathbf{a}) $, thus
\begin{eqnarray}
	\mathcal{P}[\mathbf{x}] &=&    \frac{\mathcal{Z}^0}{\mathcal{Z}(\mathbf{a})}
 W[\XP; \mathbf{a}]     \mathcal{P}^0[\mathbf{x}]   \notag \\                                &=&\frac{\mathcal{Z}^0}{\mathcal{Z}(\mathbf{a})}
 g(x_0,                          \mathbf{a})M[\XP;
                                    \mathbf{a}] 
                                    \mathcal{P}^0[\mathbf{x}]
\label{eq:pathReweighting}	
\end{eqnarray}
where the second equation defines $W[\XP; \mathbf{a}]$, 
with
\begin{align}
	g(x_0, \mathbf{a})	&= \frac{\exp(-\beta \tilde V(x_0;\mathbf{a}))}{\exp(-\beta V(x_0))}
                             = \exp(-\beta U(x_0;\mathbf{a}))\cr
	M[\XP;\mathbf{a}]	&= \frac{\prod_{i=0}^{n-1} \tilde{p}(x_{i+1} |x_i,  \mathbf{a})}{\prod_{i=0}^{n-1} p_0(x_{i+1} |x_i)}	\notag
\end{align}
where
$p_0(x_{i+1} |x_i)$ is the single-step transition probability of the current (prior) force field, and $\tilde{p}(x_{i+1} |x_i,  \mathbf{a})$ is the single-step transition probability of the new force field.
 $ \mathcal{Z}^0/\mathcal{Z}(\mathbf{a})$ is the ratio of the partition function at the current (prior) force field, $\mathcal{Z}^0$, and the partition function at the new  (posterior) force field, $\mathcal{Z}(\mathbf{a})$ (see eq.~\ref{eq:priordist}).
The ratio is linked to the free energy difference of adjusting the force field.
We treat $\mathcal{Z}(\mathbf{a})$ as a path probability normalization
constant  which guarantees that $\int \DD \XP
\mathcal{P}[\mathbf{x}]=1$.
(Note that in previous work e.g. Ref.~\onlinecite{Dellago2002} the $\mathcal{Z}^0$ was set to unity, as the partition function was implicitly embedded in the density $\rho(x_0)$  and the single step transition probabilities were considered normalised. However, in the path reweighting work of Ref.~\onlinecite{Kieninger2021} and in this work we cannot assume that anymore, and  the partition functions are explicitly taken into account.)

Inserting eq.~\ref{eq:pathReweighting} into
eq.~\ref{eq:DKL} yields for the $D_{KL}$
\begin{align}
D_{KL}  = \frac{\mathcal{Z}^0}{\mathcal{Z}(\mathbf{a})}\int \DD \XP \PP^0 [\XP]   W[\XP;\mathbf{a}] 
  \ln  \frac {\mathcal{Z}^0
  \PP^0[\XP]}{\mathcal{Z}(\mathbf{a})\PP^0[\XP]}  W[\XP;\mathbf{a}]\notag\\
=  \frac{\mathcal{Z}^0}{\mathcal{Z}(\mathbf{a})} \left(\int \DD \XP\PP^0 [\XP]   W[\XP;\mathbf{a}] 
  \ln   W[\XP;\mathbf{a}]-  \ln \frac{\mathcal{Z}(\mathbf{a})}{ \mathcal{Z}^0}\right)
 \end{align}
 where in the second equality we used  $ \frac{\mathcal{Z}^0}{\mathcal{Z}(\mathbf{a})}\int \DD \XP\PP^0 [\XP]   W[\XP;\mathbf{a}]=\int \DD \XP \mathcal{P}[\mathbf{x}] =1$.
 This equation can be written as
 \begin{align}
D_{KL}  
=  \frac{\int \DD \XP\PP^0 [\XP]   W[\XP;\mathbf{a}] 
  \ln   W[\XP;\mathbf{a}] }{\int \DD \XP\PP^0 [\XP]   W[\XP;\mathbf{a}] }   - \ln \int \DD \XP\PP^0 [\XP]   W[\XP;\mathbf{a}] 
 \end{align}

 The total Lagrange function follows then by inserting
 Eq. \ref{eq:pathReweighting} into  Eq. \ref{eq:pathLagrangian}.

\begin{align}
  \mathcal{L} &=  D_{ KL}
			  	- \mu \left( \ln \frac{ \int
                      \mathcal{D}\mathbf{x}\,W[\XP;\mathbf{a}]\mathcal{P}^0[\XP]\cdot
                      s[\XP]  } {\int
                      \mathcal{D}\mathbf{x}\,W[\XP;\mathbf{a}]\mathcal{P}^0[\XP]
                     }
- \ln s^{\mathrm{exp}}\right)
\label{eq:pathLagrangian_reweighted}					
\end{align}
where we now imposed the constraint onto the logarithm of the observable, and
the normalisation constraint is automatically obeyed.Therefore we can
leave out the third term in the Lagrange function in Eq.~\ref{eq:pathLagrangian}	

Eq.~\ref{eq:pathLagrangian_reweighted} is a central and general result of this
work.

Generalizing the Lagrangian in eq.~\ref{eq:pathLagrangian_reweighted} to multiple external constraints $(s_1^{\mathrm{exp}} \dots s_n^{\mathrm{exp}})$ that need to be satisfied simultaneously, yields
\begin{align}
    \mathcal{L} &=  D_{ KL}
			  	- \sum_{i=1}^n \mu_i \left( \ln \frac{ \int
                      \mathcal{D}\mathbf{x}\,W[\XP;\mathbf{a}]\mathcal{P}^0[\XP]\cdot
                      s_i[\XP]  } {\int
                      \mathcal{D}\mathbf{x}\,W[\XP;\mathbf{a}]\mathcal{P}^0[\XP]
                     }
- \ln s_i^{\mathrm{exp}}\right)
\end{align}

\subsection{Derivatives  of the Lagrange function $\mathcal{L}$}
To find the optimal new force field parameters $\bf a$ for a single constraint,  we determine the stationary point of  the Lagrange function,
i.e. we solve 
\begin{eqnarray}
	\frac{\partial \mathcal{L}}{\partial a_k} = 0  \qquad 	\frac{\partial \mathcal{L}}{\partial \mu} = 0 
\label{eq:dLda}	
\end{eqnarray}
for all $a_1, \dots a_m$, leading to $m+1$ constraint equations. 

Obtaining explicit expressions for these constraint equations is
easier when defining an auxiliary function $w[\XP;a] = \ln W[\XP;a]$, so that 
\begin{align}
 W[\XP;{\mathbf a}] =\exp({w[\XP;a]})
\label{eq:substitution}    
\end{align}
 The first term in the Lagrange function, $D_{KL}$ is then
 \begin{align}
D_{KL}  
=  \frac{\int \DD \XP\PP^0 [\XP]  e^{w[\XP;\mathbf{a}]} 
  w[\XP;\mathbf{a}] }{\int \DD \XP\PP^0 [\XP]   e^{w[\XP;\mathbf{a}]} }   - \ln \int \DD \XP\PP^0 [\XP]   e^{w[\XP;\mathbf{a}]} 
\label{eq:DKL_after_substitution}  
 \end{align}
 
Taking the derivative would give then 
\begin{align}
  &\frac{\partial D_{KL}   }{\partial a_k}
=  \frac{\int \DD \XP\PP^0 [\XP]  e^{w[\XP;\mathbf{a}]} w'[\XP;\mathbf{a}] 
    w[\XP;\mathbf{a}] }{\int \DD \XP\PP^0 [\XP]   e^{w[\XP;\mathbf{a}]}
    }   \cr &- \frac{\int \DD \XP\PP^0 [\XP]  e^{w[\XP;\mathbf{a}]} w'[\XP;\mathbf{a}] 
    }{\int \DD \XP\PP^0 [\XP]   e^{w[\XP;\mathbf{a}]}}\frac{\int \DD \XP\PP^0 [\XP]  e^{w[\XP;\mathbf{a}]} w[\XP;\mathbf{a}] 
    }{\int \DD \XP\PP^0 [\XP]   e^{w[\XP;\mathbf{a}]}} \cr
    \label{eq:dDKLda}
\end{align}
where  the derivative 
 $   w'[\XP; \mathbf{a}]= {\partial  w[\XP;\mathbf{a}]}/{\partial a_k}$.
Intermediate steps for this derivative are reported in appendix \ref{app:LagrangianDerviative}.

Note that the  fractions in Eq.\ref{eq:dDKLda} can be interpreted as
(path) ensemble averages, so that
\begin{align}
  \frac{\partial D_{KL}   }{\partial a_k}
&=   \left<  w'[\XP;{\mathbf a}] w[\XP;{\mathbf a}]
    \right>_{W}  
  -
 \left< w'[\XP;{\mathbf a}]
    \right>_{W}  \left< w[\XP;{\mathbf a}]
    \right>_{W} 
    \label{eq:dDKLda2}
\end{align}
where the bracket subscript $W$ indicates that path ensemble average is calculated with respect to $\mathcal{P}[\mathbf{x}]$, i.e.~the reweighted path probability density.

Taking the derivative of the total Lagrange function Eq.~\ref{eq:pathLagrangian_reweighted} with respect to $a_k$ gives
\begin{align}
	\frac{\partial \mathcal{L}}{\partial a_k}  &= \frac{\partial
  D_{KL}   }{\partial a_k} - \mu \left(
\frac{\int \DD \XP\PP^0 [\XP]  e^{w[\XP;\mathbf{a}]} w'[\XP;\mathbf{a}] 
   \cdot       s[\XP]   }{\int \DD \XP\PP^0 [\XP]   e^{w[\XP;\mathbf{a}]}
   \cdot       s[\XP]   }   \right.  \notag \\ & \left. -
  \frac{\int \DD \XP\PP^0 [\XP]  e^{w[\XP;\mathbf{a}]} w'[\XP;\mathbf{a}] 
   }{\int \DD \XP\PP^0 [\XP]   e^{w[\XP;\mathbf{a}]}
  }   
 \right),
\label{eq:dLda3}	
\end{align}
or again by using path ensemble notation
\begin{align}
	\frac{\partial \mathcal{L}}{\partial a_k}  = \frac{\partial
  D_{KL}   }{\partial a_k} 
-  \mu \left( \left<  w'[\XP;{\mathbf a}] 
    \right>_{W s} -
\left< w'[\XP;{\mathbf a}]
    \right>_{W}  \right), 
    \label{eq:dDKLda3}
\end{align}
where the $Ws$ subscript now denotes that the path ensemble averages is calculated with respect to the transformed path probability  $\mathcal{P}[\mathbf{x}]\cdot  s[\XP]$.
Note that, while the expression $\left<  w'[\XP;{\mathbf a}] \right>_{W s}$ can be evaluated for any integrable function $s[\XP]$, its interpretation as a path ensemble average is only justified if $s[\XP]$ is a positive function.

The entire expression for the derivative is then 
\begin{align}
	\frac{\partial \mathcal{L}}{\partial a_k}  
&=   \left<  w'[\XP;{\mathbf a}] w[\XP;{\mathbf a}]
    \right>_{W } 
    -
 \left< w'[\XP;{\mathbf a}]
    \right>_{W }  \left< w[\XP;{\mathbf a}]
    \right>_{W } \notag \\
&-  \mu \left( \left<  w'[\XP;{\mathbf a}] 
    \right>_{W s } -
\left< w'[\XP;{\mathbf a}]
    \right>_{W }  \right) 
    \label{eq:dDKLda4}
\end{align}
We can condense this expression even more by dropping the arguments of the
functions, which yields
\begin{align}
	\frac{\partial \mathcal{L}}{\partial a_k}  
=   \left<  w' w     \right>_{W} -
 \left< w'
    \right>_{W}  \left< w
    \right>_{W } 
-  \mu \left( \left<  w'  
    \right>_{Ws} -
\left< w'
    \right>_{W}  \right). 
    \label{eq:dDKLda5}
\end{align}

The final ingredient for optimisation is the derivative with respect
to the Lagrange multiplier $\mu$. This simply  is given by the constraint itself
\begin{align}
  \frac{\partial \mathcal{L}}{\partial \mu} = 	- \left( \ln \frac{ \int
                      \mathcal{D}\mathbf{x}\,W[\XP;\mathbf{a}]\mathcal{P}^0[\XP]\cdot
                      s[\XP]  } {\int
                      \mathcal{D}\mathbf{x}\,W[\XP;\mathbf{a}]\mathcal{P}^0[\XP]
                     }
- \ln s_{\mathrm{exp}}\right)
\end{align}
or in the condensed form by
\begin{align}
  \frac{\partial \mathcal{L}}{\partial \mu} = 	- \ln \left<    s    \right>_{W}             
  + \ln s_{\mathrm{exp}}
  \label{eq:dDKLdmu}
\end{align}
Together, Eqs. ~\ref{eq:dDKLda5}   and \ref{eq:dDKLdmu}, provide the
derivatives for finding the stationary point for the Lagrange function,
and hence the optimal force field parameters $\bf a$.

So far, the logarithm of the relative path probability $w(\mathbf{x}, \mathbf{a}) = \ln W[\mathbf{x}, \mathbf{a}]$
and the path observable $s(\mathbf{x})$ have been abstract functions. 
Next, section \ref{sec:rateEstimate} derives an expression for the term $\mu \left( \left<  w'  
    \right>_{Ws} -
\left< w'
    \right>_{W}  \right) $, 
and section \ref{sec:pathReweighting} derives expressions for $w$ and $w'$ for trajectories generated by the Euler-Maruyama integrator. 

\subsection {The rate constant estimate}
\label{sec:rateEstimate}
In the Lagrange function the experimental observable is
constrained. While this could be any dynamical observable
such a mobility, viscosity, etc., in our work it is taken to be the  kinetic observable
rate constant.
In principle the rate constant can be obtained by counting the
number of effective
transitions per unit time in a straightforward
MD simulation, but this is extremely inefficient due to the rare event
problem. Many enhanced sampling methods exist to make rate constant
computations more efficient, such as reactive flux approach~\cite{Chandler1978},
milestoning~\cite{Faradjian2004}, forward flux sampling~\cite{Allen2006}, infrequent Metadynamics~\cite{Tiwary2013} virtual interface exchange transition path sampling~\cite{Brotzakis2019f} etc~\cite{Bolhuis2009,Valsson2016}. Here, we adopt the framework of
transition path sampling \cite{Bolhuis2002} and transition interface sampling
(TIS)\cite{vanErp2003}, and in particular that of the reweighted path
ensemble (RPE)~\cite{Rogal2010}.  Defining the metastable
stable states A and B  using an order
parameter or collective variable (CV) $\lambda$, with $\lambda_{A,B}$
the boundaries of the states A and B, 
one can  compute the rate constant in the RPE framework from the following expression
\begin{align}
\label{eq:rateconst}
 k_{AB} = \frac{\phi_A  \int_{A} \DD\XP  \PP^0[\XP] \theta( \lambda_{max}[\XP] -
\lambda_B ) }{ \int_{A} \DD\XP  \PP^0[\XP] },
\end{align}
where $\int_{A} \DD\XP$ denotes a path integral over paths that leave A and go over the barrier to $B$, or return to enter $A$, after which they are terminated. 
The path length is thus flexible. 
The frequency with which these paths are sampled is determined by $\PP^0[\XP]$ and such a (reweighted) path ensemble can be obtained by e.g. TIS.
(For a brief discussion on flexible path length ensembles, see appendix \ref{app:pathEnsemble}).
The $\theta( x)$ is the Heaviside step function   and
$\lambda_{max}[\XP]$ returns the maximum value of the progress order
parameter or collective variable (CV) that is able to measure how far
the transition has proceeded. Thus, the $\theta$-function in the
integral in the numerator selects the paths that reach  the boundary
of state B,  $\lambda_B$, i.e., the reactive paths. The fraction is
thus equal to the probability of reaching B for paths that leave A.  Multiplying with the flux
$\phi_A$ through the  first interface $\lambda_0$ for paths leaving state
A, this indeed gives the rate constant. 
Using the path reweighting of Eq.\ref{eq:pathReweighting}	 we obtain
\begin{align}
\label{eq:rateconst2}
 k_{AB} = \frac{\phi_A  \int \DD\XP  \PP^0[\XP] W[\XP;\mathbf{a}]  \theta( \lambda_{max}[\XP] -
\lambda_B ) }{ \int \DD\XP \PP^0[\XP] W[\XP;\mathbf{a}]   }.
\end{align}

Setting $s[\XP] = \phi_A\theta( \lambda_{max}[\XP] - \lambda_B )$  and the experimental observable to $s^{\mathrm{exp}} = k_{AB}^{\mathrm{exp}}$ in Eq.~\ref{eq:pathLagrangian_reweighted}, we obtain the Lagrange function 
\begin{align}
  \mathcal{L} &=  D_{ KL} - 
                \mu \Big( 
			        \ln \frac{  \int \mathcal{D}\mathbf{x}\,W[\XP;\mathbf{a}]\mathcal{P}^0[\XP]\cdot
                      \phi_A \theta( \lambda_{max}[\XP] - \lambda_B )  } {\int
                      \mathcal{D}\mathbf{x}\,W[\XP;\mathbf{a}]\mathcal{P}^0[\XP] 
                     } \notag \\
            &\quad    - \ln k_{AB}^{\mathrm{exp}}
                      \Big) \, ,
\end{align}
The derivative of the Lagrangian (eqs.~\ref{eq:dLda3} and \ref{eq:dDKLda3}) is
\begin{align}
\frac{\partial \mathcal{L}  }{\partial a_k} 
    &=  \frac{\partial D_{KL}  }{\partial a_k} \cr 
    &- \mu \frac{ \phi_A \int \DD\XP \PP^0 [\XP]    e^{w [\XP;\mathbf{a}]}
  w'[\XP;\mathbf{a}] \theta( \lambda_{max}[\XP]- \lambda_B )    }{
  \phi_A \int \DD\XP  \PP^0 [\XP]  e^{w [\XP;\mathbf{a}] }\theta(
  \lambda_{max}[\XP] -  \lambda_B )}  \notag \\
   & - \mu   \frac{ \int \DD\XP  \PP^0 [\XP]  e^{w [\XP;\mathbf{a}]} w'[\XP;\mathbf{a}]  }{ \int \DD\XP  \PP^0 [\XP]  e^{w [\XP;\mathbf{a}] } } \notag\\
   &=   \frac{\partial D_{KL}  }{\partial a_k} -  \mu \left( \left<  w'  
    \right>_{AB,W} -
\left< w' \right>_{A,W}  \right) \,.
\label{eq:dDKLda6}    
\end{align}
We used the assumption that $\phi_A$ is not depending on $\bf a$, and does not depend on $\XP$, and thus $\phi_A$ cancels in the second term.  This assumption is justified if the parameters $a$ do not influence the stable state A. In general, of course the parameters can also affect the stable states, and in that case also the change in the flux need to be taken in to account. However, the effect on the flux is expected to be  small, in comparison to the change in rate constant due to the barrier height.   
In the second equality, the path ensemble average  $\left< w' \right>_{A,W}$ is calculated with respect to the reweighted path probability $\mathcal{P}[\XP]= W[\XP, \mathbf{a}]\mathcal{P}^0[\XP]$ (indicated by the subscript $W$) over all paths that leave A (indicated by the subscript $A$).
The path ensemble average $\left<  w' \right>_{AB,W}$ is calculated with respect to the path probability 
$\mathcal{P}[\XP] \theta( \lambda_{max}[\XP]- \lambda_B )$ over all paths that leave $A$.
However, since $\theta( \lambda_{max}[\XP]- \lambda_B )$ selects paths that end in $B$, one can interpret this term as a path ensemble average calculated with respect to the reweighted path probability density $\mathcal{P}[\XP]$ (indicated by the subscript $W$) over all reactive AB trajectories (indicated by the subscript $AB$)
Note the similarities to the temperature derivative of the rate constant in e.g. Ref\cite{DellagoBolhuis2004,BolhuisCsanyi2018}.

The term between brackets in the second equality of Eq.~\ref{eq:dDKLda6}    
 denotes the derivative of the
rate constant with respect to the force field parameters.

\begin{align}
\frac{\partial \ln k_{AB} }{\partial a_k}  =
 \langle w'  \rangle_{AB,W}   -  \langle w'  \rangle_{A,W}.
 \label{eq:lnkABdak}
\end{align}
This quantity can thus serve as
a first sanity check whether the reweighting approach actually works.

\subsection{Path reweighting using  the adapted force field}
\label{sec:pathReweighting}

As defined in Eq. 5, the new force field $\tilde V(x)$ differs from the current force field $V(x)$ by a perturbation $U(x;\mathbf{a})$ and the reweighting factor for the stationary density becomes
\begin{align}
	g(x_0, \mathbf{a}) 
    =\frac{\exp(-\beta (V(x_0)+ U(x_0, \mathbf{a})))}{\exp(-\beta V(x_0))}
 = \exp(-\beta U(x_0, \mathbf{a}))
\end{align} 
where we left out the normalizing partition function $Z$, which is
included in Ref\cite{Kieninger2021}, as it is already included in the
path partition normalization constant $\mathcal{Z}(\mathbf{a})$.

Following Ref\cite{Kieninger2021}, the path reweighting factor $M[\XP;\mathbf{a}]$ is
\begin{align}
   M[\XP;\mathbf{a}] &= \exp\left( -\sum_{i=0}^{n-1}
                                        \eta_i \cdot
                                        \sqrt{\frac{\Delta t}{2 k_BT
                                        \xi m}} \nabla U (x_i,
                                        \mathbf{a})\right) 
                                        \notag \\  &\times
							  \exp\left( -\sum_{i=0}^{n-1} \frac{1}{2} \cdot \frac{\Delta t}{2 k_BT \xi m} \left(\nabla U (x_i, \mathbf{a})\right)^2 \right) 
\end{align}
where we use the formulation with the random numbers $\eta$\cite{Kieninger2021}.
In this definition $\xi$ is the Langevin friction and $m$ the particle
mass, as defined in the EM integrator. 
$\eta_i$ is the random number used in the $i$th iteration of the EM integrator out of $n$ time steps (frames) and can be recorded during the simulation a the current force field.
Multiplying these two  factors gives the total path reweighting  $W[\bf x ; a ]$. Taking the logarithm yields
\begin{align}
	\ln (g(x_0,\mathbf{a}) M[\XP;\mathbf{a}] )  &=  \ln W[\XP;\mathbf{a}]  = 
                                           - \beta  U(x_0, \mathbf{a}) \notag \\
	&-\sum_{i=0}^{n-1} \eta_i \cdot \sqrt{\frac{\Delta t}{2 k_BT \xi m}} \nabla U (x_i, \mathbf{a}) \notag\\
			 & -\sum_{i=0}^{n-1} \frac{1}{2} \cdot \frac{\Delta t}{2 k_BT \xi m} \left(\nabla U (x_i, \mathbf{a})\right)^2.
\label{eq:gMP}			 
\end{align}
We might further simplify this long expression by defining
$$ 
\kappa \equiv \sqrt{\frac{\Delta t}{2 k_BT \xi m}},
$$
so that
\begin{align}
w[\XP;\mathbf{a}] &= \ln W[\XP;\mathbf{a}] 
                = 	
                - \beta  U(x_0, \mathbf{a})  \notag\\
		&-\sum_{i=0}^{n-1} \eta_i \cdot \kappa \nabla U (x_i, \mathbf{a})
			 -\sum_{i=0}^{n-1} \frac{1}{2} \cdot \left( \kappa \nabla U (x_i, \mathbf{a})\right)^2,
\label{eq:wa}			 
\end{align}
and the derivative becomes

\begin{align}
  \frac{\partial w[\XP;\mathbf{a}]  }
  {\partial a_k}  &=  {-\beta}\frac{\partial U(x_0;\mathbf{a})}{\partial a_k}  
                    -\sum_{i=0}^{n-1} \eta_i \cdot \kappa
                    \frac{\partial  \nabla U(x_i;\mathbf{a})   }{\partial a_k}  \notag\\
                  &-\sum_{i=0}^{n-1}
                    \kappa^2\left(\nabla                    U (x_i,
                    \mathbf{a})\cdot
                         \frac{\partial  \nabla U(x_i;\mathbf{a})   }{\partial a_k}  
                    \right).
                    \label{eq:derwa1}
\end{align}
The above equation can be used to compute both the caliber and the  derivatives of the Lagrange function. 
While it is possible to compute the derivative with respect to $a_k$
on the fly, it is even more efficient to be able to compute these a
posteriori. This depends on the precise functional form of  $U(x,\mathbf{a})$. In
case of a linear dependence $\widetilde{V}(x) = V(x) + a U(x)$, the perturbation forces are $a \nabla U(x)$, and the derivate with respect to $a$ is just $\nabla U(x)$. Hence it is convenient to store the sums over
$\eta_i \nabla U(x_i, \mathbf{a}) $ and $(\nabla U(x_i, \mathbf{a}))^2$ terms for each trajectory
explicitly, so that the derivatives can be computed easily for
arbitrary values of  $\bf a$. For a non-linear dependence one can still
do so, but it becomes more complicated. 

\section{Results and Discussion}
\label{sec:results}
\subsection{Testing the rate constant derivative}

Before embarking on the full problem of force field optimization, we first will check the path reweighting method, by computing  the derivative of the rate constant, i.e. Eq.~\ref{eq:lnkABdak}. 
\begin{align}
\frac{\partial \ln k_{AB} }{\partial a_k}  =
 \left<  \frac{\partial w[\XP;\mathbf{a}] }{\partial a_k}  \right>_{AB,W}   -  \left<  \frac{\partial w[\XP;\mathbf{a}] }{\partial a_k}  \right>_{A,W}  
\end{align}
Thus the rate constant derivative is equal to the difference between  two path ensembles averages.
This is a general expression and can be related to the Arrhenius law, and estimates of the activation energy from path sampling\cite{DellagoBolhuis2004,BolhuisCsanyi2018}.

\subsubsection{Diatomic system}

\begin{figure}[t]
    \centering
\includegraphics[width=8cm]{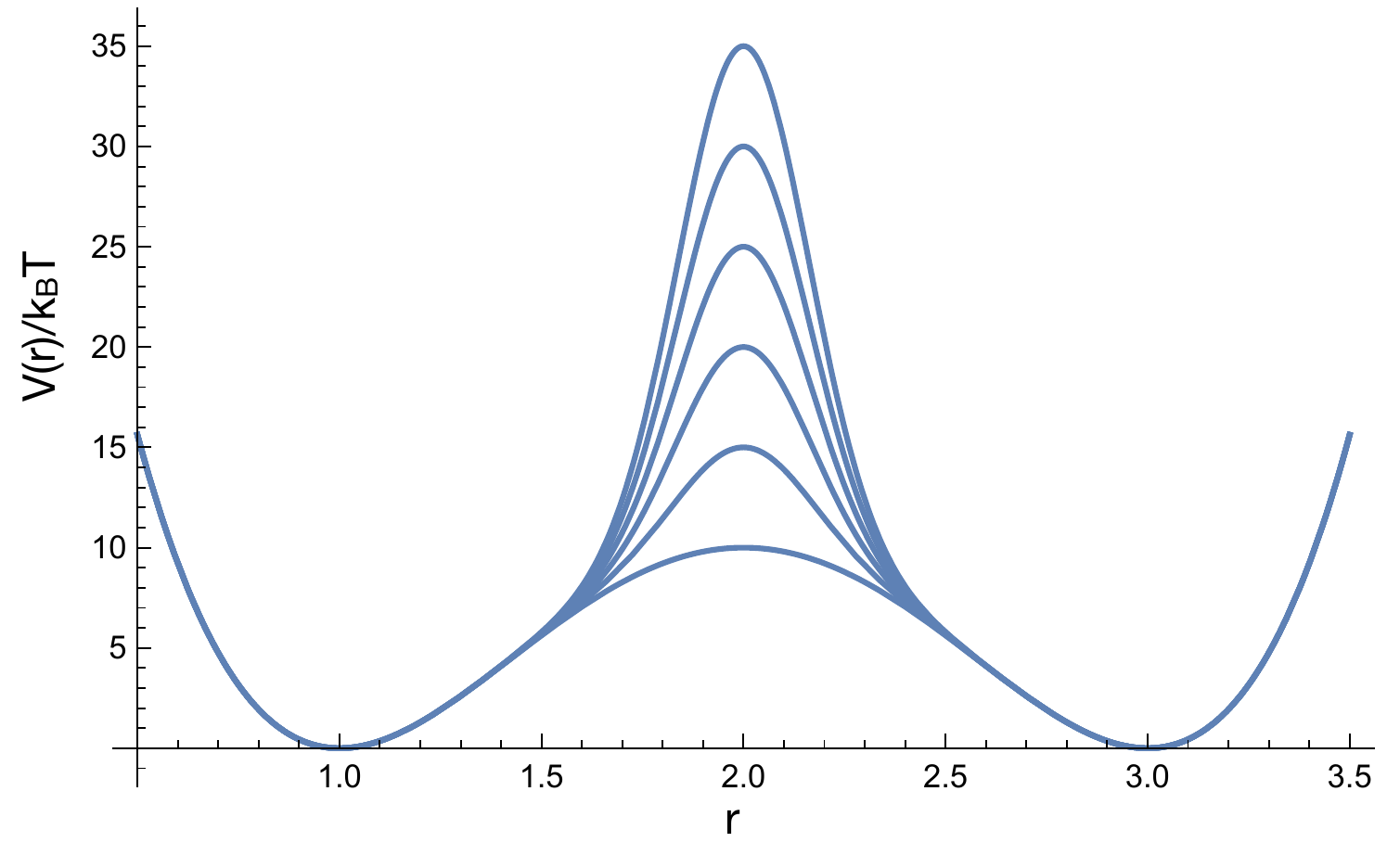}
    \caption{Total interaction potentials $\beta V(r)$ 
      in the diatomic system for different
      settings of $a$. The lowest  curve is the bistable 
      potential $\beta V_0(r)$, i.e.~$a=0$. Each curve above corresponds to an increased $a$ in steps of 5$k_BT$.}
    \label{fig:2Dpot}
  \end{figure}

Now we are ready to look at a specific system.
 We first investigate  a simple diatomic system in which the two atoms are
 held together by a  bistable potential  (in dimensionless units)
  \begin{align}
    \beta V_0(r) =   10 ( (r - 2)^2  -1) ^2   
  \end{align}
  where $r$ is the (dimensionless)  distance between the atoms. There
  are two
  minima located at $r=1$ and $r=3$. 
  We can thus interpret this
 system as a diatom that has a compact state an extended state,
 separated by  barrier with a 
 height of 10 $k_BT$. Due to the 2D nature of the system, the expanded
 state has more entropy, and is expected to be (slightly) more stable.
Next, we  add a Gaussian  to the potential 
 \begin{align}
    \beta V(r) =   10 ( (r - 2)^2  -1) ^2   +a \exp(- 20  (r - 2)^2  )
 \end{align}
 Figure \ref{fig:2Dpot} depicts this potential for several values of
$a$, ranging from $a=0$ to $a=25$.

We add a perturbation of the same Gaussian form
\begin{align}
  \beta U(r;a) = \Delta a \exp(- 20 (r - 2)^2  ),
  \label{eq:pert1D}
\end{align}
so that the perturbed total potential is 
 \begin{align}
    \beta \tilde{V}(r) =    \beta V_0(r)   +(a + \Delta a) \exp(- 20  (r - 2)^2  )
 \end{align}

As the perturbation is not affecting the stable states we can neglect
the first term in Eq ~\ref{eq:derwa1},

For this test we compute the rate constant derivatives for ${\Delta a}=0$, that is, a zero perturbation. In particular, then the gradient of the force in the last term of Eq. \ref{eq:derwa1} vanishes, leaving 
\begin{align}
\label{eq:derwa3}
  \left(\frac{\partial  w[\XP;{a}]}{\partial a} \right)_{\Delta a =0}\  =
  -\sum_{i=0}^{n-1} \eta_i \cdot \kappa     \frac{\partial  \nabla U(x_i,{a})   }{\partial a_k}.  
\end{align}

  \begin{table}[b]
   \centering
    \caption{Logarithmic rate constant,  path action derivative and fluxes for
      the diatom  system. The first set of result are for the
      forward expansion transition, the second set for the backward
    contraction transition. Note that the second set shows a
    systematically lower rate constant because the expanded state is
    slight more stable due to a higher entropy.}
    \label{tab:bistablegauss}
    \begin{tabular}{rrrrrrr}
    a &  ln k & dlnk/da &  flux $\phi_A$ \\
    \hline
  1 & -9.01672 &-0.774092 &   0.002443 \\
2 & -9.83885& -0.82772  &    0.002391 \\
3 & -10.716 &-0.881064  &    0.002415 \\
4 &-11.5969 &-0.895259&     0.002351 \\
5 &-12.5809 &-0.909444  &   0.002406 \\
10& -17.1724 &-0.937627 &   0.002406 \\
15 &-22.231 &-0.944568  &   0.002425 \\
20 &-26.728 &-0.966014  &   0.002405\\
\hline
1  &-9.89133 &-0.777035  &  0.001879 \\
2 &-10.6125 &-0.813796   &  0.001862 \\
3 &-11.4701 &-0.861863   &  0.001870 \\
4 &-12.3586 &-0.891026   &  0.001825 \\
5 &-13.287 &-0.912954    &  0.001892 \\
10& -18.2126 &-0.934554  &  0.001881 \\
15& -22.9438 &-0.947677  &  0.001864 \\
20& -27.7684 &-0.955044  &  0.001874 \\
        \end{tabular}
\end{table}

\begin{figure}[b]
    \centering
\includegraphics[width=8.3cm]{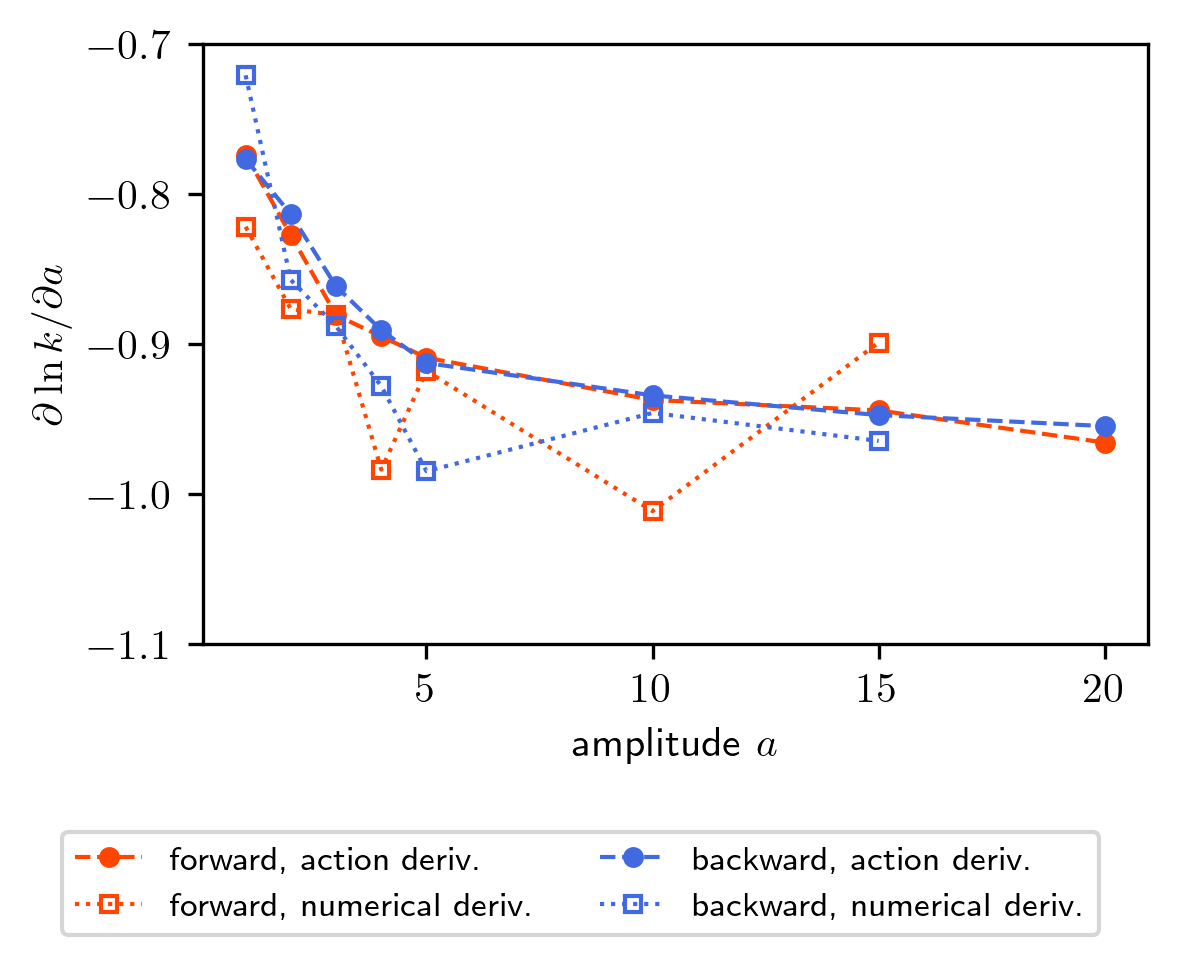}
    \caption{The derivative of the (log) rate constant with respect to
      the amplitude  $a$ of the Gaussian perturbation for the
      diatomic system. The path action based
      derivative is compared to the numerical derivative. }
    \label{fig:ratesdiatom}
    
  \end{figure}

We can now sample the transition over the barrier in this diatomic
system using SRTIS\cite{Du2013}. The stable states are defined as $\lambda_A =1$
and $\lambda_B=3$, while the interfaces
were put at $\lambda= \{
1.30, 1.35, 1.40, 1.45, 1.50, 1.55, 1.60, 1.65, 
1.70, 1.72, 1.74,$ $ 1.76, 1.78,$
$ 1.80, 1.82, 1.84, 
1.86, 1.88,1.90, 1.92, 1.94, 1.96,$ $ 1.98\}$.
The Langevin settings are $\gamma= 2.5$ and $dt=0.001$. Integration is
via the EM algorithm. Sampling 10000 cycles with SRTIS \cite{Du2013}, where each
cycle consisted of 100 shots, 100 interface exchanges, and 100 state
swaps,  resulted in  a path ensemble for each interface. The measured crossing
histograms were joined with WHAM~\cite{Ferrenberg1989},  which together with 
the effective positive flux through
the first interface\cite{vanErp2003,Bolhuis2009,Du2013}  leads to rate constant estimates over the barrier. The
WHAM also allowed to assign a weight to each trajectory in the path
ensemble. Each of these trajectories  can be evaluated in terms of the path action and its derivatives.

The results are given in Table \ref{tab:bistablegauss}. For different
values of $a$ the (log) rate constant $k$ is given, as well as the rate
derivative,
and the flux through the first interface. The rate constant is
here without the flux term, so the true rate constant is  $k$ times
the flux.  The rate constant for the forward (expansion of the dimer)
and reverse (contraction of the dimer) processes are slightly
different, caused by the difference in stability, arising from
the larger  entropy in the expanded state.

Note that the barrier height scales with the parameter $a$. In principle,
the logarithm of the rate constant therefore  should roughly follow $v_{bar}= 10 + a $, in our case. Clearly, this is not
strictly obeyed, even when taking the flux into account. This could be due to 
the used  integrator, but more likely because 
the diffusive barrier crossing is best described by
Kramers' theory, which has a dependency on the curvature of the barrier,
which is increasing with $a$. Hence, we expect the rate constant behave as $\ln k =
c_0 + c_1 a + c_2 \ln a$ with $c_i$ some fit parameters, which indeed seems to be the case. 

We can now also compare the derivative of the rate constant $d \ln k/da $, as computed from the path action averaged over the path ensemble, directly with the numerical derivative of the measured rates. This is shown in figure \ref{fig:ratesdiatom}. 
Clearly the agreement is good, especially considering the 
different origins of the two data sets.  In particular the numerical
derivative is prone to large errors. 
Note that the rate constant derivatives for the forward and backward process are
(almost) equal, because the stable states are not affected by changing
$a$. Thus, the rate constant is affected in exactly by the same factor by the change
in the barrier potential.

    \begin{figure}[b]
    \centering
\includegraphics[width=8cm]{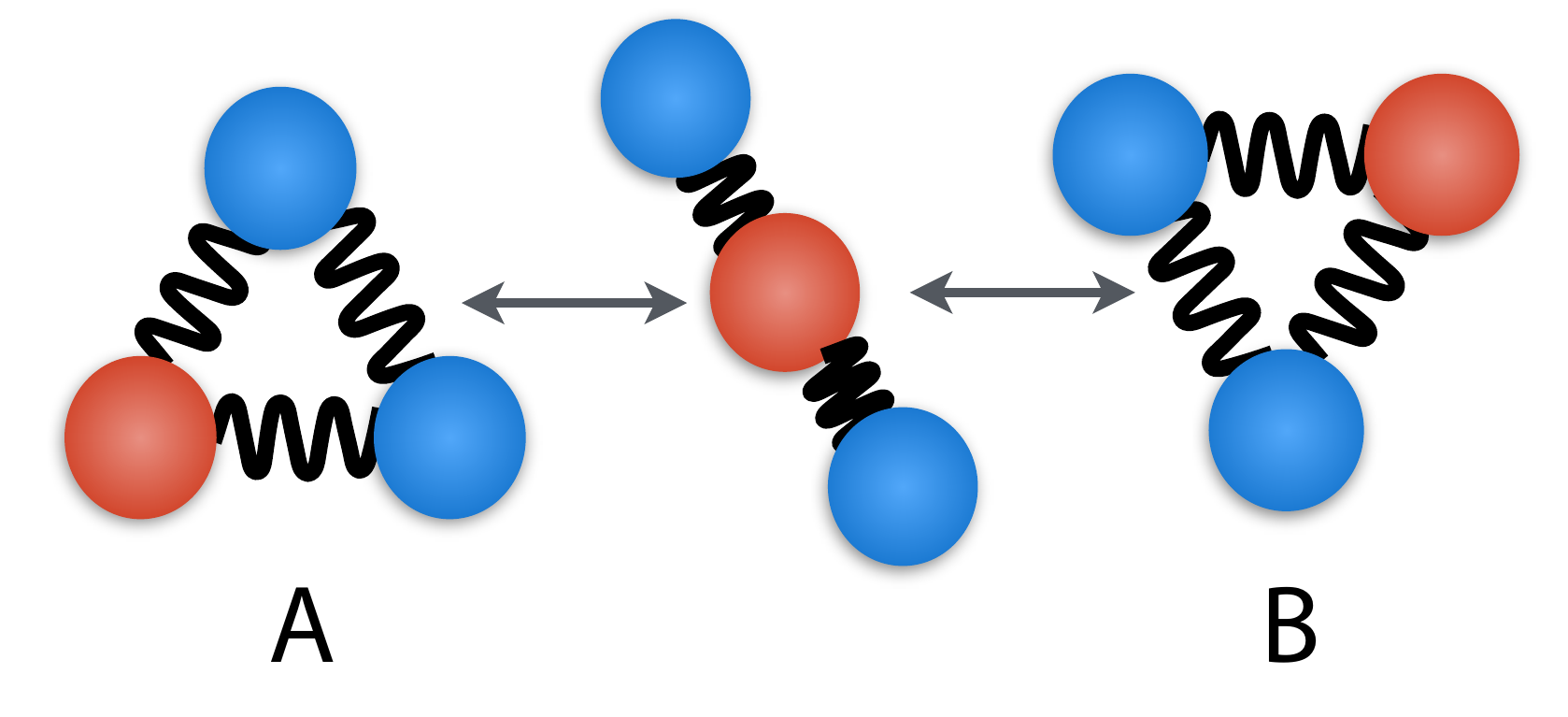}
    \caption{Cartoon of a 2D triatomic system held together by harmonic springs. The trimer can isomerise as indicated in the figure. For  illustrative purposes one particle is colored red. Note that we focus on only one of the 3 possible equivalent transition channels.  \flabel{fig:trimercartoon}}
\end{figure}

\subsubsection{Triatomic system}
    
Next, we consider a 2D triatomic system in which three atoms
interacting with WCA potentials~\cite{Weeks1971} are bound by a harmonic potential that has a minimum at a certain equilibrium bond distance: $V(r) = \frac{1}{2}a (r-r_{eq})^2$. This trimer can undergo an isomerisation transition where one particle passes between the other two, hence changing from a clock wise to anticlockwise arrangement of the (labeled) particles (see \fref{fig:trimercartoon}). Note that this system was also studied in Ref\cite{Dellago2002}.
We look at isomerisation rate constants for this trimer and measure the rate constants and its derivative as function of the  force constant $a$, and the equilibrium distance $r_{eq}$. 
In fact, since we are interested in the changes from a reference system we define the perturbed system as 
\begin{align}
    \tilde{V}(r) = \frac{1}{2}(a +\Delta a)  (r-(r_{eq} + \Delta r_{eq}))^2 .
\end{align}
We can again look at the derivatives to the rate constants $$ \left( \frac{\partial
    \ln k_{AB} }{ \partial   a } \right)_{\Delta a =0},\ $$ and
$$ \left( \frac{\partial \ln k_{AB} }{ \partial   r_{eq}} \right)_{\Delta r_{eq} =0}. $$
Also here the path action derivative does not contain the second term in
Eq.\ref{eq:derwa1}, as we set the perturbation $\Delta a=0$: 
\begin{align}
\label{eq:derwa2}
  \left(\frac{\partial  w[\XP;{a}]}{\partial a} \right)_{ \Delta a =0}\  =
  -\sum_{i=0}^{n-1} \eta_i \cdot \kappa     \frac{\partial  \nabla U(x_i,{a})   }{\partial a_k}.  
\end{align}
\begin{figure}[b]
    \centering
\includegraphics[width=8.3cm]{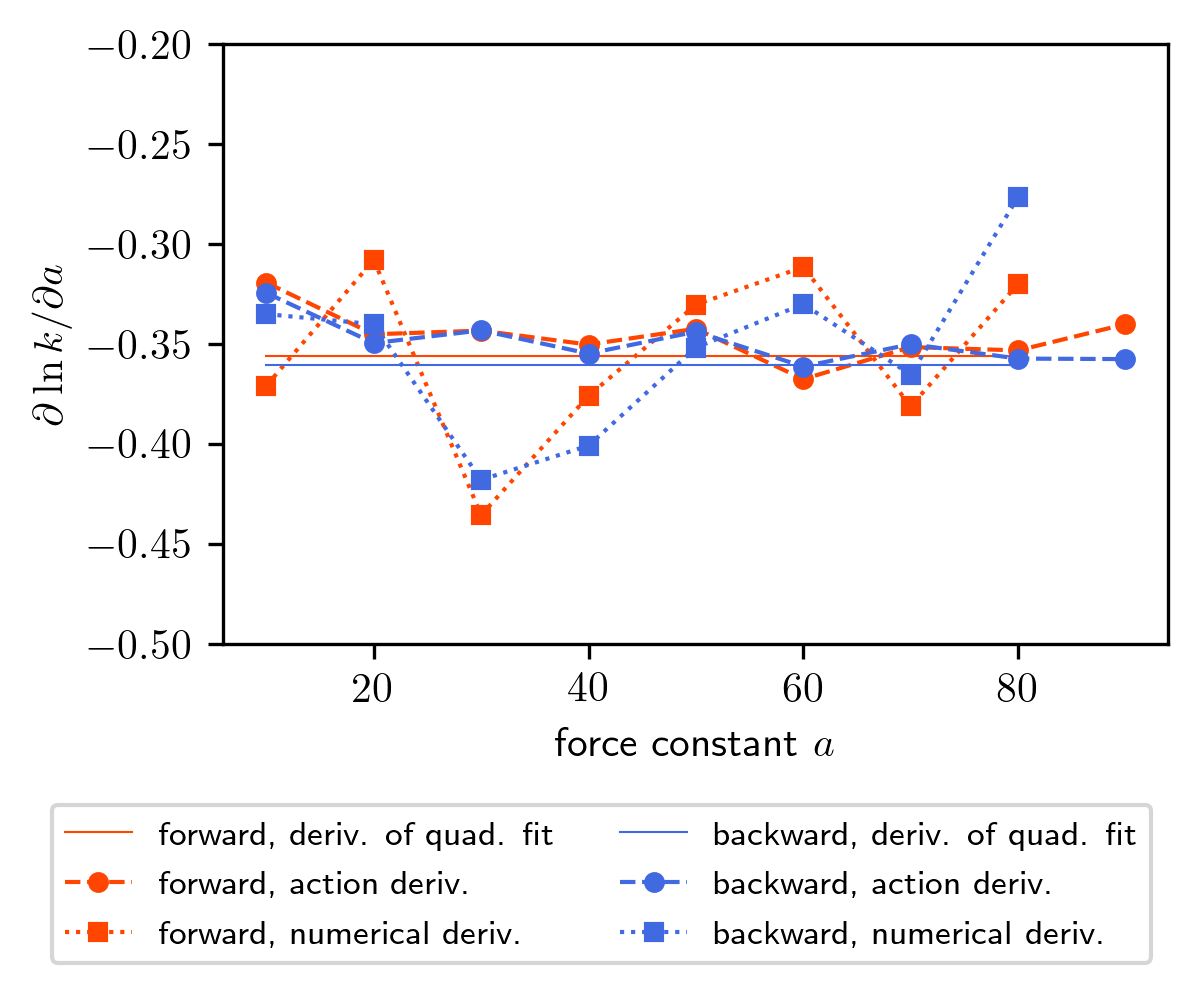}
    \caption{The derivative of the (log) rate constant with respect to
      the force constant  $a$ of the harmonic potential  for the
      diatomic system. The path action based
      derivative is compared to the numerical derivative. 
      The derivative of the (linear) fit to the log rate constant is also shown.  }
    \label{fig:ratestriatom_a}
\end{figure}

    \begin{figure}
    \centering
\includegraphics[width=8.3cm]{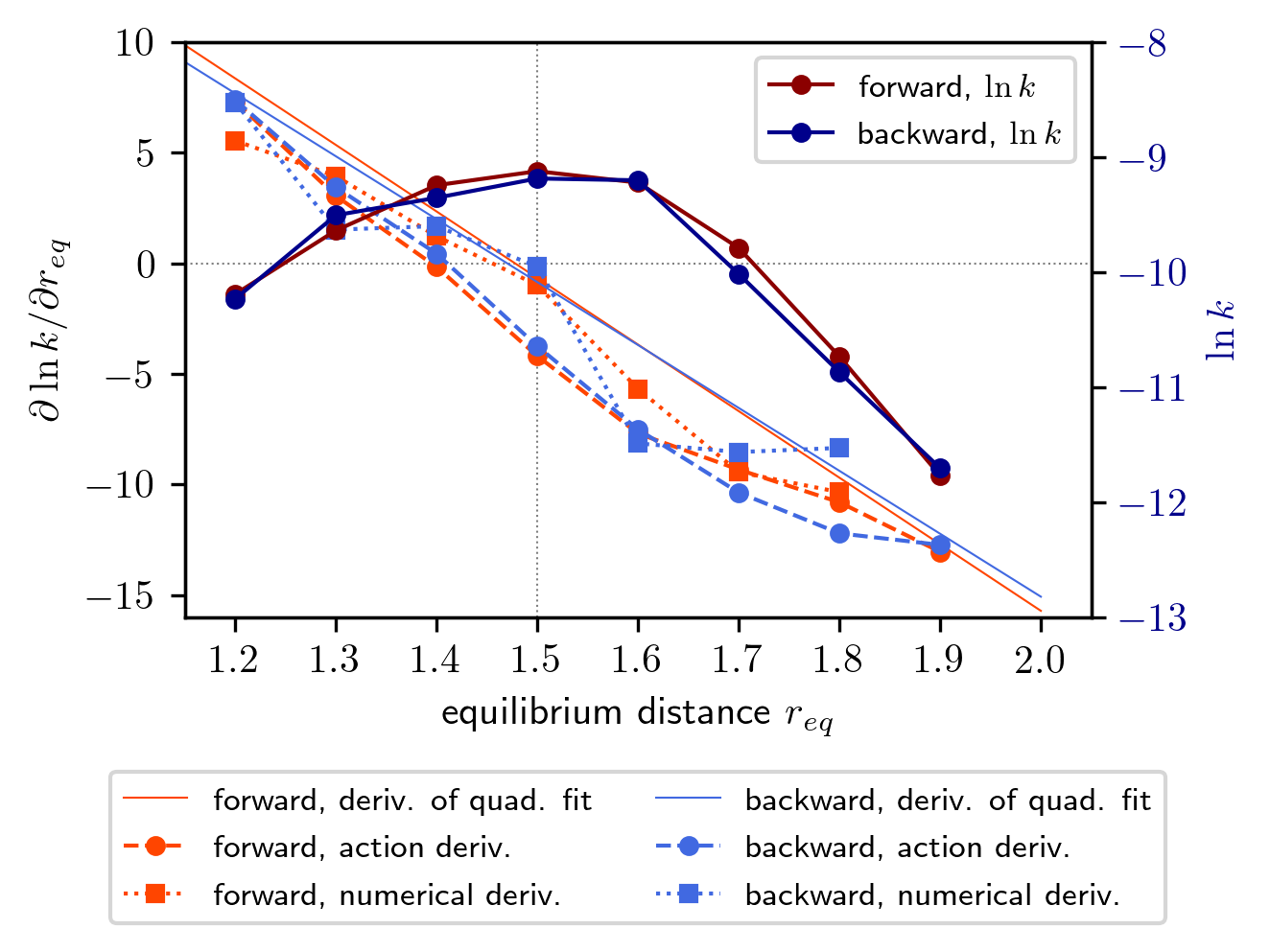}
    \caption{The derivative of the (log) rate constant with respect to
      the equilibrium distance   $r_{eq}$ of the harmonic potential  for the
      diatomic system. The path action based
      derivative is compared to the numerical derivative. Also shown
      is the derivative of the (quadratic) fit to the rate constant.
      The (log)
    rate constant is also included, and shows a maximum around $r_{eq}=1.5$ }
    \label{fig:ratestriatom_r}
\end{figure}

We can sample the transition over the barrier in the triatomic
system using SRTIS. The collective variable used to defined  stable
states and interfaces is the  shortest distance $r_{perp}$ of the hopping particle to the
axis between the two  remaining particles. This distance is negative
or positive depending on which side the hopping particles is located. the CV is then
transformed as
$$
\lambda= \frac{3}{2} \left( \frac{  r_{perp}} {r_{eq}} +1 \right) 
$$
The stable state A is defined as   $\lambda< \lambda_A =0$. The stable
state B is defined from the other side and is also  $\lambda_0
<\lambda_B =0$. Note that since the reference state is different, both
definitions  refer to different states, even if the value of $\lambda$
is equal. Interfaces were defined at  $\lambda = 0.4,
0.45, 0.50, 0.55, 0.60, 0.65, 0.70, 0.75, 0.80,
0.85, 0.90,$ $ 0.95, 1.00, 1.05, 1.10, 1.15, 1.20,
1.25, 1.30, 1.35, 1.40, 1.45,$ $  1.50, 1.55, 1.60,
1.65, 1.70, 1.75, 1.80
$.
The Langevin settings are $\gamma= 2.5$ and $dt=0.001$. Integration is done
via the EM algorithm.
Sampling was done using the same settings as for the diatom system, leading,
after WHAM analysis of the crossing probability histograms to 
the   rate constant estimates over the barrier, as well
as to an ensemble of weighted paths, that can be evaluated in terms of the path action and its derivatives.

The derivatives are shown in Figs.\ref{fig:ratestriatom_a} and
\ref{fig:ratestriatom_r}. The derivative of the (log) rate constant with respect to $a$
is  flat, as expected since $a$ is a prefactor to the
perturbation.  The value of the derivative  is very close to the
analytical value of $-0.375$.
For the derivative with respect to the equilibrium
distance $r_{eq}$ the situation is very different. The rate constant varies in a non-monotonic way. 
At  shorter distance
$r_{eq}$ the particles are forced on top each other and repel each
other again, lowering the rate constant again for low values. At larger distance
the particles have to travel more before they can overcome the barrier,
leading also to higher barriers and lower rates. The maximum rate constant 
translates as a change of sign in the derivative. The values of the action derivative
agree with the rate constant derivatives, although not as good as those for the $a$
parameter, possibly because 
$r_{eq}$ affects the
path ensemble much stronger than $a$ does.  Also the non-linearity of
the rate dependence on $r_{eq}$  can play a role here.

\subsection{Optimisation of the Lagrange function}

Now that the rate constant derivatives are tested, and 
well-predicted by our path reweighting approach, we can turn to  the optimisation of
the force field parameters. Here we focus first on the trimer system, since this has
two parameters to optimize, which both can be tweaked to reproduce the
imposed rate.
To optimise the Lagrange function Eq.\ref{eq:pathLagrangian_reweighted},			
we have to be able to compute the Lagrange function 
as a function of parameters not only at zero perturbation, but also for the reweighted
path ensemble $W[\XP;a]$. Moreover, we need to be able to take its
derivatives at nonzero perturbation.

To do so, we keep track of several variables for each path  that
are appearing in the polynomial  expansion of the path action. 
To be precise, we compute the path action as 
\begin{align}
   & w[a,r;\Delta a, \Delta r]  = \notag\\
   &-\beta  ( \Delta a \, u_{0,da2}  +
                                (a+ \Delta a) \Delta r  \, u_{0,dadr} +
                                \frac{3}{2}  (a+ \Delta a)\Delta r^2 )  \notag
  \\ &+         \Delta a \, \eta f_{da}
	+  (a+ \Delta a) \Delta r \,  \eta f_{dr}
	-      \Delta a^2   f^2_{da^2} \notag  \\
	&-              2 \Delta a  (a+ \Delta a)  \Delta r \, f^2_{dadr}
	-      \Delta r^2 (a+ \Delta a)^2 \, f^2_{dr2}, 
	\label{eq:war}
\end{align}
    \begin{figure}[b]
    \centering
\includegraphics[width=8cm]{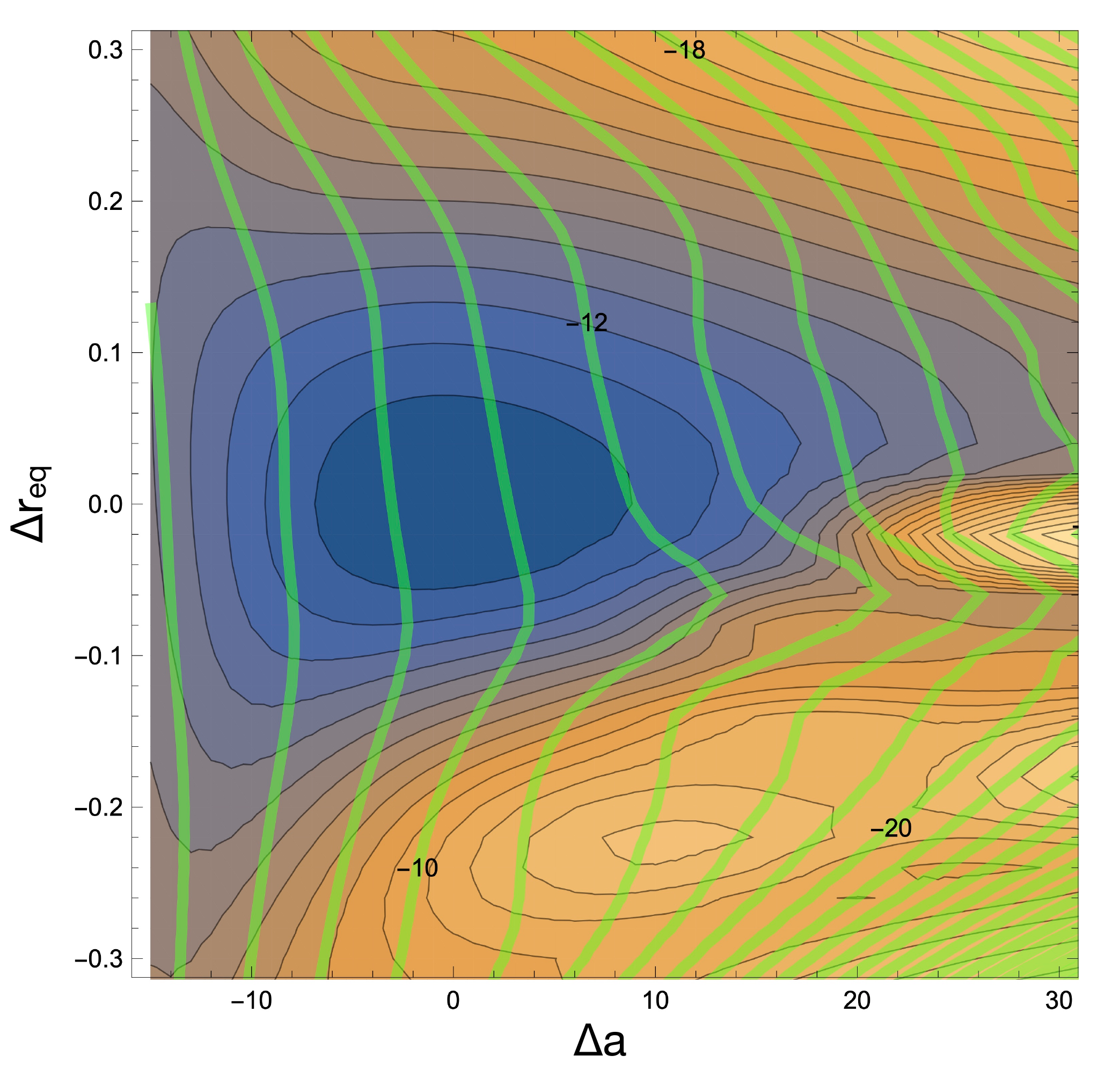}
    \caption{Contour plot of $D_{KL}$ (shaded contour plot) for the
      triatomic system as function of $\Delta r, \Delta a$ for a
      reference path ensemble obtain at $ r_{eq} =1.5, a =20$
      $k_BT$. Note that the minimum value of $D_{KL}$ is at  the origin,
      as expected.
    On top of the shaded  contour plot is a green line contours plot
    of the predicted rate constant $\ln k_{AB}$, for the settings. Several
    numerical values of the contours are indicated.  The
      graphical solution
      to optimisation problem is to pick an imposed green contour and
      minimize the  $D_{KL}$ along this contour.  }
    \label{fig:DKLtriatom}
\end{figure}
where the different $u$ and $\eta f$ terms  refer to the specific contribution to
the energy in the first time slice 0 and the random number force (gradients),
respectively (see Eq.~\ref{eq:derwa1}). The $f^2$  terms refer to the gradient square terms.
Using these quantities it is easy to compute the  derivatives $d w[a,r;\Delta a, \Delta r] /da$ and  $d
w[a,r;\Delta a, \Delta r] /dr$. From these  we compute the path ensemble
averages $\langle w' w' \rangle_ W$,   $\langle w \rangle_ W$  and
$\langle w' \rangle_ W$.
Finally, we construct the reweighted rate constant   $\ln k[a,r;\Delta a, \Delta r]$
$D_{KL}$, and the Langrange function $\mathcal{L}$, from
the above equations.

For a particular set of values of $a =20 $ and $r_{eq}=1.5$
 we computed the path ensemble. In figure \ref{fig:DKLtriatom}, we
 show the $D_{KL}$, and the log rate constant  $\ln k[a,r;\Delta a, \Delta r]$ on top of
 each other as a function of $\Delta a$  and $\Delta r$.
 As expected, the $D_{KL}$ is minimal for the reference value $\Delta
 a=0, \Delta r=0$, that is, at zero perturbation.

The reweighted rate constant  is identical to  the predicted rate constant from the prior path
ensemble for  $\Delta
 a=0, \Delta r=0$, but clearly varies if the system is
 perturbed. Note that $\Delta
 a$ seems to have the strongest effect in changing the rate. In contrast, for a similar change in $D_{KL}$ varying $\Delta r$ also changes the
 rate constant but not as dramatically. Indeed, the $D_{KL}$ is very sensitive to
 the $\Delta r$, already indicating that it is probably better to
 adjust $a$ than $ r_{eq}$.
 
  When we change the rate constant from the observed value of $\ln k_{AB} = -9$ to the
 new value $\ln k^{exp}$ we need to compute the derivatives and
 optimise the Lagrange function. We apply the method of Ref.
 \cite{Platt1988}.
This iterative method starts at certain initial values and slowly
converges to  the solution.
 The  solution for  the most optimal set of $\{ \Delta a, \Delta
 r \}$ is shown in table  \ref{tab:triatomoptimize}.
 In this table we show several sets of optimal solutions, as a function
 of imposed  $k_{AB}^{exp}$. The first set shows the optimal solution for both
 $\Delta a$ and $\Delta r$. The other two sets  we optimised for only one of the two
 parameters, and set the other to zero. From the value of $D_{KL}$   (which is identical to the negative of the Lagrange
 function $\mathcal{L}$ when the constraints are obeyed) it is clear that this is always less optimal than the two
 parameter  solution, thus illustrating the need for multi-parameter force field optimization.
 Also note that the further the imposed $\ln k_{AB^{exp}}$ is from  -9, the
 more $\mathcal{L}$, and thus $D_{KL}$ deviates.

\begin{table}[t]
    \centering
  \caption{Results of the optimization procedure for different
      imposed log rates. The first set allows both parameters to vary;
      the second set varies only $\Delta a$, and the third only $\Delta r$ }
    \label{tab:triatomoptimize}
    \begin{tabular}{rrrrrrr}
    $\ln k_{AB}^{exp}$ &	$\mu$	&$\Delta a$ &	$\Delta r$	&$D_{KL}$ \\
\hline 
-8	&-0.162561	&-3.03452 -	&0.00290183	&0.0831778\\
-12	&0.27589	&8.43988	&0.01461	&0.454862\\
-16	&0.303009	&19.0579	&0.0440781	&1.68525\\
-20	&0.280855	&29.1066	&0.0659962	&2.87969\\
	   \hline			
-8	&-0.164467	&-3.06428	&0	&0.0840814\\
-12	&0.317654	&8.84244	&0	&0.489055\\
-16	&1.00039	&19.8675	&0	&2.98142\\
-20	&0.57278	&28.7734	&0	&6.95817\\
	   \hline			
-12&	0.587054	&0	&0.258986	&4.18148\\
-16	&0.286304	&0	&0.385952	&5.617\\
-20	&0.547232	&0	&0.490967	&6.96569\\
     \end{tabular}
\end{table}

\subsection{Dissociation from a LJ cluster, a model for ligand-protein dissociation}

Having shown that we can apply our framework to model systems for molecular
reaction, we explore in this section  a slightly more elaborate system, which
also has interesting physical properties, namely particle dissociation
 from a cluster of LJ particles. Such a process can be viewed as analogous to
the ligand unbinding, which is an important problem in biophysics\cite{Copeland2006,Tonge2018a}.
Moreover, it can be seen as dissociation for small
nano-clusters\cite{Romano2011,Oh2019,Wang2019b,Zhang2020,Fisher2020,Mitchell2021}.

    \begin{figure}[b]
    \centering
\includegraphics[width=8cm]{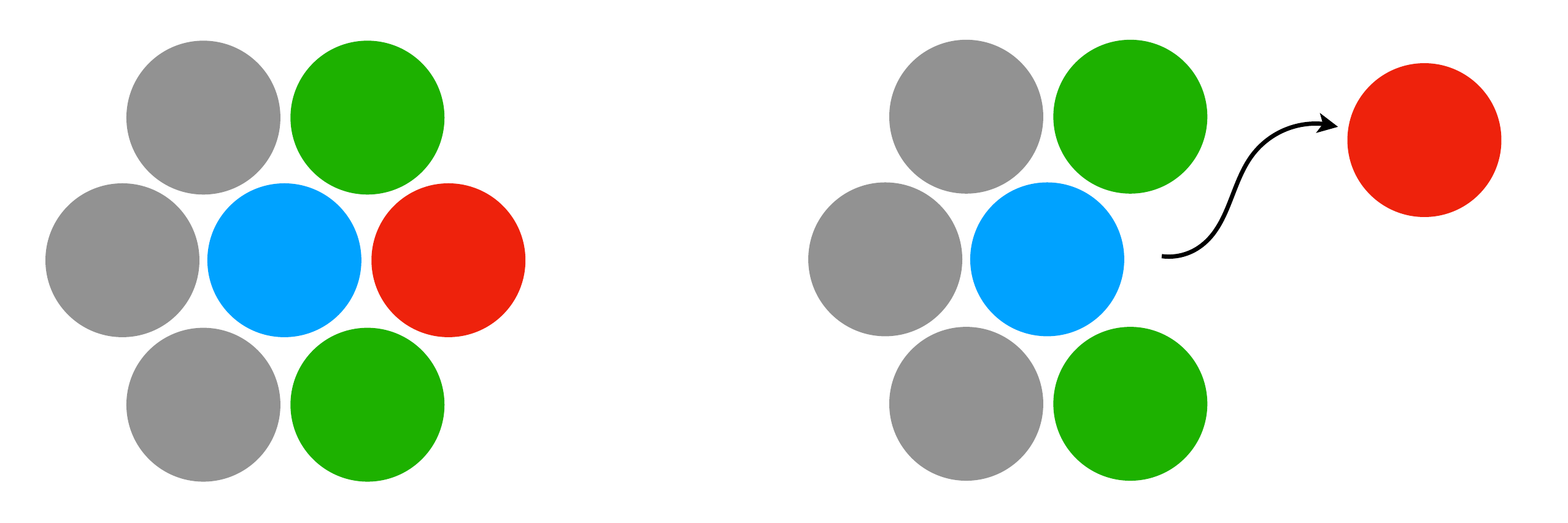}
    \caption{Cartoon of the 2D dissociation model. The red particle, originally bound to the central (blue) and outer (green) particles, can escape into the bulk.  \flabel{fig:cartoon2Dunb}}
\end{figure}

\subsubsection{Model}

We describe the dissociation transition using a model of Lennard-Jones (LJ)-like
particles. We first consider a 7 particle setup in two dimensions, indicated in
\fref{fig:cartoon2Dunb}. In the simplest instance of this model all particles are kept fixed except the 
red  particle, which can dissociate into the bulk. This
 red particle  interacts with both the central blue particle and
 the green 
 particles via an attractive LJ-like  interaction potentials with 
 adjustable depths.  The interaction with the gray particles is
purely  repulsive 
and given by a standard repulsive WCA potential~\cite{Weeks1971}.
Setting the particle diameter as the unit of length $\sigma=1$, and denoting the red particle with index 0, the total energy is  thus
\begin{align}
V_{tot} = V_{\epsilon_1} (r_{0c}) + \sum_{j\in green} V_{\epsilon_2}(r_{0j})  +
  \sum_{j\in gray}V_{wca}(r_{0j}),  
\end{align}
where the (adjustable) potentials  $V_{\epsilon} (r)$, in units of
$k_BT$, are given by
\begin{align}
  &\beta V_{\epsilon} (r) = \begin{cases}
      4 \epsilon_0 (r^{-12} - r^{-6})    +v_{\epsilon} \,\, &   r< 2^{1/6}\\
      4 \epsilon (r^{-12} - r^{-6}  + v_{s})\,\, & 2^{1/6} < r
      < r_c \\
       0  \,\, &  r>r_c
    \end{cases}
\end{align}
Here, the  constants $v_s$ and  $v_{\epsilon}$ shift the potential such
that the potential is zero at the cutoff $r_c =1.5 $, and
continuous  at
the minimum  $r=2^{1/6}$. $\epsilon$ is the (reduced/dimensionless) depth of the potential, while $\epsilon_0 =1$ is the standard reference value for the potential. 
For this particular cutoff it follows  $ v_{s}
=r_c^{-6} - r_c^{-12} =  0.0800841$, and  $ v_{\epsilon}  =  (1-\epsilon  )  +   4 \epsilon   v_s$;

The switching at the minimum of the potential is done to avoid
problems with the rather steep 
repulsive part of the potential arising  for the high values of $\epsilon_1$ required  to bind the particle. 
Therefore, we chose to keep the repulsive part
of the potential  equal to the standard WCA potential, i.e. not scale
with $\epsilon$.
In this way, only when the particles are in the
attractive part of the potential they contribute to the path
action derivatives, making evaluation of the path action more robust. Note that this
does not change the generality of our approach. 
 
 \begin{figure}[t]
    \centering
\includegraphics[width=3.7cm]{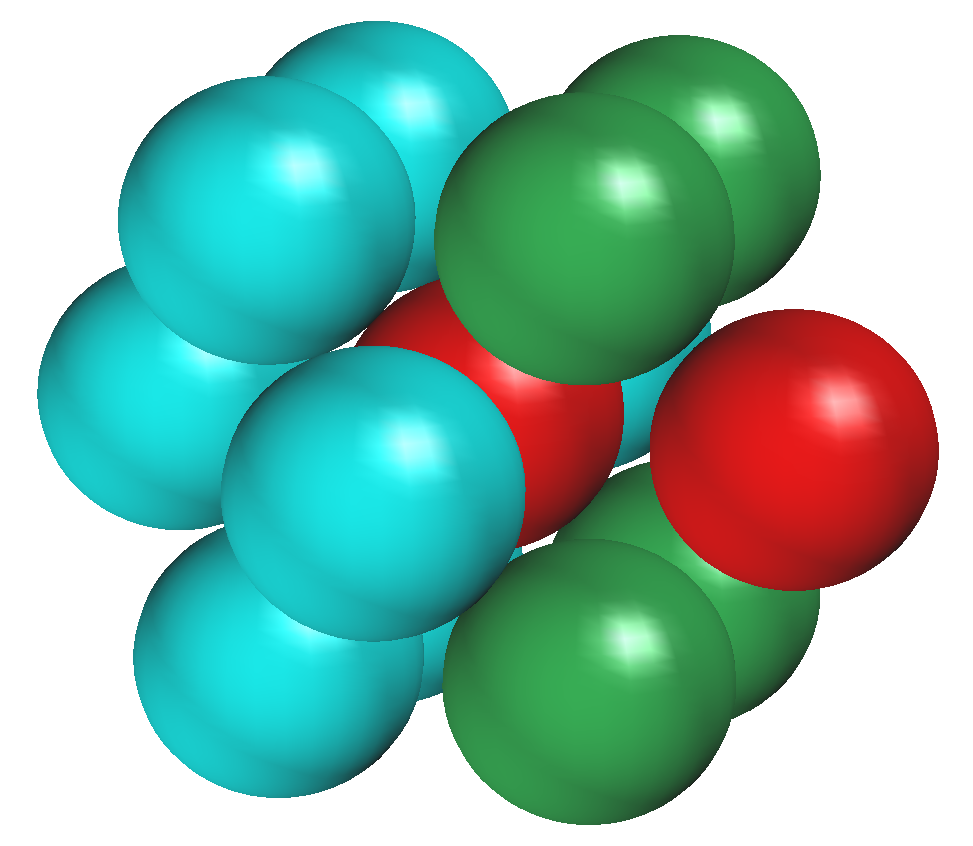}
    \caption{Cartoon of the 3D dissociation model. The escaping outer red particle is initially bound to the central particle (also red) and 4 outer green particles.  \flabel{fig:cartoon3Dclust}}
\end{figure}

In the first model, we keep the all  particles fixed except the red
dissociating particle, so the
only important interactions are that of the red particle with the 6
other particles.
 In the second model, we allow the other particles to move as well. To keep
the cluster together, we apply an additional potential that binds the
non-red particles  to the central blue particle by an additional strong LJ
interaction $V_{\epsilon_3} (r)$ , with $\epsilon_3=20$. While this
keeps the cluster intact, rearrangements are still possible. We  avoid
these by imposing an additional 
weak  harmonic spring between neighbouring  particles with a spring
constant $k=1$ (this of course excludes the red particle).

The result is a fluctuating cluster of 6 particles, that can expel the
red particle. During the dissociation, the green particles can move
closer to each other, gaining in entropy.
We can interpret this simple model as representing a ligand unbinding
reaction, e.g. of  a protein,  in which the protein binding pocket
slightly rearranges upon (un)binding.

Finally, in order to show that our methodology  easily extends to 3D systems,  we consider a 3D version of model with 13 LJ particles as
depicted in \fref{fig:cartoon3Dclust}. Here 4 outer green particles
bind the ligand  with an attraction $\epsilon_2$.

    \begin{figure*}[t]
    \centering
    \includegraphics[width=\textwidth]{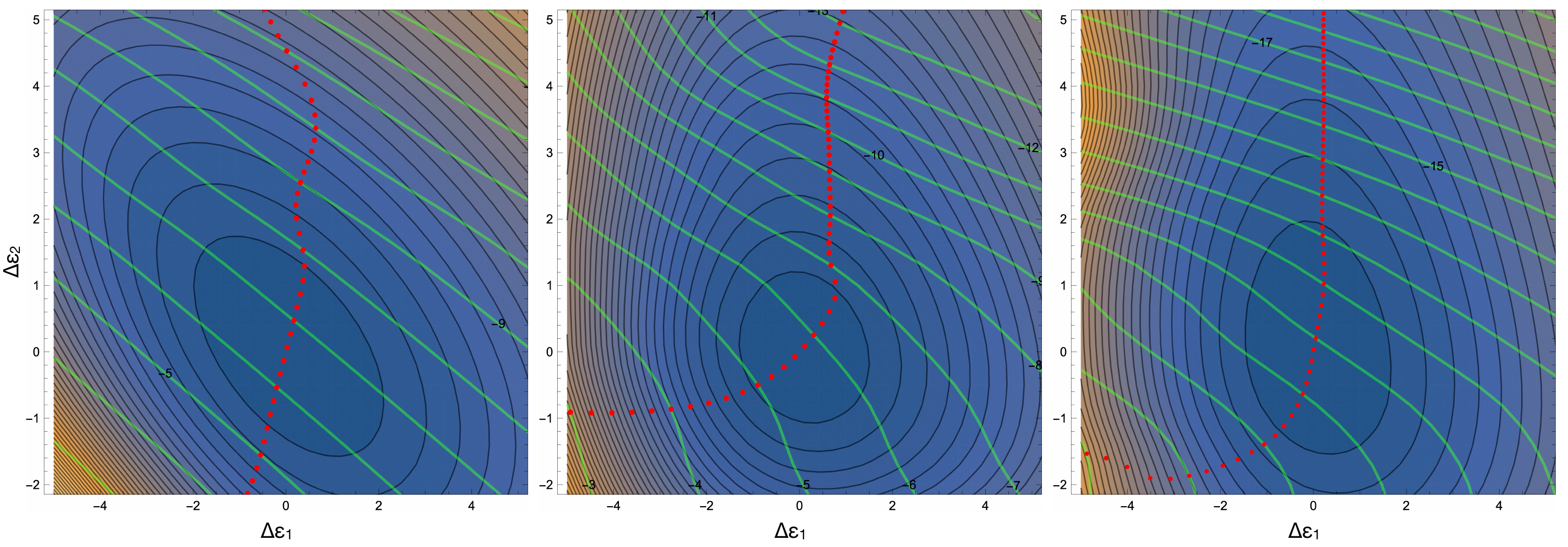}
    \caption{Contour plot of $D_{KL}$ (shaded contour plot) for
      the unbinding transition in the
      7-particle  system as function of $\Delta \epsilon_1, \Delta \epsilon_2$ for a
      reference path ensemble. 
      Note that the minimum value of $D_{KL}$ is at  the origin.
    On top of the shaded  contour plot is a green line contours plot
    of the predicted rate constant $\ln k_{AB}$, for the settings. Several
    numerical values of the contours are indicated.  The
      graphical solution
      to optimisation problem is to pick an imposed green contour and
      minimize the  $D_{KL}$ along this contour. The red points depict
      this optimal solution.  A) fixed cluster, reference path ensemble 
      obtained at $\epsilon_1   = 10, \epsilon_2
      = 1.0       k_BT$.
B) flexible cluster, reference path ensemble. 
      obtained at $ \epsilon_1   = 12 , \epsilon_2
      = 2.0       k_BT$.      
C) flexible cluster, reference path ensemble. 
      obtained at $ \epsilon_1   = 15, \epsilon_2
      = 5.0       k_BT$.     
    \flabel{fig:DKLunb2D}}
\end{figure*}

\subsubsection{Path ensembles}

We first start exploring the fixed 7-particle model.
Setting the central particle interaction to $\epsilon_1=10$, and
the green particle interaction to  $\epsilon_2=1$, we sample unbinding
transitions using SRTIS. The order parameter $\lambda$ used is the center-to-center distance between the red and blue particles. Stable states
were defined as $r<2^{1/6}$ and $r>2.5$ for the initial and final states
respectively (Note that beyond $r=2.5$ the ligand cannot yet be considered escaped to the bulk. While, it is possible to take this  into account see e.g. Ref\cite{Vijaykumar2016,Vijaykumar2017},  we assume here for simplicity that the ligand is dissociated).
Interfaces were positioned at $
\lambda= 0.15, 
0.15$, 
$0.20, 
0.25, 
0.30, 
0.35, 
0.40, 
0.50, 
0.60, 
0.70, 
0.80, 
1.00$, with respect to the minimum distance $r_A=2^{1/6}$.
We integrate the equations of motion using the EM integrator with a
time step of $dt=0.001$ and a friction of $2.5$. 
In total we perform $10^5$ shooting and replica exchange moves. Acceptance
ratios for the shooting move ranges  from 0.4  for the first interface to
0.15  for the last interface. Replica exchange moves where accepted around
50\%.
Path lengths vary from  50 timesteps for the first interface to  a few
thousand for the last interface.  
The crossing probability of the final interface (obtained from WHAM) is  $\ln P(\lambda_B
|\lambda_1)= -6.567$. The flux of the first interface is $\phi = 0.003550 $
 The total rate constant is  thus $k_{AB} = 5\times 10^{-6}$ per in unit of
 time step.
   Note that,  when optimising the rate  constant, we assume that the fluxes
  are not altered much (as above), and we only have to consider the crossing probability.

We performed also runs for the flexible cluster. The simulation for the flexible 2D cluster is similar to the fixed case, but the results will be very different, as shown below.

  We can now use our framework to look for the best set of new
  parameters $\epsilon_1' = \epsilon_1 + \Delta \epsilon_1$ and
  $\epsilon_2' = \epsilon_2 + \Delta \epsilon_2$.
   The path action is  given by 
\begin{align}
&    w[\epsilon_1,\epsilon_2,;\Delta \epsilon_1, \Delta \epsilon_2]  =
  -\beta  ( \Delta \epsilon_1 \, u_{0,\epsilon_1}  +\Delta \epsilon_2
  \, u_{0, \epsilon_2}  ) \notag  \\                     
   &+         \Delta \epsilon_1 \, \eta f_{\Delta \epsilon_1}
       +         \Delta \epsilon_2 \, \eta f_{\Delta \epsilon_2 } \notag\\
      & -      \Delta \epsilon_1^2   f^2_{\Delta \epsilon_1^2 }
       -      \Delta \epsilon_1  \Delta \epsilon_2   f^2_{\Delta \epsilon_1 \Delta \epsilon_2 }
    -      \Delta \epsilon_2^2   f^2_{\Delta \epsilon_2^2  de2^2} 
\end{align}
where the $u_0$, $\eta f$-functions and $f$-functions involve the potential energy of the first slice, and the gradient of the potential energy, c.f.~Eq.\ref{eq:wa} and Eq.\ref{eq:war}.
Note that as mentioned above only the attractive part of the potential has to be taken
into account.

\subsubsection{Caliber/$D_{KL}$ and rate constant predictions}

Our framework then gives the rate constant predictions for the altered parameters, as
well as the caliber or  $D_{KL}$.     \fref{fig:DKLunb2D}  
  presents  both
predictions as
contour plots. The green contours delimit   the (log of the)  dissociation rate
constant (in fact, the crossing probability)
predictions, while the blue-ochre contours depict the $D_{KL}$.  
Several conclusions follow  from this figure.
The first is that for the fixed cluster the $D_{KL}$ contours show some anti-correlation in the two parameters. This indicates that the path ensemble is least disturbed when  a increase in $\epsilon_2$ is compensated by a  decrease of $\epsilon_1$.
The $D_{KL}$ contours also are slightly asymmetric, showing a larger sensitivity to negative values of 
 $\epsilon_2$. As this  parameter
is set to the relatively low value of  $\epsilon_2=1$, reduction below $\Delta
\epsilon_2<-1$, will therefore  reverse the sign of the attractive interaction,
which of course completely alters the systems. 
The flexible cluster \fref{fig:DKLunb2D}B  shows no such anti-correlation in $D_{KL}$, indicating that the compensating effect has largely disappeared.

The second observation is that  the green rate constant contours are roughly linear with  a negative slope 
This indicates that both increasing $\epsilon_1$ and
$\epsilon_1$ have similar effects for a large variety of values.
So, to have a similar
decrease in rate constant
one could choose  either to increase $\epsilon_1$ or  $\epsilon_2$.
Note that the slope of the contour is roughly  $-0.5$, as there are two outer particles
(green) and only 1 central particle, so
changing their interaction strengths $\epsilon_2$  has therefore twice the effect.
This observation indicates that $\epsilon_1$ and
$\epsilon_2$ are more or less interchangeable, and one can choose many 
combinations for arriving at the same rate constant.

In case of a flexible cluster, 
as shown in \fref{fig:DKLunb2D}B,  the predicted rate constant contours  become nonlinear.
In \fref{fig:DKLunb2D}C we show a case where the outer and central
particles have equal attraction strength.

\subsubsection{Optimal parameters for target rate constants}

Next, we would like to find the optimal choice for the force field
parameters $\epsilon_{1,2}$. As before, optimising the parameters for  a  given rate constant amounts to
following a green rate constant contour until the  $D_{KL}$  is minimal.
Using the optimisation procedure outlined in the previous section, we
arrive at a prediction given by the red points in     \fref{fig:DKLunb2D}. This prediction corresponds thus to
the most optimal force field parameters, which minimises the change in the
path ensemble with respect to the original force field. 

In all cases in \fref{fig:DKLunb2D}, a trivial 
observation is that the curve passes through the origin, as there the
original force field reproduces the original rate constant most optimally.
For the fixed cluster  \fref{fig:DKLunb2D}A, the optimal curve is roughly vertical, with only a relatively  small deviation in $\epsilon_1$. This means that that whether enhancing or reducing the rate, it is always better to change $\epsilon_2$   rather than $\epsilon_1$. Remarkably, this trend even holds for $\Delta \epsilon_2 <-1$, which changes the interaction from attraction to repulsion. This is likely a consequence of the immobility of the particles in this case.

\begin{figure}[b]
 \includegraphics[width=8.4cm]{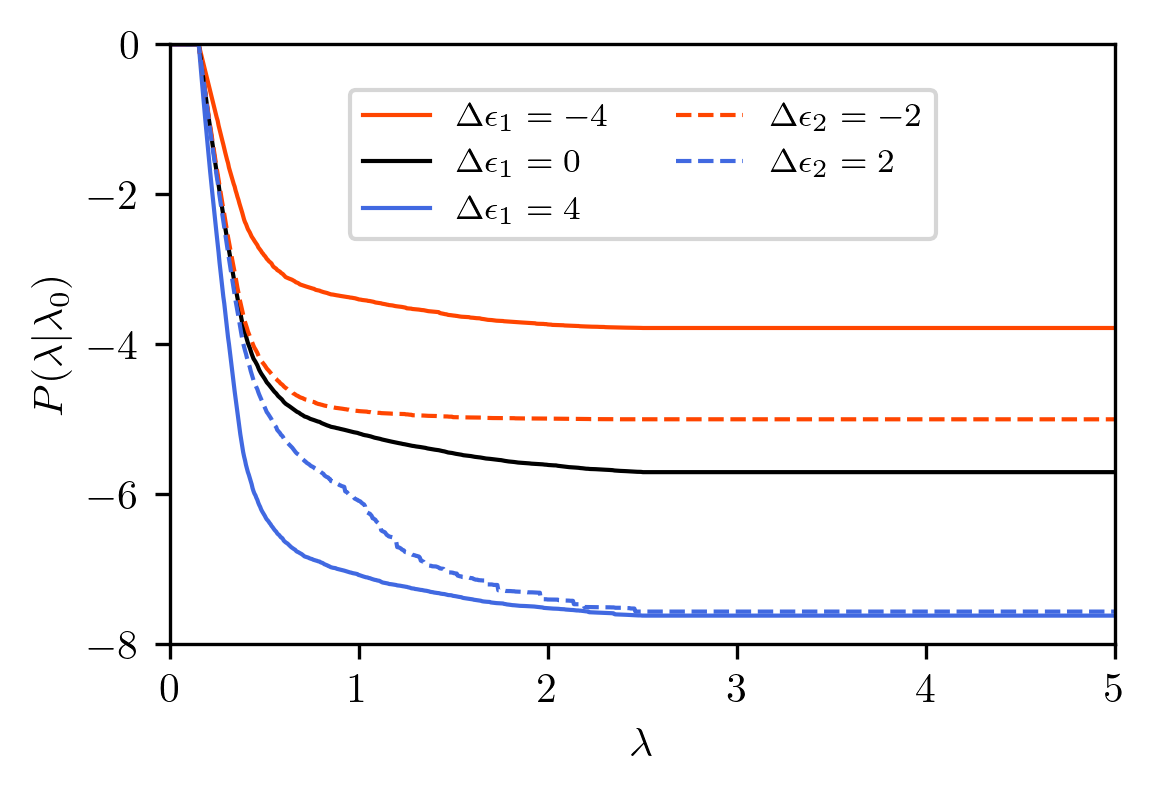}
\caption{Reweighted crossing probabilities for the flexible 2D cluster
      obtained at $ \epsilon_1   = 12 , \epsilon_2
      = 2.0        k_BT$. The  black solid line is the reference crossing probability. The red and blue  curves are reweighed crossing probabilities changing either $\epsilon_1 $ (solid)  or $\epsilon_2 $ (dashed) . Clearly  changing  $\epsilon_1$  has most effect on the beginning of the crossing probability curves, while changing $\epsilon_2$ in the positive direction has more influence on the latter part, also compared to changing in the negative direction, thus explaining the asymmetry in the binding problem.  
     \label{fig:reweighted_tcp}}
\end{figure}

\begin{figure*}[t]
    \centering
\includegraphics[width=\textwidth]{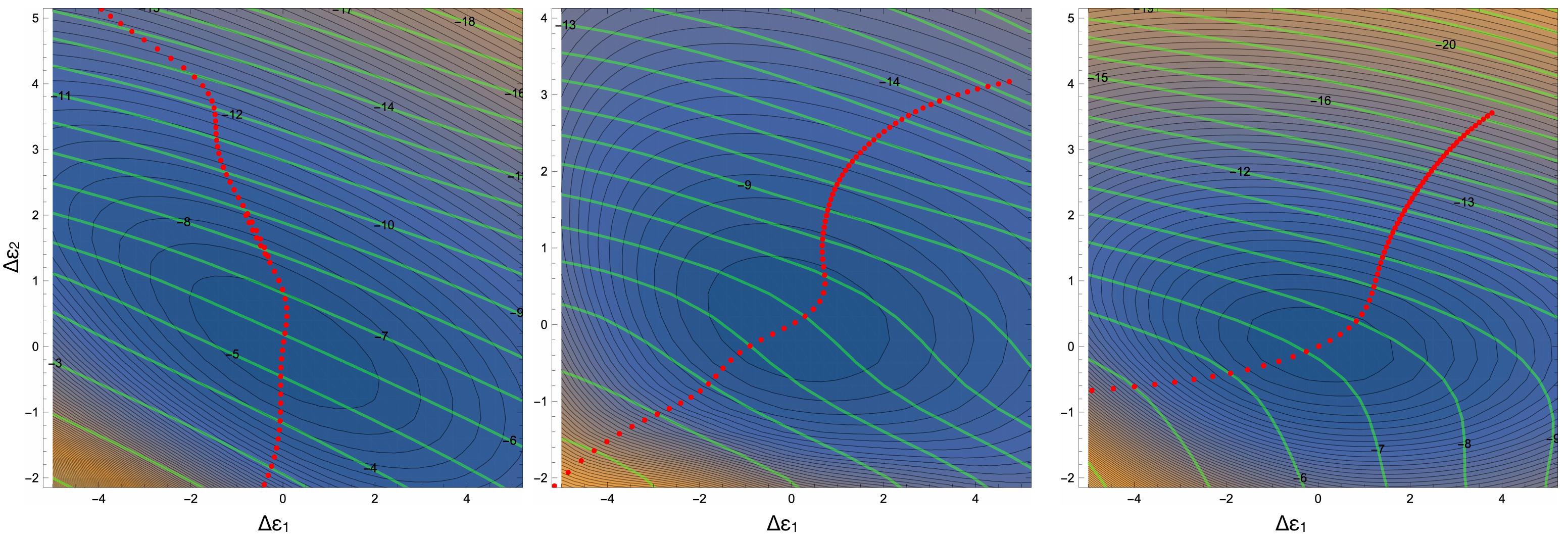}
    \caption{Contour plot of $D_{KL}$ (shaded contour plot) for
      the unbinding transition in the
     3D  13-particle  system as function of $\Delta \epsilon_1, \Delta \epsilon_2$ for a
      reference path ensemble. 
      Note that the minimum value of $D_{KL}$ is at  the origin.
    On top of the shaded  contour plot is a green line contours plot
    of the predicted rate $\ln k_{AB}$, for the settings. Several
    numerical values of the contours are indicated.  The
      graphical solution
      to optimisation problem is to pick an imposed green contour and
      minimize the  $D_{KL}$ along this contour. The red points depict
      this optimal solution. 
A) fixed cluster, reference path ensemble. 
      obtained at $\epsilon_1   = 12, \epsilon_2
      = 1.0        k_BT$.
B) flexible cluster, reference path ensemble. 
      obtained at $ \epsilon_1   = 12 , \epsilon_2
      = 2.0        k_BT$.      
C) flexible cluster, reference path ensemble. 
      obtained at $ \epsilon_1   = 20, \epsilon_2
      = 1.0        k_BT$.
    \flabel{fig:DKLunb3D}}
\end{figure*}

For the flexible cluster cases in \fref{fig:DKLunb2D}B,C, the most striking feature is perhaps the L-shaped curve, indicating a asymmetry concerning  reducing the  rate
or enhancing the rate. For enhancing of the rate,  e.g. by one order  natural log  unit, the red curve shows roughly a linear behavior, that is,
$\epsilon_{1,2}$ are contributing to the rate constant in equal
proportions. However, when reducing 
the rate constant further, the curve bends over more vertically, indicating that it
is better to change  $\epsilon_2$ instead of $\epsilon_1$. 
In contrast, when increasing the rate, the curve bends over horizontally, indicating that it is now better to change  $\epsilon_1$ instead of $\epsilon_2$. 
This conclusion  also holds for \fref{fig:DKLunb2D}C, where $\epsilon_1=15$, and $\epsilon_2=5$.

We can interpret this behavior as follows. The best parameters are
those that perturb the path ensemble as little as possible. When
reducing the rate constant, it is better  to adjust the $\epsilon_2$  parameter
than the $\epsilon_1$ parameter, even if they both can lead to the
same rate constant predictions. This can be interpreted by  realising that changing 
the central particle  interaction   $\epsilon_1$
will alter the entire reweighted path ensemble: all interface
ensembles will be affected. In contrast, change of $\epsilon_2$  will
mostly affect only the interface ensembles further out. Since distant
interfaces have a (much) lower weight in the ensemble, the
perturbation, as measured by the  $D_{KL}$ /caliber will be smaller.
In contrast, for increasing the rate, changing  $\epsilon_2$ will not get you very far, and substantial change of $\epsilon_1$ is also necessary. These effects can be shown by plotting in the total crossing probability for the different settings, in Figure \ref{fig:reweighted_tcp}. Here it is clear that when looking e.g. at the curves for $\Delta \epsilon_1 =4$ and $\Delta \epsilon_2 =2$ the final rate constant predictions (crossing probs) are almost equal, but the intermediate crossing probability, and hence the path ensembles, are  very different. Clearly, the $\Delta \epsilon_2 =2$ case is much closer to the original data set (black solid line), especially in the beginning of the crossing probability where the path ensemble is  most dominant.

In \fref{fig:DKLunb3D} we plot the results for the 3D systems. They
are remarkably similar to the 2D systems, showing the robustness of the
results.
One striking difference is the slope of the green rate constant contours,
which is now -0.25  (for the flexible 3D case), 
as we now have 4 outer particles. So a change in $\epsilon_{2}$ has
a 4-fold effect on the rate constant. Also, the $D_{KL}$ contours appear different, and a bit
more skewed compared  to  the more circular ones in the 2D case.
Remarkably, the optimal solution for the parameters (the red curves) look
again qualitatively similar to those  of the 2D cases.
Only for very strong interaction $\Delta \epsilon_2 >2$ of the outer particles, shown in
\fref{fig:DKLunb3D}B,C the red curve  bends over.

\subsubsection{Comparison with rate constant predictions}

While the prediction of the optimal parameters is already providing valuable
insight, the ultimate goal is to establish a better force field model.
To assess the quality of the predictions, 
we can compare  the predicted rate constants with a independent calculations at
these different force field parameters. \fref{fig:2Dunbcompare} shows
this comparison for the flexible 2D unbinding.  The agreement is good, especially up to two $k_BT$ from the reference point ($ \epsilon_1   = 12 , \epsilon_2      = 2.0 $).

\subsubsection{Physical insight from the optimization}

These results also reveal several physical aspects:
The central particle is more  important for increasing the rate,
whereas  outer particles are more important for decreasing the rate.
This is explained by the fact that the entire path ensemble is mostly influenced by the central particle's interaction change, while 
outer particles only affect the barrier region. 
Outer particles therefore are prime targets for modulating when reducing the
dissociation rate constants. Translating to real proteins this amounts
to engineering mutations or post-translational modifications ~\cite{Winter2020} of binding pocket residues close to the surface or modulation of the ligand chemistry in order to bind better to encounter complex sites at the surface.
Of course, this extrapolation to realistic systems is currently no more than a hypothesis, that requires further testing. Yet, the general principle is likely to be robust.

    \begin{figure}[t]
    \centering
    \includegraphics[width=8.3cm]{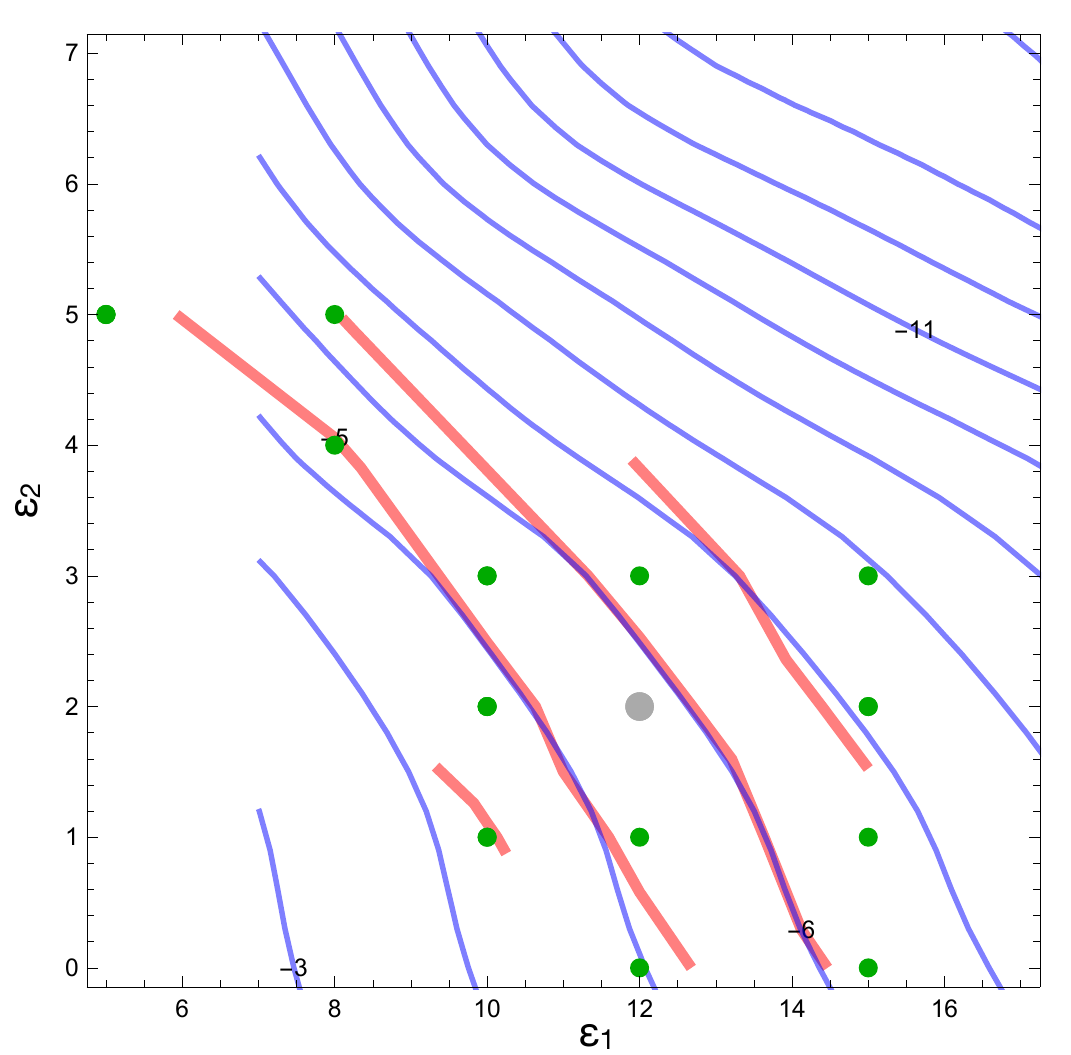}
    \caption{Contour plot of rate constant predictions (blue) based on the
      reference ensemble of  the flexible 2D  cluster
      obtained at $ \epsilon_1   = 12 , \epsilon_2
      = 2.0        k_BT$ (grey point), compared with red contours interpolated from the true rate constants computed at the indicated green points. 
    \flabel{fig:2Dunbcompare}}
\end{figure}

\section{Conclusion}

In this paper we have introduced a MaxCal based method to optimise force field
parameters in order to impose an dynamical constraint, in particular a rate constant in a complex molecular transition.
Without any computationally expensive recalculation of the kinetics, the method yields the optimal change in the parameters that leads to an imposed rate constant after  the path reweighting, while making the least possible
perturbation to the prior trajectory ensemble as measured by the caliber or KL divergence.  

We show  that the path reweighting leads to meaningful prediction of the rate constants, which agrees with  direct calculation for the new  force field,  even up to more than an order of magnitude suppression of the rate constant in case of the unbinding model. While in this work we develop the methodology and  applied it only to simple models, we expect the method to be generally applicable.

Besides a corrected force field, we find that the optimization for multiple parameters allows an interpretation that gives physical insight in which parts of the path ensemble are affected by the parameter changes.
Thus, the method provides a powerful tool to inspect rare event trajectory ensembles, and extract valuable information from them.
Trajectory ensembles provide us not
only with mechanisms, rate constants and transition states for rare event
dynamics, but they can also inform on  how 
such properties change with the model
parameters. Importantly, they can predict the change in rate
constants, and moreover, point us in the direction of parameters that
least affect the original path ensembles. This gives the possibility  to extend the optimization of force fields that  are already optimised for
thermodynamics to kinetics.  
And moreover, gives us pointers to what parts of the systems are most sensitive and thus most sensible to adapt or mutate. 

Of course several directions for further research can be considered.

First of all, we have used only the simplest of path actions, i.e. the Onsager-Machlup action for the EM integrator. It is known that EM is not a great integrator, and an obvious extension of the methods would be to extend it to underdamped Langevin integrators \cite{Kieninger2021}.

Second, we only have considered up to two parameters. We envision that extension to many parameters is in principle straightforward, but some bookkeeping issues arise, e.g.  keeping track of all the cross terms in the square gradient.  Nevertheless, the method should be equally effective for optimization in higher dimensions. 

Third, the method can and should be applied in a realistic force field, and included in a MD engine. This is non-trivial, since the evaluation of the path action should be done in the integrator.  Nevertheless, we believe that this is a worthwhile  and necessary research direction, to make our approach useful. 
One of the issues is the functional form of the potential perturbation. To start, one could try simple functional forms, such as dihedral terms, and Lennard-Jones-like interactions. In a more advanced exploration one could try to use machine learning to learn the functional form of the perturbation. 

Fourth, the methods can be pushed beyond the standard atomistic force fields into coarse-grained force fields, for which rate constants are notoriously difficult to reproduce. Such an approach would go beyond a simple adjustment of the diffusion constant, or friction.

Fifth, the method should be also applicable to other dynamical observable, such as diffusion and viscosity. In fact, all observables based on time-correlation functions of a trajectory ensemble should be treatable.  

Sixth, parameters beyond the force field might be optimized, including the diffusion, effective mass, and even molecular topology. 
The holy grail in chemistry, material science and biology is to accurately determine the link between microscopic degrees of freedom such as the chemistry, the microscopic mechanisms, transition rates, structure populations with macroscopic thermodynamic or kinetic properties related to function. Having an efficient algorithm that is able to provide the link between kinetic/thermodynamic macroscopic properties and force field parameters reporting on the underlying physics could be potentially very useful in protein engineering   and design of  small molecules  and of material  with tunable macroscopic properties~\cite{Copeland2006,Tonge2018a,Romano2011,Oh2019,Wang2019b,Zhang2020,Fisher2020,Mitchell2021}. 
An example of such a design application,
are the coarse-grained molecular force fields where sole parameters are the identity of the aminoacid, which are a.o.~employed in  protein aggregation and protein folding~\cite{Dignon2018,Best2016}. For such problems one could apply our approach to suggest changes in the aminoacid sequence in the direction of promoting or depressing the rate constant of e.g aggregation, protein folding, or liquid/liquid phase separation. In this way one could optimize the design of novel (bio)material to be further tested in the wet lab, thus bypassing expensive and timely optimization protocols in the wet lab.

Moreover, this method could be particularly useful in chemical reactions, by correcting the kinetics of imperfect while computationally cheap calculations by reactive force fields~\cite{Senftle2016} instead of doing calculations with accurate albeit computationally expensive density functional theory potentials. Another utility is the generality of the method, giving the possibility to combine it with deep learning potentials that are known to be very flexible and have already been used when calibrating force fields for thermodynamics in material science~\cite{Zhang2018,rogal2019}.

Finally, our  methodology of constrained-optimization based on the Maximum-Entropy is not limited to molecular systems and could even potentially contribute in various problems of time-series, where the underlying dynamics can be modelled as stochastic, undergoing rare events and with a potential energy function reporting on interactions of agents, such as in modeling of stock-market, fluid-flows and  health-pandemics. ~\cite{Bouchaud2000,Haworth1986,Jones2020}.

\section*{Acknowledgements}

The authors thank Christoph Dellago and Pieter Rein ten Wolde for useful feedback and helpful discussions..
BGK acknowledges funding by Deutsche Forschungsgemeinschaft (DFG, German Research Foundation) through SFB 1449 – 431232613, sub-project C02.
The ideas for this contribution were initiated during the workshop
"Accelerating the Understanding of Rare Events" (6-10 September 2021) at the Lorentz Center in Leiden, NL.

\appendix
\section{Domain of the path integral and path ensemble}
\label{app:pathEnsemble}

\subsection{Fixed path length $\mathcal{T}$}
Let $x_i \in \Gamma$ be a point in the configuration or phase space of the molecule.
$\mathbf{x}_n = (x_0, x_1 \dots x_n)$ is a time-discretized path of length $\mathcal{T} = n \Delta t$. 
$\Omega_n$ is the space of all paths with this specific length. 
A domain $\mathcal{S}$ within the path space $\Omega_n$ is constructed as a product of subsets
of the phase space $S_i \subset \Gamma$, i.e. 
\begin{align}
	\mathcal{S}_n &= S_0 \times S_1 \dots \times S_n\, .
\label{eq:pathSpaceDomain}	
\end{align}
where the subset $S_i$ represents the phase space volume in which $x_i$ may be found.
The path integral over domain $\mathcal{S}$ then is defined as
\begin{align}
	\int_{\mathcal{S}_n} \mathcal{D} \mathbf{x}_n \, s[\mathbf{x}_n] 
	&= \int_{S_0} \int_{S_1} \dots \int_{S_n} \mathrm{d}x_0 \,  \mathrm{d}x_1  \dots \mathrm{d}x_n \, s[\mathbf{x}_n] \, ,
\end{align}
where $s: \Omega_n \rightarrow \mathbb{R}$ is a path space function \cite{Donati2017}.
The path probability density $\mathcal{P}: \Omega_n \rightarrow \mathbb{R}_{\ge 0}$ for paths of fixed length $\mathcal{T}$ is normalized as
\begin{align}
	\int_{\mathcal{S}_n} \mathcal{D} \mathbf{x}_n \, \mathcal{P}[\mathbf{x}_n]  
	&= 1\, ,
\end{align}
and a path ensemble average is given as
\begin{align}
	\langle s \rangle_{S_n} &= \int_{\mathcal{S}_n} \mathcal{D} \mathbf{x}_n \, \mathcal{P}[\mathbf{x}_n] s[\mathbf{x}_n]\, ,
\end{align}

Time-lagged correlation functions $C_{kl}(\mathcal{T})$ between
phase space functions
$\chi_k: \Gamma \rightarrow \mathbb{R}$ and 
$\chi_l: \Gamma \rightarrow \mathbb{R}$, 
can be written as such a path ensemble average 
\begin{align}
	\quad C_{kl}(\mathcal{T}) 
	&= \langle \chi_k\chi_l \rangle_{S_n} \cr
	&= \int_{\mathcal{S}_n = \Omega_n} \mathcal{D} \mathbf{x}_n \, \chi_k(x_0) \mathcal{P}(\mathbf{x}_n)\chi_l(x_n) 
\end{align}
where the path space function only evaluates the initial and the final state of the path $s[\XP] =\chi_k(x_0)\chi_l(x_n)$, and the integration is carried out over the entire path space $\mathcal{S}_n = \Omega_n$.
This is equation is used when reweighting Markov state models ~\cite{Donati2017, Donati2018, Kieninger2021}.

%
%
\subsection{Activated paths with fixed path length $\mathcal{T}$}
Instead of integrating over all paths with length $\mathcal{T}$, we can restrict the domain of the path integral to activated paths from state $A \subset \Gamma$ to state $B \subset \Gamma$  (and $A \cap B  = \emptyset$). 
An activated path starts in $A$ at time 
$t  = 0$, then leaves $A$ at $t=\Delta t$, samples the transition region $\Gamma \backslash (A \cup B)$ from $t= \Delta t$ until $t = (n-1)\Delta t$, 
and then enters either $A$ or $B$.
The requirement that the path only enters $A$ or $B$ at the very last step is equal to stating that the path is terminated after it has entered either of the two states.
Thus, the domain of the path integral is
$A_n = S_0 \times S_1 \dots \times S_n$ with
\begin{align}
	S_ i &= 	\begin{cases}
				A							&i = 0 \cr
				\Gamma \backslash (A \cup B)		&1 \le i \le n-1 \cr
				A \cup B						&i = n \, .
			\end{cases}
\label{eq:subsets_activedPaths}			
\end{align}

\subsection{Transition path ensemble}
When calculating rate constants according to eq.~\ref{eq:rateconst}, $\int_A \mathcal{D} \XP$ denotes a path integral over activated paths of arbitrary length $\mathcal{T}$, i.e.
\begin{align}
    \int_A \mathcal{D} \XP \, s[\XP] 
	 &= \sum_{n = 3}^{\infty} \int_{A_n} \mathcal{D}\XP \, s[\mathbf{x}_n] \, ,
\label{eq:integral_activedPaths}		 
\end{align} 
where $n=3$ is the smallest path length that allows an activated path.
In eq.~\ref{eq:integral_activedPaths} we assume that the path function $s[\mathbf{x}_n]$ can suitably be defined for arbitrary path lengths.
The normalization of the path probability for the path integral in eq.~\ref{eq:integral_activedPaths} is constructed as follows
\begin{align}
    \int_A \mathcal{D} \XP \, \mathcal{P}[\XP] 
	 &= \sum_{n = 3}^{\infty} \int_{A_n} \mathcal{D}\XP \, \mathcal{P}[\mathbf{x}_n] 
	 = \sum_{n = 3}^{\infty} P_{A,n} = 1\, .
\label{eq:pathProbability_activatedPaths}	 
\end{align} 
$\int_{A_n} \mathcal{D}\XP \, \mathcal{P}[\mathbf{x}_n] = P_{A.n}$ is the probability of an activated path within the ensemble of paths with length $\mathcal{T} = n \Delta t$.
Since all paths will eventually enter either $A$ or $B$, the probability of activated paths decreases with increasing $n$, and we can assume that the sum converges.
In practice, it is sufficient to evaluate the sum up to a maximum path length $n_{\mathrm{max}}$.

An path ensemble average for eq.~\ref{eq:integral_activedPaths} (as e.g.~in eq.~\ref{eq:dDKLda6}) is constructed as follows
\begin{align}
    \langle s \rangle_A
     &= \int_A \mathcal{D} \XP \, \mathcal{P}[\XP] \cdot s[\XP]
	 = \sum_{n = 3}^{\infty} \int_{A_n} \mathcal{D}\XP \, \mathcal{P}[\mathbf{x}_n] \cdot s[\XP_n]\, .
\end{align} 
In transition path sampling (and transition interface sampling) a transition path ensemble refers to set of activated paths that have been sampled according to eq.~\ref{eq:pathProbability_activatedPaths}.

\medskip
We can reconcile this view with a fixed length $L$ path ensemble,  by realising  that we can
always introduce an additional  integral to a path ensemble $S_n$ of the remaining
length $L-n$ time slices which normalizes to unity (assuming that all single step probabilities are normalized). Inserting this
into the integrals does not change the final outcome.

\section{From eq.~\ref{eq:DKL_after_substitution} to eq.~\ref{eq:dDKLda}}
\label{app:LagrangianDerviative}
Eq.~\ref{eq:DKL_after_substitution} can be written as
\begin{eqnarray}
D_{KL}  
&=&  \frac{\mathcal{F}(\mathbf{a})}{\mathcal{Z}(\mathbf{a})}-  \ln \mathcal{Z}(\mathbf{a}) + \ln \mathcal{Z}^0  
\end{eqnarray}
with
$
 \mathcal{F}(\mathbf{a})  = \mathcal{Z}^0 \int \mathcal{D} \mathbf{x}\,  \mathcal{P}^0 [\mathbf{x}]    e^{w[\mathbf{x}, \mathbf{a}]} w[\mathbf{x}, \mathbf{a}]
$.
The derivative of $D_{KL} $ with respect to a parameter $a_k$ then is
\begin{align}
\frac{\partial}{\partial a_k} D_{KL}  
&=  \frac{\mathcal{F}'(\mathbf{a})\mathcal{Z}(\mathbf{a}) + \mathcal{F}(\mathbf{a})\mathcal{Z}'(\mathbf{a})}{\mathcal{Z}^2(\mathbf{a})}-  \frac{\mathcal{Z}'(\mathbf{a})}{\mathcal{Z}(\mathbf{a})} 
\end{align}
where
\begin{subequations}
\begin{align}
\mathcal{Z}'(\mathbf{a}) 
    &= \frac{\partial}{\partial a_k} \mathcal{Z}(\mathbf{a}) 
	= \mathcal{Z}^0 \int \mathcal{D} \mathbf{x}\,  \mathcal{P}^0 [\mathbf{x}] e^{w[\mathbf{x}, \mathbf{a}]} w' [\mathbf{x}, \mathbf{a}] \label{eq:Z_prime}\\
\mathcal{F}'(\mathbf{a}) 
	&= \frac{\partial}{\partial a_k} \mathcal{F}(\mathbf{a}) \cr
	& = \int \mathcal{D} \mathbf{x}\,  \mathcal{P}^0 [\mathbf{x}]  e^{w[\mathbf{x}, \mathbf{a}]} w' [\mathbf{x}, \mathbf{a}]  w [\mathbf{x}, \mathbf{a}] + \cr
	&\quad \int \mathcal{D} \mathbf{x}\,  \mathcal{P}^0 [\mathbf{x}] e^{w[\mathbf{x}, \mathbf{a}]} w' [\mathbf{x}, \mathbf{a}] \cr
	& = \int \mathcal{D} \mathbf{x}\,  \mathcal{P}^0 [\mathbf{x}]  e^{w[\mathbf{x}, \mathbf{a}]} w' [\mathbf{x}, \mathbf{a}]  w [\mathbf{x}, \mathbf{a}] + \mathcal{Z}'(\mathbf{a})\, . \cr
	\label{eq:F_prime} 
\end{align}
\end{subequations}
Thus,
\begin{align}
\frac{\partial}{\partial a_k} D_{KL}  
&=  \frac{\mathcal{Z}^0 \int \mathcal{D} \mathbf{x}\,  \mathcal{P}^0 [\mathbf{x}]  e^{w[\mathbf{x}, \mathbf{a}]} w' [\mathbf{x}, \mathbf{a}]  w [\mathbf{x}, \mathbf{a}]}{\mathcal{Z}(\mathbf{a})} + \cr
&\quad \frac{\mathcal{Z}'(\mathbf{a})}{\mathcal{Z}(\mathbf{a})}
+\frac{\mathcal{F}(\mathbf{a})\mathcal{Z}'(\mathbf{a})}{\mathcal{Z}^2(\mathbf{a})}-  \frac{\mathcal{Z}'(\mathbf{a})}{\mathcal{Z}(\mathbf{a})} 
\end{align}
The two terms $\mathcal{Z}'(\mathbf{a})/\mathcal{Z}(\mathbf{a})$ cancel, and reinserting $\mathcal{F}(\mathbf{a})$, $\mathcal{Z}(\mathbf{a})$, and $\mathcal{Z}'(\mathbf{a})$ yields eq.~\ref{eq:dDKLda}.

\bibliography{ref.bib}

\begin{thebibliography}{78}%
\makeatletter
\providecommand \@ifxundefined [1]{%
 \@ifx{#1\undefined}
}%
\providecommand \@ifnum [1]{%
 \ifnum #1\expandafter \@firstoftwo
 \else \expandafter \@secondoftwo
 \fi
}%
\providecommand \@ifx [1]{%
 \ifx #1\expandafter \@firstoftwo
 \else \expandafter \@secondoftwo
 \fi
}%
\providecommand \natexlab [1]{#1}%
\providecommand \enquote  [1]{``#1''}%
\providecommand \bibnamefont  [1]{#1}%
\providecommand \bibfnamefont [1]{#1}%
\providecommand \citenamefont [1]{#1}%
\providecommand \href@noop [0]{\@secondoftwo}%
\providecommand \href [0]{\begingroup \@sanitize@url \@href}%
\providecommand \@href[1]{\@@startlink{#1}\@@href}%
\providecommand \@@href[1]{\endgroup#1\@@endlink}%
\providecommand \@sanitize@url [0]{\catcode `\\12\catcode `\$12\catcode
  `\&12\catcode `\#12\catcode `\^12\catcode `\_12\catcode `\%12\relax}%
\providecommand \@@startlink[1]{}%
\providecommand \@@endlink[0]{}%
\providecommand \url  [0]{\begingroup\@sanitize@url \@url }%
\providecommand \@url [1]{\endgroup\@href {#1}{\urlprefix }}%
\providecommand \urlprefix  [0]{URL }%
\providecommand \Eprint [0]{\href }%
\providecommand \doibase [0]{http://dx.doi.org/}%
\providecommand \selectlanguage [0]{\@gobble}%
\providecommand \bibinfo  [0]{\@secondoftwo}%
\providecommand \bibfield  [0]{\@secondoftwo}%
\providecommand \translation [1]{[#1]}%
\providecommand \BibitemOpen [0]{}%
\providecommand \bibitemStop [0]{}%
\providecommand \bibitemNoStop [0]{.\EOS\space}%
\providecommand \EOS [0]{\spacefactor3000\relax}%
\providecommand \BibitemShut  [1]{\csname bibitem#1\endcsname}%
\let\auto@bib@innerbib\@empty
\bibitem [{\citenamefont {Br{\"{u}}nger}\ \emph {et~al.}(1998)\citenamefont
  {Br{\"{u}}nger}, \citenamefont {Adams}, \citenamefont {Clore}, \citenamefont
  {Delano}, \citenamefont {Gros}, \citenamefont {Grossekunstleve},
  \citenamefont {Jiang}, \citenamefont {Kuszewski}, \citenamefont {Nilges},
  \citenamefont {Pannu}, \citenamefont {Read}, \citenamefont {Rice},
  \citenamefont {Simonson},\ and\ \citenamefont {Warren}}]{Brunger1998}%
  \BibitemOpen
  \bibfield  {author} {\bibinfo {author} {\bibfnamefont {A.~T.}\ \bibnamefont
  {Br{\"{u}}nger}}, \bibinfo {author} {\bibfnamefont {P.~D.}\ \bibnamefont
  {Adams}}, \bibinfo {author} {\bibfnamefont {G.~M.}\ \bibnamefont {Clore}},
  \bibinfo {author} {\bibfnamefont {W.~L.}\ \bibnamefont {Delano}}, \bibinfo
  {author} {\bibfnamefont {P.}~\bibnamefont {Gros}}, \bibinfo {author}
  {\bibfnamefont {R.~W.}\ \bibnamefont {Grossekunstleve}}, \bibinfo {author}
  {\bibfnamefont {J.~S.}\ \bibnamefont {Jiang}}, \bibinfo {author}
  {\bibfnamefont {J.}~\bibnamefont {Kuszewski}}, \bibinfo {author}
  {\bibfnamefont {M.}~\bibnamefont {Nilges}}, \bibinfo {author} {\bibfnamefont
  {N.~S.}\ \bibnamefont {Pannu}}, \bibinfo {author} {\bibfnamefont {R.~J.}\
  \bibnamefont {Read}}, \bibinfo {author} {\bibfnamefont {L.~M.}\ \bibnamefont
  {Rice}}, \bibinfo {author} {\bibfnamefont {T.}~\bibnamefont {Simonson}}, \
  and\ \bibinfo {author} {\bibfnamefont {G.~L.}\ \bibnamefont {Warren}},\
  }\href@noop {} {\bibfield  {journal} {\bibinfo  {journal} {Acta Crystallogr.
  Sect. D Biol. Crystallogr.}\ }\textbf {\bibinfo {volume} {54}},\ \bibinfo
  {pages} {905} (\bibinfo {year} {1998})}\BibitemShut {NoStop}%
\bibitem [{\citenamefont {Bonomi}\ \emph {et~al.}(2017)\citenamefont {Bonomi},
  \citenamefont {Heller}, \citenamefont {Camilloni},\ and\ \citenamefont
  {Vendruscolo}}]{Bonomi2017}%
  \BibitemOpen
  \bibfield  {author} {\bibinfo {author} {\bibfnamefont {M.}~\bibnamefont
  {Bonomi}}, \bibinfo {author} {\bibfnamefont {G.~T.}\ \bibnamefont {Heller}},
  \bibinfo {author} {\bibfnamefont {C.}~\bibnamefont {Camilloni}}, \ and\
  \bibinfo {author} {\bibfnamefont {M.}~\bibnamefont {Vendruscolo}},\ }\href
  {\doibase 10.1016/j.sbi.2016.12.004} {\bibfield  {journal} {\bibinfo
  {journal} {Curr. Opin. Struct. Biol.}\ }\textbf {\bibinfo {volume} {42}},\
  \bibinfo {pages} {106} (\bibinfo {year} {2017})}\BibitemShut {NoStop}%
\bibitem [{\citenamefont {Capelli}\ \emph {et~al.}(2018)\citenamefont
  {Capelli}, \citenamefont {Tiana},\ and\ \citenamefont
  {Camilloni}}]{Capelli2018}%
  \BibitemOpen
  \bibfield  {author} {\bibinfo {author} {\bibfnamefont {R.}~\bibnamefont
  {Capelli}}, \bibinfo {author} {\bibfnamefont {G.}~\bibnamefont {Tiana}}, \
  and\ \bibinfo {author} {\bibfnamefont {C.}~\bibnamefont {Camilloni}},\
  }\href@noop {} {\bibfield  {journal} {\bibinfo  {journal} {J. Chem. Phys.}\
  }\textbf {\bibinfo {volume} {148}},\ \bibinfo {pages} {184114} (\bibinfo
  {year} {2018})}\BibitemShut {NoStop}%
\bibitem [{\citenamefont {Brotzakis}\ \emph {et~al.}(2021)\citenamefont
  {Brotzakis}, \citenamefont {Vendruscolo},\ and\ \citenamefont
  {Bolhuis}}]{Brotzakis2021}%
  \BibitemOpen
  \bibfield  {author} {\bibinfo {author} {\bibfnamefont {Z.~F.}\ \bibnamefont
  {Brotzakis}}, \bibinfo {author} {\bibfnamefont {M.}~\bibnamefont
  {Vendruscolo}}, \ and\ \bibinfo {author} {\bibfnamefont {P.~G.}\ \bibnamefont
  {Bolhuis}},\ }\href {\doibase 10.1073/pnas.2012423118} {\bibfield  {journal}
  {\bibinfo  {journal} {Proceedings of the National Academy of Sciences}\
  }\textbf {\bibinfo {volume} {118}},\ \bibinfo {pages} {e2012423118} (\bibinfo
  {year} {2021})}\BibitemShut {NoStop}%
\bibitem [{\citenamefont {Alder}\ and\ \citenamefont
  {Wainwright}(1957)}]{Alder1957}%
  \BibitemOpen
  \bibfield  {author} {\bibinfo {author} {\bibfnamefont {B.~J.}\ \bibnamefont
  {Alder}}\ and\ \bibinfo {author} {\bibfnamefont {T.~E.}\ \bibnamefont
  {Wainwright}},\ }\href {\doibase 10.1063/1.1743957} {\bibfield  {journal}
  {\bibinfo  {journal} {J. Chem. Phys.}\ }\textbf {\bibinfo {volume} {27}},\
  \bibinfo {pages} {1208} (\bibinfo {year} {1957})}\BibitemShut {NoStop}%
\bibitem [{\citenamefont {Frenkel}\ and\ \citenamefont
  {Smit}(2001)}]{Frenkel2001}%
  \BibitemOpen
  \bibfield  {author} {\bibinfo {author} {\bibfnamefont {D.}~\bibnamefont
  {Frenkel}}\ and\ \bibinfo {author} {\bibfnamefont {B.}~\bibnamefont {Smit}},\
  }\href@noop {} {\emph {\bibinfo {title} {{Understanding Molecular
  Simulation}}}},\ \bibinfo {edition} {2nd}\ ed.\ (\bibinfo  {publisher}
  {Academic Press, Inc.},\ \bibinfo {address} {Orlando, FL, USA},\ \bibinfo
  {year} {2001})\BibitemShut {NoStop}%
\bibitem [{\citenamefont {Shaw}\ \emph {et~al.}(2008)\citenamefont {Shaw},
  \citenamefont {Deneroff}, \citenamefont {Dror}, \citenamefont {Kuskin},
  \citenamefont {Larson}, \citenamefont {Salmon}, \citenamefont {Young},
  \citenamefont {Batson}, \citenamefont {Bowers}, \citenamefont {Chao},
  \citenamefont {Eastwood}, \citenamefont {Gagliardo},\ and\ \citenamefont
  {Gross}}]{Shaw08}%
  \BibitemOpen
  \bibfield  {author} {\bibinfo {author} {\bibfnamefont {D.~E.}\ \bibnamefont
  {Shaw}}, \bibinfo {author} {\bibfnamefont {M.~M.}\ \bibnamefont {Deneroff}},
  \bibinfo {author} {\bibfnamefont {R.~O.}\ \bibnamefont {Dror}}, \bibinfo
  {author} {\bibfnamefont {J.~S.}\ \bibnamefont {Kuskin}}, \bibinfo {author}
  {\bibfnamefont {R.~H.}\ \bibnamefont {Larson}}, \bibinfo {author}
  {\bibfnamefont {J.~K.}\ \bibnamefont {Salmon}}, \bibinfo {author}
  {\bibfnamefont {C.}~\bibnamefont {Young}}, \bibinfo {author} {\bibfnamefont
  {B.}~\bibnamefont {Batson}}, \bibinfo {author} {\bibfnamefont {K.~J.}\
  \bibnamefont {Bowers}}, \bibinfo {author} {\bibfnamefont {J.~C.}\
  \bibnamefont {Chao}}, \bibinfo {author} {\bibfnamefont {M.~P.}\ \bibnamefont
  {Eastwood}}, \bibinfo {author} {\bibfnamefont {J.}~\bibnamefont {Gagliardo}},
  \ and\ \bibinfo {author} {\bibfnamefont {S.~C.}\ \bibnamefont {Gross}},\
  }\href {\doibase 10.1145/1364782.1364802} {\bibfield  {journal} {\bibinfo
  {journal} {Commun. ACM}\ }\textbf {\bibinfo {volume} {51}},\ \bibinfo {pages}
  {91} (\bibinfo {year} {2008})}\BibitemShut {NoStop}%
\bibitem [{\citenamefont {Stone}\ \emph {et~al.}(2007)\citenamefont {Stone},
  \citenamefont {Phillips}, \citenamefont {Freddolino}, \citenamefont {Hardy},
  \citenamefont {Trabuco},\ and\ \citenamefont {Schulten}}]{Stone2007}%
  \BibitemOpen
  \bibfield  {author} {\bibinfo {author} {\bibfnamefont {J.~E.}\ \bibnamefont
  {Stone}}, \bibinfo {author} {\bibfnamefont {J.~C.}\ \bibnamefont {Phillips}},
  \bibinfo {author} {\bibfnamefont {P.~L.}\ \bibnamefont {Freddolino}},
  \bibinfo {author} {\bibfnamefont {D.~J.}\ \bibnamefont {Hardy}}, \bibinfo
  {author} {\bibfnamefont {L.~G.}\ \bibnamefont {Trabuco}}, \ and\ \bibinfo
  {author} {\bibfnamefont {K.}~\bibnamefont {Schulten}},\ }\href@noop {}
  {\bibfield  {journal} {\bibinfo  {journal} {J Comput Chem}\ }\textbf
  {\bibinfo {volume} {28}},\ \bibinfo {pages} {2618} (\bibinfo {year}
  {2007})}\BibitemShut {NoStop}%
\bibitem [{\citenamefont {Ciccotti}\ \emph {et~al.}(2022)\citenamefont
  {Ciccotti}, \citenamefont {Dellago}, \citenamefont {Ferrario}, \citenamefont
  {Hern{\'{a}}ndez},\ and\ \citenamefont {Tuckerman}}]{Ciccotti2022}%
  \BibitemOpen
  \bibfield  {author} {\bibinfo {author} {\bibfnamefont {G.}~\bibnamefont
  {Ciccotti}}, \bibinfo {author} {\bibfnamefont {C.}~\bibnamefont {Dellago}},
  \bibinfo {author} {\bibfnamefont {M.}~\bibnamefont {Ferrario}}, \bibinfo
  {author} {\bibfnamefont {E.~R.}\ \bibnamefont {Hern{\'{a}}ndez}}, \ and\
  \bibinfo {author} {\bibfnamefont {M.~E.}\ \bibnamefont {Tuckerman}},\ }\href
  {\doibase 10.1140/epjb/s10051-021-00249-x} {\bibfield  {journal} {\bibinfo
  {journal} {European Physical Journal B}\ }\textbf {\bibinfo {volume} {95}}
  (\bibinfo {year} {2022}),\ 10.1140/epjb/s10051-021-00249-x}\BibitemShut
  {NoStop}%
\bibitem [{\citenamefont {Hollingsworth}\ and\ \citenamefont
  {Dror}(2018)}]{Hollingsworth2018}%
  \BibitemOpen
  \bibfield  {author} {\bibinfo {author} {\bibfnamefont {S.~A.}\ \bibnamefont
  {Hollingsworth}}\ and\ \bibinfo {author} {\bibfnamefont {R.~O.}\ \bibnamefont
  {Dror}},\ }\href {\doibase 10.1016/j.neuron.2018.08.011} {\bibfield
  {journal} {\bibinfo  {journal} {Neuron}\ }\textbf {\bibinfo {volume} {99}},\
  \bibinfo {pages} {1129} (\bibinfo {year} {2018})}\BibitemShut {NoStop}%
\bibitem [{\citenamefont {Eyring}(1935)}]{Eyring1935}%
  \BibitemOpen
  \bibfield  {author} {\bibinfo {author} {\bibfnamefont {H.}~\bibnamefont
  {Eyring}},\ }\href@noop {} {\bibfield  {journal} {\bibinfo  {journal} {J.
  Chem. Phys.}\ }\textbf {\bibinfo {volume} {3}},\ \bibinfo {pages} {107}
  (\bibinfo {year} {1935})}\BibitemShut {NoStop}%
\bibitem [{\citenamefont {Chandler}(1987)}]{Chandlerbook}%
  \BibitemOpen
  \bibfield  {author} {\bibinfo {author} {\bibfnamefont {D.}~\bibnamefont
  {Chandler}},\ }\href {http://opac.inria.fr/record=b1081336} {\emph {\bibinfo
  {title} {{Introduction to modern statistical mechanics}}}}\ (\bibinfo
  {publisher} {Oxford University Press},\ \bibinfo {year} {1987})\BibitemShut
  {NoStop}%
\bibitem [{\citenamefont {Bolhuis}\ and\ \citenamefont
  {Dellago}(2010)}]{Bolhuis2009}%
  \BibitemOpen
  \bibfield  {author} {\bibinfo {author} {\bibfnamefont {P.~G.}\ \bibnamefont
  {Bolhuis}}\ and\ \bibinfo {author} {\bibfnamefont {C.}~\bibnamefont
  {Dellago}},\ }\href {\doibase 10.1002/9780470890905.ch3} {\bibfield
  {journal} {\bibinfo  {journal} {Rev. Comput. Chem.}\ }\textbf {\bibinfo
  {volume} {27}},\ \bibinfo {pages} {1} (\bibinfo {year} {2010})}\BibitemShut
  {NoStop}%
\bibitem [{\citenamefont {Valsson}\ \emph {et~al.}(2016)\citenamefont
  {Valsson}, \citenamefont {Tiwary},\ and\ \citenamefont
  {Parrinello}}]{Valsson2016}%
  \BibitemOpen
  \bibfield  {author} {\bibinfo {author} {\bibfnamefont {O.}~\bibnamefont
  {Valsson}}, \bibinfo {author} {\bibfnamefont {P.}~\bibnamefont {Tiwary}}, \
  and\ \bibinfo {author} {\bibfnamefont {M.}~\bibnamefont {Parrinello}},\
  }\href {\doibase 10.1146/annurev-physchem-040215-112229} {\bibfield
  {journal} {\bibinfo  {journal} {Annu. Rev. Phys. Chem.}\ }\textbf {\bibinfo
  {volume} {67}},\ \bibinfo {pages} {159} (\bibinfo {year} {2016})}\BibitemShut
  {NoStop}%
\bibitem [{\citenamefont {Harrison}\ \emph {et~al.}(2018)\citenamefont
  {Harrison}, \citenamefont {Schall}, \citenamefont {Maskey}, \citenamefont
  {Mikulski}, \citenamefont {Knippenberg},\ and\ \citenamefont
  {Morrow}}]{Harrison2018}%
  \BibitemOpen
  \bibfield  {author} {\bibinfo {author} {\bibfnamefont {J.~A.}\ \bibnamefont
  {Harrison}}, \bibinfo {author} {\bibfnamefont {J.~D.}\ \bibnamefont
  {Schall}}, \bibinfo {author} {\bibfnamefont {S.}~\bibnamefont {Maskey}},
  \bibinfo {author} {\bibfnamefont {P.~T.}\ \bibnamefont {Mikulski}}, \bibinfo
  {author} {\bibfnamefont {M.~T.}\ \bibnamefont {Knippenberg}}, \ and\ \bibinfo
  {author} {\bibfnamefont {B.~H.}\ \bibnamefont {Morrow}},\ }\href {\doibase
  10.1063/1.5020808} {\bibfield  {journal} {\bibinfo  {journal} {Applied
  Physics Reviews}\ }\textbf {\bibinfo {volume} {5}},\ \bibinfo {pages}
  {031104} (\bibinfo {year} {2018})}\BibitemShut {NoStop}%
\bibitem [{\citenamefont {Piana}\ \emph {et~al.}(2011)\citenamefont {Piana},
  \citenamefont {Lindorff-Larsen},\ and\ \citenamefont {Shaw}}]{Piana2011}%
  \BibitemOpen
  \bibfield  {author} {\bibinfo {author} {\bibfnamefont {S.}~\bibnamefont
  {Piana}}, \bibinfo {author} {\bibfnamefont {K.}~\bibnamefont
  {Lindorff-Larsen}}, \ and\ \bibinfo {author} {\bibfnamefont {D.~E.}\
  \bibnamefont {Shaw}},\ }\href {\doibase 10.1016/j.bpj.2011.03.051} {\bibfield
   {journal} {\bibinfo  {journal} {Biophysical journal}\ }\textbf {\bibinfo
  {volume} {100}},\ \bibinfo {pages} {L47} (\bibinfo {year}
  {2011})}\BibitemShut {NoStop}%
\bibitem [{\citenamefont {van~der Spoel}(2021)}]{VanderSpoel2021}%
  \BibitemOpen
  \bibfield  {author} {\bibinfo {author} {\bibfnamefont {D.}~\bibnamefont
  {van~der Spoel}},\ }\href {\doibase 10.1016/j.sbi.2020.08.006} {\bibfield
  {journal} {\bibinfo  {journal} {Current Opinion in Structural Biology}\
  }\textbf {\bibinfo {volume} {67}},\ \bibinfo {pages} {18} (\bibinfo {year}
  {2021})}\BibitemShut {NoStop}%
\bibitem [{\citenamefont {Tkatchenko}\ and\ \citenamefont
  {Scheffler}(2009)}]{Tkatchenko2009}%
  \BibitemOpen
  \bibfield  {author} {\bibinfo {author} {\bibfnamefont {A.}~\bibnamefont
  {Tkatchenko}}\ and\ \bibinfo {author} {\bibfnamefont {M.}~\bibnamefont
  {Scheffler}},\ }\href {\doibase 10.1103/PhysRevLett.102.073005} {\bibfield
  {journal} {\bibinfo  {journal} {Phys. Rev. Lett.}\ }\textbf {\bibinfo
  {volume} {102}},\ \bibinfo {pages} {6} (\bibinfo {year} {2009})}\BibitemShut
  {NoStop}%
\bibitem [{\citenamefont {Zhang}\ \emph {et~al.}(2018)\citenamefont {Zhang},
  \citenamefont {Han}, \citenamefont {Wang}, \citenamefont {Car},\ and\
  \citenamefont {Weinan}}]{Zhang2018}%
  \BibitemOpen
  \bibfield  {author} {\bibinfo {author} {\bibfnamefont {L.}~\bibnamefont
  {Zhang}}, \bibinfo {author} {\bibfnamefont {J.}~\bibnamefont {Han}}, \bibinfo
  {author} {\bibfnamefont {H.}~\bibnamefont {Wang}}, \bibinfo {author}
  {\bibfnamefont {R.}~\bibnamefont {Car}}, \ and\ \bibinfo {author}
  {\bibfnamefont {E.}~\bibnamefont {Weinan}},\ }\href {\doibase
  10.1103/PhysRevLett.120.143001} {\bibfield  {journal} {\bibinfo  {journal}
  {Phys. Rev. Let.}\ }\textbf {\bibinfo {volume} {120}},\ \bibinfo {pages}
  {143001} (\bibinfo {year} {2018})},\ \Eprint
  {http://arxiv.org/abs/1707.09571} {arXiv:1707.09571} \BibitemShut {NoStop}%
\bibitem [{\citenamefont {Bonati}\ and\ \citenamefont
  {Parrinello}(2018)}]{Bonati2018}%
  \BibitemOpen
  \bibfield  {author} {\bibinfo {author} {\bibfnamefont {L.}~\bibnamefont
  {Bonati}}\ and\ \bibinfo {author} {\bibfnamefont {M.}~\bibnamefont
  {Parrinello}},\ }\href {\doibase 10.1103/PhysRevLett.121.265701} {\bibfield
  {journal} {\bibinfo  {journal} {Physical Review Letters}\ }\textbf {\bibinfo
  {volume} {121}},\ \bibinfo {pages} {265701} (\bibinfo {year} {2018})},\
  \Eprint {http://arxiv.org/abs/1809.11088} {arXiv:1809.11088} \BibitemShut
  {NoStop}%
\bibitem [{\citenamefont {Singraber}\ \emph {et~al.}(2019)\citenamefont
  {Singraber}, \citenamefont {Behler},\ and\ \citenamefont
  {Dellago}}]{Singraber2019}%
  \BibitemOpen
  \bibfield  {author} {\bibinfo {author} {\bibfnamefont {A.}~\bibnamefont
  {Singraber}}, \bibinfo {author} {\bibfnamefont {J.}~\bibnamefont {Behler}}, \
  and\ \bibinfo {author} {\bibfnamefont {C.}~\bibnamefont {Dellago}},\ }\href
  {\doibase 10.1021/acs.jctc.8b00770} {\bibfield  {journal} {\bibinfo
  {journal} {J. Chem. Theory Comput.}\ }\textbf {\bibinfo {volume} {15}},\
  \bibinfo {pages} {1827} (\bibinfo {year} {2019})}\BibitemShut {NoStop}%
\bibitem [{\citenamefont {Lindorff-Larsen}\ \emph {et~al.}(2012)\citenamefont
  {Lindorff-Larsen}, \citenamefont {Maragakis}, \citenamefont {Piana},
  \citenamefont {Eastwood}, \citenamefont {Dror},\ and\ \citenamefont
  {Shaw}}]{Lindorff2012}%
  \BibitemOpen
  \bibfield  {author} {\bibinfo {author} {\bibfnamefont {K.}~\bibnamefont
  {Lindorff-Larsen}}, \bibinfo {author} {\bibfnamefont {P.}~\bibnamefont
  {Maragakis}}, \bibinfo {author} {\bibfnamefont {S.}~\bibnamefont {Piana}},
  \bibinfo {author} {\bibfnamefont {M.~P.}\ \bibnamefont {Eastwood}}, \bibinfo
  {author} {\bibfnamefont {R.~O.}\ \bibnamefont {Dror}}, \ and\ \bibinfo
  {author} {\bibfnamefont {D.~E.}\ \bibnamefont {Shaw}},\ }\href@noop {}
  {\bibfield  {journal} {\bibinfo  {journal} {PloS one}\ }\textbf {\bibinfo
  {volume} {7}},\ \bibinfo {pages} {e32131} (\bibinfo {year}
  {2012})}\BibitemShut {NoStop}%
\bibitem [{\citenamefont {Vitalini}\ \emph {et~al.}(2015)\citenamefont
  {Vitalini}, \citenamefont {Mey}, \citenamefont {No{\'e}},\ and\ \citenamefont
  {Keller}}]{Vitalini2015}%
  \BibitemOpen
  \bibfield  {author} {\bibinfo {author} {\bibfnamefont {F.}~\bibnamefont
  {Vitalini}}, \bibinfo {author} {\bibfnamefont {A.~S.}\ \bibnamefont {Mey}},
  \bibinfo {author} {\bibfnamefont {F.}~\bibnamefont {No{\'e}}}, \ and\
  \bibinfo {author} {\bibfnamefont {B.~G.}\ \bibnamefont {Keller}},\
  }\href@noop {} {\bibfield  {journal} {\bibinfo  {journal} {The Journal of
  Chemical Physics}\ }\textbf {\bibinfo {volume} {142}},\ \bibinfo {pages}
  {02B611\_1} (\bibinfo {year} {2015})}\BibitemShut {NoStop}%
\bibitem [{\citenamefont {Cavalli}\ \emph {et~al.}(2013)\citenamefont
  {Cavalli}, \citenamefont {Camilloni},\ and\ \citenamefont
  {Vendruscolo}}]{Cavalli2013a}%
  \BibitemOpen
  \bibfield  {author} {\bibinfo {author} {\bibfnamefont {A.}~\bibnamefont
  {Cavalli}}, \bibinfo {author} {\bibfnamefont {C.}~\bibnamefont {Camilloni}},
  \ and\ \bibinfo {author} {\bibfnamefont {M.}~\bibnamefont {Vendruscolo}},\
  }\href@noop {} {\bibfield  {journal} {\bibinfo  {journal} {J. Chem. Phys.}\
  }\textbf {\bibinfo {volume} {138}},\ \bibinfo {pages} {094112} (\bibinfo
  {year} {2013})}\BibitemShut {NoStop}%
\bibitem [{\citenamefont {Boomsma}\ \emph {et~al.}(2014)\citenamefont
  {Boomsma}, \citenamefont {Ferkinghoff-Borg},\ and\ \citenamefont
  {Lindorff-Larsen}}]{Boomsma2014}%
  \BibitemOpen
  \bibfield  {author} {\bibinfo {author} {\bibfnamefont {W.}~\bibnamefont
  {Boomsma}}, \bibinfo {author} {\bibfnamefont {J.}~\bibnamefont
  {Ferkinghoff-Borg}}, \ and\ \bibinfo {author} {\bibfnamefont
  {K.}~\bibnamefont {Lindorff-Larsen}},\ }\href {\doibase
  10.1371/journal.pcbi.1003406} {\bibfield  {journal} {\bibinfo  {journal}
  {PLoS Comput. Biol.}\ }\textbf {\bibinfo {volume} {10}},\ \bibinfo {pages}
  {1} (\bibinfo {year} {2014})}\BibitemShut {NoStop}%
\bibitem [{\citenamefont {Bonomi}\ \emph {et~al.}(2015)\citenamefont {Bonomi},
  \citenamefont {Camilloni}, \citenamefont {Cavalli},\ and\ \citenamefont
  {Vendruscolo}}]{Bonomi2015}%
  \BibitemOpen
  \bibfield  {author} {\bibinfo {author} {\bibfnamefont {M.}~\bibnamefont
  {Bonomi}}, \bibinfo {author} {\bibfnamefont {C.}~\bibnamefont {Camilloni}},
  \bibinfo {author} {\bibfnamefont {A.}~\bibnamefont {Cavalli}}, \ and\
  \bibinfo {author} {\bibfnamefont {M.}~\bibnamefont {Vendruscolo}},\
  }\href@noop {} {\bibfield  {journal} {\bibinfo  {journal} {Sci. Adv.}\
  }\textbf {\bibinfo {volume} {2}},\ \bibinfo {pages} {1} (\bibinfo {year}
  {2015})}\BibitemShut {NoStop}%
\bibitem [{\citenamefont {Bolhuis}\ \emph {et~al.}(2021)\citenamefont
  {Bolhuis}, \citenamefont {Brotzakis},\ and\ \citenamefont
  {Vendruscolo}}]{Bolhuis2021}%
  \BibitemOpen
  \bibfield  {author} {\bibinfo {author} {\bibfnamefont {P.~G.}\ \bibnamefont
  {Bolhuis}}, \bibinfo {author} {\bibfnamefont {Z.~F.}\ \bibnamefont
  {Brotzakis}}, \ and\ \bibinfo {author} {\bibfnamefont {M.}~\bibnamefont
  {Vendruscolo}},\ }\href {\doibase 10.1140/epjb/s10051-021-00154-3} {\bibfield
   {journal} {\bibinfo  {journal} {The European Physical Journal B}\ }\textbf
  {\bibinfo {volume} {94}} (\bibinfo {year} {2021}),\
  10.1140/epjb/s10051-021-00154-3}\BibitemShut {NoStop}%
\bibitem [{\citenamefont {Tsai}\ \emph {et~al.}(2022)\citenamefont {Tsai},
  \citenamefont {Fields},\ and\ \citenamefont {Tiwary}}]{Tsai2022}%
  \BibitemOpen
  \bibfield  {author} {\bibinfo {author} {\bibfnamefont {S.-T.}\ \bibnamefont
  {Tsai}}, \bibinfo {author} {\bibfnamefont {E.}~\bibnamefont {Fields}}, \ and\
  \bibinfo {author} {\bibfnamefont {P.}~\bibnamefont {Tiwary}},\ }\href
  {http://arxiv.org/abs/2203.00597} {\ ,\ \bibinfo {pages} {1} (\bibinfo {year}
  {2022})},\ \Eprint {http://arxiv.org/abs/2203.00597} {arXiv:2203.00597}
  \BibitemShut {NoStop}%
\bibitem [{\citenamefont {Cesari}\ \emph {et~al.}(2019)\citenamefont {Cesari},
  \citenamefont {Bottaro}, \citenamefont {Lindorff-Larsen}, \citenamefont
  {Ban{\'{a}}{\v{s}}}, \citenamefont {{\v{S}}poner},\ and\ \citenamefont
  {Bussi}}]{Cesari2019}%
  \BibitemOpen
  \bibfield  {author} {\bibinfo {author} {\bibfnamefont {A.}~\bibnamefont
  {Cesari}}, \bibinfo {author} {\bibfnamefont {S.}~\bibnamefont {Bottaro}},
  \bibinfo {author} {\bibfnamefont {K.}~\bibnamefont {Lindorff-Larsen}},
  \bibinfo {author} {\bibfnamefont {P.}~\bibnamefont {Ban{\'{a}}{\v{s}}}},
  \bibinfo {author} {\bibfnamefont {J.}~\bibnamefont {{\v{S}}poner}}, \ and\
  \bibinfo {author} {\bibfnamefont {G.}~\bibnamefont {Bussi}},\ }\href
  {\doibase 10.1021/acs.jctc.9b00206} {\bibfield  {journal} {\bibinfo
  {journal} {Journal of Chemical Theory and Computation}\ }\textbf {\bibinfo
  {volume} {15}},\ \bibinfo {pages} {3425} (\bibinfo {year}
  {2019})}\BibitemShut {NoStop}%
\bibitem [{\citenamefont {Tesei}\ \emph {et~al.}(2021)\citenamefont {Tesei},
  \citenamefont {Schulze}, \citenamefont {Crehuet},\ and\ \citenamefont
  {Lindorff-Larsen}}]{Tesei2021}%
  \BibitemOpen
  \bibfield  {author} {\bibinfo {author} {\bibfnamefont {G.}~\bibnamefont
  {Tesei}}, \bibinfo {author} {\bibfnamefont {T.~K.}\ \bibnamefont {Schulze}},
  \bibinfo {author} {\bibfnamefont {R.}~\bibnamefont {Crehuet}}, \ and\
  \bibinfo {author} {\bibfnamefont {K.}~\bibnamefont {Lindorff-Larsen}},\
  }\href {\doibase 10.1073/pnas.2111696118} {\bibfield  {journal} {\bibinfo
  {journal} {Proceedings of the National Academy of Sciences}\ }\textbf
  {\bibinfo {volume} {118}} (\bibinfo {year} {2021}),\
  10.1073/pnas.2111696118}\BibitemShut {NoStop}%
\bibitem [{\citenamefont {Fr{\"{o}}hlking}\ \emph {et~al.}(2020)\citenamefont
  {Fr{\"{o}}hlking}, \citenamefont {Bernetti}, \citenamefont {Calonaci},\ and\
  \citenamefont {Bussi}}]{Frohlking2020}%
  \BibitemOpen
  \bibfield  {author} {\bibinfo {author} {\bibfnamefont {T.}~\bibnamefont
  {Fr{\"{o}}hlking}}, \bibinfo {author} {\bibfnamefont {M.}~\bibnamefont
  {Bernetti}}, \bibinfo {author} {\bibfnamefont {N.}~\bibnamefont {Calonaci}},
  \ and\ \bibinfo {author} {\bibfnamefont {G.}~\bibnamefont {Bussi}},\ }\href
  {\doibase 10.1063/5.0011346} {\bibfield  {journal} {\bibinfo  {journal} {J.
  Chem. Phys.}\ }\textbf {\bibinfo {volume} {152}} (\bibinfo {year} {2020}),\
  10.1063/5.0011346},\ \Eprint {http://arxiv.org/abs/2004.01630}
  {arXiv:2004.01630} \BibitemShut {NoStop}%
\bibitem [{\citenamefont {Varela-Rial}\ \emph {et~al.}(2022)\citenamefont
  {Varela-Rial}, \citenamefont {Maryanow}, \citenamefont {Majewski},
  \citenamefont {Doerr}, \citenamefont {Schapin}, \citenamefont
  {Jim{\'{e}}nez-Luna},\ and\ \citenamefont {{De
  Fabritiis}}}]{Varela-Rial2022}%
  \BibitemOpen
  \bibfield  {author} {\bibinfo {author} {\bibfnamefont {A.}~\bibnamefont
  {Varela-Rial}}, \bibinfo {author} {\bibfnamefont {I.}~\bibnamefont
  {Maryanow}}, \bibinfo {author} {\bibfnamefont {M.}~\bibnamefont {Majewski}},
  \bibinfo {author} {\bibfnamefont {S.}~\bibnamefont {Doerr}}, \bibinfo
  {author} {\bibfnamefont {N.}~\bibnamefont {Schapin}}, \bibinfo {author}
  {\bibfnamefont {J.}~\bibnamefont {Jim{\'{e}}nez-Luna}}, \ and\ \bibinfo
  {author} {\bibfnamefont {G.}~\bibnamefont {{De Fabritiis}}},\ }\href
  {\doibase 10.1021/acs.jcim.1c00691} {\bibfield  {journal} {\bibinfo
  {journal} {Journal of Chemical Information and Modeling}\ }\textbf {\bibinfo
  {volume} {62}},\ \bibinfo {pages} {225} (\bibinfo {year} {2022})}\BibitemShut
  {NoStop}%
\bibitem [{\citenamefont {Qiu}\ \emph {et~al.}(2021)\citenamefont {Qiu},
  \citenamefont {Smith}, \citenamefont {Boothroyd}, \citenamefont {Jang},
  \citenamefont {Hahn}, \citenamefont {Wagner}, \citenamefont {Bannan},
  \citenamefont {Gokey}, \citenamefont {Lim}, \citenamefont {Stern},
  \citenamefont {Rizzi}, \citenamefont {Tjanaka}, \citenamefont {Tresadern},
  \citenamefont {Lucas}, \citenamefont {Shirts}, \citenamefont {Gilson},
  \citenamefont {Chodera}, \citenamefont {Bayly}, \citenamefont {Mobley},\ and\
  \citenamefont {Wang}}]{Qiu2021}%
  \BibitemOpen
  \bibfield  {author} {\bibinfo {author} {\bibfnamefont {Y.}~\bibnamefont
  {Qiu}}, \bibinfo {author} {\bibfnamefont {D.~G.~A.}\ \bibnamefont {Smith}},
  \bibinfo {author} {\bibfnamefont {S.}~\bibnamefont {Boothroyd}}, \bibinfo
  {author} {\bibfnamefont {H.}~\bibnamefont {Jang}}, \bibinfo {author}
  {\bibfnamefont {D.~F.}\ \bibnamefont {Hahn}}, \bibinfo {author}
  {\bibfnamefont {J.}~\bibnamefont {Wagner}}, \bibinfo {author} {\bibfnamefont
  {C.~C.}\ \bibnamefont {Bannan}}, \bibinfo {author} {\bibfnamefont
  {T.}~\bibnamefont {Gokey}}, \bibinfo {author} {\bibfnamefont {V.~T.}\
  \bibnamefont {Lim}}, \bibinfo {author} {\bibfnamefont {C.~D.}\ \bibnamefont
  {Stern}}, \bibinfo {author} {\bibfnamefont {A.}~\bibnamefont {Rizzi}},
  \bibinfo {author} {\bibfnamefont {B.}~\bibnamefont {Tjanaka}}, \bibinfo
  {author} {\bibfnamefont {G.}~\bibnamefont {Tresadern}}, \bibinfo {author}
  {\bibfnamefont {X.}~\bibnamefont {Lucas}}, \bibinfo {author} {\bibfnamefont
  {M.~R.}\ \bibnamefont {Shirts}}, \bibinfo {author} {\bibfnamefont {M.~K.}\
  \bibnamefont {Gilson}}, \bibinfo {author} {\bibfnamefont {J.~D.}\
  \bibnamefont {Chodera}}, \bibinfo {author} {\bibfnamefont {C.~I.}\
  \bibnamefont {Bayly}}, \bibinfo {author} {\bibfnamefont {D.~L.}\ \bibnamefont
  {Mobley}}, \ and\ \bibinfo {author} {\bibfnamefont {L.-P.}\ \bibnamefont
  {Wang}},\ }\href {\doibase 10.1021/acs.jctc.1c00571} {\bibfield  {journal}
  {\bibinfo  {journal} {J. Chem. Theory Comput.}\ }\textbf {\bibinfo {volume}
  {17}},\ \bibinfo {pages} {6262} (\bibinfo {year} {2021})}\BibitemShut
  {NoStop}%
\bibitem [{\citenamefont {Rose}\ \emph {et~al.}(2021)\citenamefont {Rose},
  \citenamefont {Mair},\ and\ \citenamefont {Garrahan}}]{Rose2021}%
  \BibitemOpen
  \bibfield  {author} {\bibinfo {author} {\bibfnamefont {D.~C.}\ \bibnamefont
  {Rose}}, \bibinfo {author} {\bibfnamefont {J.~F.}\ \bibnamefont {Mair}}, \
  and\ \bibinfo {author} {\bibfnamefont {J.~P.}\ \bibnamefont {Garrahan}},\
  }\href {\doibase 10.1088/1367-2630/abd7bd} {\bibfield  {journal} {\bibinfo
  {journal} {New Journal of Physics}\ }\textbf {\bibinfo {volume} {23}},\
  \bibinfo {pages} {013013} (\bibinfo {year} {2021})}\BibitemShut {NoStop}%
\bibitem [{\citenamefont {Das}\ \emph {et~al.}(2021)\citenamefont {Das},
  \citenamefont {Rose}, \citenamefont {Garrahan},\ and\ \citenamefont
  {Limmer}}]{Das2021}%
  \BibitemOpen
  \bibfield  {author} {\bibinfo {author} {\bibfnamefont {A.}~\bibnamefont
  {Das}}, \bibinfo {author} {\bibfnamefont {D.~C.}\ \bibnamefont {Rose}},
  \bibinfo {author} {\bibfnamefont {J.~P.}\ \bibnamefont {Garrahan}}, \ and\
  \bibinfo {author} {\bibfnamefont {D.~T.}\ \bibnamefont {Limmer}},\ }\href
  {\doibase 10.1063/5.0057323} {\bibfield  {journal} {\bibinfo  {journal} {The
  Journal of Chemical Physics}\ }\textbf {\bibinfo {volume} {155}},\ \bibinfo
  {pages} {134105} (\bibinfo {year} {2021})}\BibitemShut {NoStop}%
\bibitem [{\citenamefont {Das}\ \emph {et~al.}(2022)\citenamefont {Das},
  \citenamefont {Kuznets-Speck},\ and\ \citenamefont {Limmer}}]{Das2022}%
  \BibitemOpen
  \bibfield  {author} {\bibinfo {author} {\bibfnamefont {A.}~\bibnamefont
  {Das}}, \bibinfo {author} {\bibfnamefont {B.}~\bibnamefont {Kuznets-Speck}},
  \ and\ \bibinfo {author} {\bibfnamefont {D.~T.}\ \bibnamefont {Limmer}},\
  }\href {\doibase 10.1103/PhysRevLett.128.028005} {\bibfield  {journal}
  {\bibinfo  {journal} {Phys. Rev. Lett.}\ }\textbf {\bibinfo {volume} {128}},\
  \bibinfo {pages} {028005} (\bibinfo {year} {2022})}\BibitemShut {NoStop}%
\bibitem [{\citenamefont {Donati}\ \emph {et~al.}(2017)\citenamefont {Donati},
  \citenamefont {Hartmann},\ and\ \citenamefont {Keller}}]{Donati2017}%
  \BibitemOpen
  \bibfield  {author} {\bibinfo {author} {\bibfnamefont {L.}~\bibnamefont
  {Donati}}, \bibinfo {author} {\bibfnamefont {C.}~\bibnamefont {Hartmann}}, \
  and\ \bibinfo {author} {\bibfnamefont {B.~G.}\ \bibnamefont {Keller}},\
  }\href@noop {} {\bibfield  {journal} {\bibinfo  {journal} {The Journal of
  chemical physics}\ }\textbf {\bibinfo {volume} {146}},\ \bibinfo {pages}
  {244112} (\bibinfo {year} {2017})}\BibitemShut {NoStop}%
\bibitem [{\citenamefont {Donati}\ and\ \citenamefont
  {Keller}(2018)}]{Donati2018}%
  \BibitemOpen
  \bibfield  {author} {\bibinfo {author} {\bibfnamefont {L.}~\bibnamefont
  {Donati}}\ and\ \bibinfo {author} {\bibfnamefont {B.~G.}\ \bibnamefont
  {Keller}},\ }\href@noop {} {\bibfield  {journal} {\bibinfo  {journal} {The
  Journal of chemical physics}\ }\textbf {\bibinfo {volume} {149}},\ \bibinfo
  {pages} {072335} (\bibinfo {year} {2018})}\BibitemShut {NoStop}%
\bibitem [{\citenamefont {Kieninger}\ and\ \citenamefont
  {Keller}(2021)}]{Kieninger2021}%
  \BibitemOpen
  \bibfield  {author} {\bibinfo {author} {\bibfnamefont {S.}~\bibnamefont
  {Kieninger}}\ and\ \bibinfo {author} {\bibfnamefont {B.~G.}\ \bibnamefont
  {Keller}},\ }\href {\doibase 10.1063/5.0038408} {\bibfield  {journal}
  {\bibinfo  {journal} {The Journal of Chemical Physics}\ }\textbf {\bibinfo
  {volume} {154}},\ \bibinfo {pages} {094102} (\bibinfo {year}
  {2021})}\BibitemShut {NoStop}%
\bibitem [{\citenamefont {Bolhuis}\ \emph {et~al.}(2002)\citenamefont
  {Bolhuis}, \citenamefont {Chandler}, \citenamefont {Dellago},\ and\
  \citenamefont {Geissler}}]{Bolhuis2002}%
  \BibitemOpen
  \bibfield  {author} {\bibinfo {author} {\bibfnamefont {P.~G.}\ \bibnamefont
  {Bolhuis}}, \bibinfo {author} {\bibfnamefont {D.}~\bibnamefont {Chandler}},
  \bibinfo {author} {\bibfnamefont {C.}~\bibnamefont {Dellago}}, \ and\
  \bibinfo {author} {\bibfnamefont {P.~L.}\ \bibnamefont {Geissler}},\ }\href
  {\doibase 10.1146/annurev.physchem.53.082301.113146} {\bibfield  {journal}
  {\bibinfo  {journal} {Annu. Rev. Phys. Chem.}\ }\textbf {\bibinfo {volume}
  {53}},\ \bibinfo {pages} {291} (\bibinfo {year} {2002})}\BibitemShut
  {NoStop}%
\bibitem [{\citenamefont {Dellago}\ \emph {et~al.}(2002)\citenamefont
  {Dellago}, \citenamefont {Bolhuis},\ and\ \citenamefont
  {Geissler}}]{Dellago2002}%
  \BibitemOpen
  \bibfield  {author} {\bibinfo {author} {\bibfnamefont {C.}~\bibnamefont
  {Dellago}}, \bibinfo {author} {\bibfnamefont {P.}~\bibnamefont {Bolhuis}}, \
  and\ \bibinfo {author} {\bibfnamefont {P.}~\bibnamefont {Geissler}},\
  }\href@noop {} {\bibfield  {journal} {\bibinfo  {journal} {Advances in
  Chemical Physics}\ }\textbf {\bibinfo {volume} {123}} (\bibinfo {year}
  {2002})}\BibitemShut {NoStop}%
\bibitem [{\citenamefont {Du}\ and\ \citenamefont {Bolhuis}(2013)}]{Du2013}%
  \BibitemOpen
  \bibfield  {author} {\bibinfo {author} {\bibfnamefont {W.-N.}\ \bibnamefont
  {Du}}\ and\ \bibinfo {author} {\bibfnamefont {P.~G.}\ \bibnamefont
  {Bolhuis}},\ }\href {\doibase 10.1063/1.4813777} {\bibfield  {journal}
  {\bibinfo  {journal} {The Journal of Chemical Physics}\ }\textbf {\bibinfo
  {volume} {139}},\ \bibinfo {pages} {044105} (\bibinfo {year}
  {2013})}\BibitemShut {NoStop}%
\bibitem [{\citenamefont {Allen}\ \emph {et~al.}(2005)\citenamefont {Allen},
  \citenamefont {Warren},\ and\ \citenamefont {ten Wolde}}]{Allen2005}%
  \BibitemOpen
  \bibfield  {author} {\bibinfo {author} {\bibfnamefont {R.~J.}\ \bibnamefont
  {Allen}}, \bibinfo {author} {\bibfnamefont {P.~B.}\ \bibnamefont {Warren}}, \
  and\ \bibinfo {author} {\bibfnamefont {P.~R.}\ \bibnamefont {ten Wolde}},\
  }\href {\doibase 10.1103/physrevlett.94.018104} {\bibfield  {journal}
  {\bibinfo  {journal} {Physical Review Letters}\ }\textbf {\bibinfo {volume}
  {94}} (\bibinfo {year} {2005}),\ 10.1103/physrevlett.94.018104}\BibitemShut
  {NoStop}%
\bibitem [{\citenamefont {Zuckerman}\ and\ \citenamefont
  {Chong}(2017)}]{Zuckerman2017}%
  \BibitemOpen
  \bibfield  {author} {\bibinfo {author} {\bibfnamefont {D.~M.}\ \bibnamefont
  {Zuckerman}}\ and\ \bibinfo {author} {\bibfnamefont {L.~T.}\ \bibnamefont
  {Chong}},\ }\href {\doibase 10.1146/annurev-biophys-070816-033834} {\bibfield
   {journal} {\bibinfo  {journal} {Annual Review of Biophysics}\ }\textbf
  {\bibinfo {volume} {46}},\ \bibinfo {pages} {43} (\bibinfo {year}
  {2017})}\BibitemShut {NoStop}%
\bibitem [{\citenamefont {Leimkuhler}\ and\ \citenamefont
  {Matthews}(2015)}]{Leimkuhler2015}%
  \BibitemOpen
  \bibfield  {author} {\bibinfo {author} {\bibfnamefont {B.}~\bibnamefont
  {Leimkuhler}}\ and\ \bibinfo {author} {\bibfnamefont {C.}~\bibnamefont
  {Matthews}},\ }\href@noop {} {\bibfield  {journal} {\bibinfo  {journal}
  {Interdisciplinary applied mathematics}\ }\textbf {\bibinfo {volume} {36}}
  (\bibinfo {year} {2015})}\BibitemShut {NoStop}%
\bibitem [{\citenamefont {{\O}ksendal}(2003)}]{Oksendal2003}%
  \BibitemOpen
  \bibfield  {author} {\bibinfo {author} {\bibfnamefont {B.}~\bibnamefont
  {{\O}ksendal}},\ }in\ \href@noop {} {\emph {\bibinfo {booktitle} {Stochastic
  differential equations}}}\ (\bibinfo  {publisher} {Springer},\ \bibinfo
  {year} {2003})\ pp.\ \bibinfo {pages} {65--84}\BibitemShut {NoStop}%
\bibitem [{\citenamefont {Kloeden}\ and\ \citenamefont
  {Platen}(1992)}]{Platen1992}%
  \BibitemOpen
  \bibfield  {author} {\bibinfo {author} {\bibfnamefont {P.~E.}\ \bibnamefont
  {Kloeden}}\ and\ \bibinfo {author} {\bibfnamefont {E.}~\bibnamefont
  {Platen}},\ }\href@noop {} {\emph {\bibinfo {title} {{Numerical Solution of
  Stochastic Differential Equations.}}}},\ \bibinfo {edition} {1st}\ ed.\
  (\bibinfo  {publisher} {Springer, Berlin, Heidelberg},\ \bibinfo {year}
  {1992})\BibitemShut {NoStop}%
\bibitem [{\citenamefont {Hazoglou}\ \emph {et~al.}(2015)\citenamefont
  {Hazoglou}, \citenamefont {Walther}, \citenamefont {Dixit},\ and\
  \citenamefont {Dill}}]{Hazoglou2015}%
  \BibitemOpen
  \bibfield  {author} {\bibinfo {author} {\bibfnamefont {M.~J.}\ \bibnamefont
  {Hazoglou}}, \bibinfo {author} {\bibfnamefont {V.}~\bibnamefont {Walther}},
  \bibinfo {author} {\bibfnamefont {P.~D.}\ \bibnamefont {Dixit}}, \ and\
  \bibinfo {author} {\bibfnamefont {K.~A.}\ \bibnamefont {Dill}},\ }\href@noop
  {} {\bibfield  {journal} {\bibinfo  {journal} {The Journal of Chemical
  Physics}\ }\textbf {\bibinfo {volume} {143}},\ \bibinfo {pages} {051104}
  (\bibinfo {year} {2015})}\BibitemShut {NoStop}%
\bibitem [{\citenamefont {Chandler}(1978)}]{Chandler1978}%
  \BibitemOpen
  \bibfield  {author} {\bibinfo {author} {\bibfnamefont {D.}~\bibnamefont
  {Chandler}},\ }\href@noop {} {\bibfield  {journal} {\bibinfo  {journal} {J.
  Chem. Phys.}\ }\textbf {\bibinfo {volume} {68}},\ \bibinfo {pages} {2959}
  (\bibinfo {year} {1978})}\BibitemShut {NoStop}%
\bibitem [{\citenamefont {Faradjian}\ and\ \citenamefont
  {Elber}(2004)}]{Faradjian2004}%
  \BibitemOpen
  \bibfield  {author} {\bibinfo {author} {\bibfnamefont {A.~K.}\ \bibnamefont
  {Faradjian}}\ and\ \bibinfo {author} {\bibfnamefont {R.}~\bibnamefont
  {Elber}},\ }\href {\doibase 10.1063/1.1738640} {\bibfield  {journal}
  {\bibinfo  {journal} {Journal of Chemical Physics}\ }\textbf {\bibinfo
  {volume} {120}},\ \bibinfo {pages} {10880} (\bibinfo {year}
  {2004})}\BibitemShut {NoStop}%
\bibitem [{\citenamefont {Allen}\ \emph {et~al.}(2006)\citenamefont {Allen},
  \citenamefont {Frenkel},\ and\ \citenamefont {{Ten Wolde}}}]{Allen2006}%
  \BibitemOpen
  \bibfield  {author} {\bibinfo {author} {\bibfnamefont {R.~J.}\ \bibnamefont
  {Allen}}, \bibinfo {author} {\bibfnamefont {D.}~\bibnamefont {Frenkel}}, \
  and\ \bibinfo {author} {\bibfnamefont {P.~R.}\ \bibnamefont {{Ten Wolde}}},\
  }\href {\doibase 10.1063/1.2140273} {\bibfield  {journal} {\bibinfo
  {journal} {Journal of Chemical Physics}\ }\textbf {\bibinfo {volume} {124}}
  (\bibinfo {year} {2006}),\ 10.1063/1.2140273},\ \Eprint
  {http://arxiv.org/abs/0509499} {arXiv:0509499 [cond-mat]} \BibitemShut
  {NoStop}%
\bibitem [{\citenamefont {Tiwary}\ and\ \citenamefont
  {Parrinello}(2013)}]{Tiwary2013}%
  \BibitemOpen
  \bibfield  {author} {\bibinfo {author} {\bibfnamefont {P.}~\bibnamefont
  {Tiwary}}\ and\ \bibinfo {author} {\bibfnamefont {M.}~\bibnamefont
  {Parrinello}},\ }\href {\doibase 10.1103/PhysRevLett.111.230602} {\bibfield
  {journal} {\bibinfo  {journal} {Physical Review Letters}\ }\textbf {\bibinfo
  {volume} {111}},\ \bibinfo {pages} {230602} (\bibinfo {year}
  {2013})}\BibitemShut {NoStop}%
\bibitem [{\citenamefont {Brotzakis}\ and\ \citenamefont
  {Bolhuis}(2019)}]{Brotzakis2019f}%
  \BibitemOpen
  \bibfield  {author} {\bibinfo {author} {\bibfnamefont {Z.~F.}\ \bibnamefont
  {Brotzakis}}\ and\ \bibinfo {author} {\bibfnamefont {P.~G.}\ \bibnamefont
  {Bolhuis}},\ }\href@noop {} {\bibfield  {journal} {\bibinfo  {journal}
  {Journal of Chemical Physics}\ }\textbf {\bibinfo {volume} {151}},\ \bibinfo
  {pages} {174111} (\bibinfo {year} {2019})}\BibitemShut {NoStop}%
\bibitem [{\citenamefont {van Erp}\ \emph {et~al.}(2003)\citenamefont {van
  Erp}, \citenamefont {Moroni},\ and\ \citenamefont {Bolhuis}}]{vanErp2003}%
  \BibitemOpen
  \bibfield  {author} {\bibinfo {author} {\bibfnamefont {T.}~\bibnamefont {van
  Erp}}, \bibinfo {author} {\bibfnamefont {D.}~\bibnamefont {Moroni}}, \ and\
  \bibinfo {author} {\bibfnamefont {P.}~\bibnamefont {Bolhuis}},\ }\href
  {http://scitation.aip.org/content/aip/journal/jcp/118/17/10.1063/1.1562614}
  {\bibfield  {journal} {\bibinfo  {journal} {J of chem phys}\ }\textbf
  {\bibinfo {volume} {118}},\ \bibinfo {pages} {7762} (\bibinfo {year}
  {2003})}\BibitemShut {NoStop}%
\bibitem [{\citenamefont {Rogal}\ \emph {et~al.}(2010)\citenamefont {Rogal},
  \citenamefont {Lechner}, \citenamefont {Juraszek}, \citenamefont {Ensing},\
  and\ \citenamefont {Bolhuis}}]{Rogal2010}%
  \BibitemOpen
  \bibfield  {author} {\bibinfo {author} {\bibfnamefont {J.}~\bibnamefont
  {Rogal}}, \bibinfo {author} {\bibfnamefont {W.}~\bibnamefont {Lechner}},
  \bibinfo {author} {\bibfnamefont {J.}~\bibnamefont {Juraszek}}, \bibinfo
  {author} {\bibfnamefont {B.}~\bibnamefont {Ensing}}, \ and\ \bibinfo {author}
  {\bibfnamefont {P.~G.}\ \bibnamefont {Bolhuis}},\ }\href {\doibase
  10.1063/1.3491817} {\bibfield  {journal} {\bibinfo  {journal} {J. Chem.
  Phys.}\ }\textbf {\bibinfo {volume} {133}},\ \bibinfo {pages} {174109}
  (\bibinfo {year} {2010})}\BibitemShut {NoStop}%
\bibitem [{\citenamefont {Dellago}\ and\ \citenamefont
  {Bolhuis}(2004)}]{DellagoBolhuis2004}%
  \BibitemOpen
  \bibfield  {author} {\bibinfo {author} {\bibfnamefont {C.}~\bibnamefont
  {Dellago}}\ and\ \bibinfo {author} {\bibfnamefont {P.~G.}\ \bibnamefont
  {Bolhuis}},\ }\href {\doibase 10.1080/08927020412331294869} {\bibfield
  {journal} {\bibinfo  {journal} {Molecular Simulation}\ }\textbf {\bibinfo
  {volume} {30}},\ \bibinfo {pages} {795} (\bibinfo {year} {2004})}\BibitemShut
  {NoStop}%
\bibitem [{\citenamefont {Bolhuis}\ and\ \citenamefont
  {Cs{\'{a}}nyi}(2018)}]{BolhuisCsanyi2018}%
  \BibitemOpen
  \bibfield  {author} {\bibinfo {author} {\bibfnamefont {P.~G.}\ \bibnamefont
  {Bolhuis}}\ and\ \bibinfo {author} {\bibfnamefont {G.}~\bibnamefont
  {Cs{\'{a}}nyi}},\ }\href {\doibase 10.1103/PhysRevLett.120.250601} {\bibfield
   {journal} {\bibinfo  {journal} {Physical Review Letters}\ }\textbf {\bibinfo
  {volume} {120}},\ \bibinfo {pages} {1} (\bibinfo {year} {2018})}\BibitemShut
  {NoStop}%
\bibitem [{\citenamefont {{Ferrenberg, Alan M.; Swendsen,
  Robert}}(1989)}]{Ferrenberg1989}%
  \BibitemOpen
  \bibfield  {author} {\bibinfo {author} {\bibfnamefont {H.}~\bibnamefont
  {{Ferrenberg, Alan M.; Swendsen, Robert}}},\ }\href {\doibase
  10.1103/PhysRevE.67.067102} {\bibfield  {journal} {\bibinfo  {journal} {Phys.
  Rev. Lett.}\ }\textbf {\bibinfo {volume} {63}},\ \bibinfo {pages} {1195}
  (\bibinfo {year} {1989})},\ \Eprint {http://arxiv.org/abs/0302123}
  {arXiv:0302123 [cond-mat]} \BibitemShut {NoStop}%
\bibitem [{\citenamefont {Weeks}\ \emph {et~al.}(1971)\citenamefont {Weeks},
  \citenamefont {Chandler},\ and\ \citenamefont {Andersen}}]{Weeks1971}%
  \BibitemOpen
  \bibfield  {author} {\bibinfo {author} {\bibfnamefont {J.~D.}\ \bibnamefont
  {Weeks}}, \bibinfo {author} {\bibfnamefont {D.}~\bibnamefont {Chandler}}, \
  and\ \bibinfo {author} {\bibfnamefont {H.~C.}\ \bibnamefont {Andersen}},\
  }\href {\doibase 10.1063/1.1674820} {\bibfield  {journal} {\bibinfo
  {journal} {The Journal of Chemical Physics}\ }\textbf {\bibinfo {volume}
  {54}},\ \bibinfo {pages} {5237} (\bibinfo {year} {1971})}\BibitemShut
  {NoStop}%
\bibitem [{\citenamefont {Platt}\ and\ \citenamefont {Barr}(1987)}]{Platt1988}%
  \BibitemOpen
  \bibfield  {author} {\bibinfo {author} {\bibfnamefont {J.~C.}\ \bibnamefont
  {Platt}}\ and\ \bibinfo {author} {\bibfnamefont {A.~H.}\ \bibnamefont
  {Barr}},\ }\href {https://resolver.caltech.edu/CaltechCSTR:1988.cs-tr-88-17}
  {\bibfield  {journal} {\bibinfo  {journal} {NIPS'87: Proceedings of the 1987
  International Conference on Neural Information Processing Systems}\ ,\
  \bibinfo {pages} {612–621}} (\bibinfo {year} {1987})}\BibitemShut {NoStop}%
\bibitem [{\citenamefont {Copeland}\ \emph {et~al.}(2006)\citenamefont
  {Copeland}, \citenamefont {Pompliano},\ and\ \citenamefont
  {Meek}}]{Copeland2006}%
  \BibitemOpen
  \bibfield  {author} {\bibinfo {author} {\bibfnamefont {R.~A.}\ \bibnamefont
  {Copeland}}, \bibinfo {author} {\bibfnamefont {D.~L.}\ \bibnamefont
  {Pompliano}}, \ and\ \bibinfo {author} {\bibfnamefont {T.~D.}\ \bibnamefont
  {Meek}},\ }\href@noop {} {\bibfield  {journal} {\bibinfo  {journal} {Nat Rev
  Drug Discov}\ }\textbf {\bibinfo {volume} {5}},\ \bibinfo {pages} {730}
  (\bibinfo {year} {2006})}\BibitemShut {NoStop}%
\bibitem [{\citenamefont {Tonge}(2018)}]{Tonge2018a}%
  \BibitemOpen
  \bibfield  {author} {\bibinfo {author} {\bibfnamefont {P.~J.}\ \bibnamefont
  {Tonge}},\ }\href {\doibase 10.1021/acschemneuro.7b00185} {\bibfield
  {journal} {\bibinfo  {journal} {ACS Chemical Neuroscience}\ }\textbf
  {\bibinfo {volume} {9}},\ \bibinfo {pages} {29} (\bibinfo {year}
  {2018})}\BibitemShut {NoStop}%
\bibitem [{\citenamefont {Romano}\ and\ \citenamefont
  {Sciortino}(2011)}]{Romano2011}%
  \BibitemOpen
  \bibfield  {author} {\bibinfo {author} {\bibfnamefont {F.}~\bibnamefont
  {Romano}}\ and\ \bibinfo {author} {\bibfnamefont {F.}~\bibnamefont
  {Sciortino}},\ }\href {\doibase 10.1038/nmat2975} {\bibfield  {journal}
  {\bibinfo  {journal} {Nature Materials}\ }\textbf {\bibinfo {volume} {10}},\
  \bibinfo {pages} {171} (\bibinfo {year} {2011})}\BibitemShut {NoStop}%
\bibitem [{\citenamefont {Oh}\ \emph {et~al.}(2019)\citenamefont {Oh},
  \citenamefont {Lee}, \citenamefont {Glotzer}, \citenamefont {Yi},\ and\
  \citenamefont {Pine}}]{Oh2019}%
  \BibitemOpen
  \bibfield  {author} {\bibinfo {author} {\bibfnamefont {J.~S.}\ \bibnamefont
  {Oh}}, \bibinfo {author} {\bibfnamefont {S.}~\bibnamefont {Lee}}, \bibinfo
  {author} {\bibfnamefont {S.~C.}\ \bibnamefont {Glotzer}}, \bibinfo {author}
  {\bibfnamefont {G.~R.}\ \bibnamefont {Yi}}, \ and\ \bibinfo {author}
  {\bibfnamefont {D.~J.}\ \bibnamefont {Pine}},\ }\href {\doibase
  10.1038/s41467-019-11915-1} {\bibfield  {journal} {\bibinfo  {journal}
  {Nature Communications}\ }\textbf {\bibinfo {volume} {10}} (\bibinfo {year}
  {2019}),\ 10.1038/s41467-019-11915-1}\BibitemShut {NoStop}%
\bibitem [{\citenamefont {Wang}\ \emph {et~al.}(2019)\citenamefont {Wang},
  \citenamefont {Wang}, \citenamefont {Li}, \citenamefont {Cheung},
  \citenamefont {Tian}, \citenamefont {Kim}, \citenamefont {Yi}, \citenamefont
  {Ducrot},\ and\ \citenamefont {Wang}}]{Wang2019b}%
  \BibitemOpen
  \bibfield  {author} {\bibinfo {author} {\bibfnamefont {Z.}~\bibnamefont
  {Wang}}, \bibinfo {author} {\bibfnamefont {Z.}~\bibnamefont {Wang}}, \bibinfo
  {author} {\bibfnamefont {J.}~\bibnamefont {Li}}, \bibinfo {author}
  {\bibfnamefont {S.~T.~H.}\ \bibnamefont {Cheung}}, \bibinfo {author}
  {\bibfnamefont {C.}~\bibnamefont {Tian}}, \bibinfo {author} {\bibfnamefont
  {S.~H.}\ \bibnamefont {Kim}}, \bibinfo {author} {\bibfnamefont {G.~R.}\
  \bibnamefont {Yi}}, \bibinfo {author} {\bibfnamefont {E.}~\bibnamefont
  {Ducrot}}, \ and\ \bibinfo {author} {\bibfnamefont {Y.}~\bibnamefont
  {Wang}},\ }\href {\doibase 10.1021/jacs.9b07785} {\bibfield  {journal}
  {\bibinfo  {journal} {Journal of the American Chemical Society}\ }\textbf
  {\bibinfo {volume} {141}},\ \bibinfo {pages} {14853} (\bibinfo {year}
  {2019})}\BibitemShut {NoStop}%
\bibitem [{\citenamefont {Zhang}\ \emph {et~al.}(2020)\citenamefont {Zhang},
  \citenamefont {Alberstein}, \citenamefont {{De Yoreo}},\ and\ \citenamefont
  {Tezcan}}]{Zhang2020}%
  \BibitemOpen
  \bibfield  {author} {\bibinfo {author} {\bibfnamefont {S.}~\bibnamefont
  {Zhang}}, \bibinfo {author} {\bibfnamefont {R.~G.}\ \bibnamefont
  {Alberstein}}, \bibinfo {author} {\bibfnamefont {J.~J.}\ \bibnamefont {{De
  Yoreo}}}, \ and\ \bibinfo {author} {\bibfnamefont {F.~A.}\ \bibnamefont
  {Tezcan}},\ }\href {\doibase 10.1038/s41467-020-17562-1} {\bibfield
  {journal} {\bibinfo  {journal} {Nature Communications}\ }\textbf {\bibinfo
  {volume} {11}},\ \bibinfo {pages} {1} (\bibinfo {year} {2020})}\BibitemShut
  {NoStop}%
\bibitem [{\citenamefont {Fisher}\ and\ \citenamefont
  {Elbaum-Garfinkle}(2020)}]{Fisher2020}%
  \BibitemOpen
  \bibfield  {author} {\bibinfo {author} {\bibfnamefont {R.~S.}\ \bibnamefont
  {Fisher}}\ and\ \bibinfo {author} {\bibfnamefont {S.}~\bibnamefont
  {Elbaum-Garfinkle}},\ }\href {\doibase 10.1038/s41467-020-18224-y} {\bibfield
   {journal} {\bibinfo  {journal} {Nature Communications}\ }\textbf {\bibinfo
  {volume} {11}} (\bibinfo {year} {2020}),\
  10.1038/s41467-020-18224-y}\BibitemShut {NoStop}%
\bibitem [{\citenamefont {Mitchell}\ \emph {et~al.}(2021)\citenamefont
  {Mitchell}, \citenamefont {Billingsley}, \citenamefont {Haley}, \citenamefont
  {Wechsler}, \citenamefont {Peppas},\ and\ \citenamefont
  {Langer}}]{Mitchell2021}%
  \BibitemOpen
  \bibfield  {author} {\bibinfo {author} {\bibfnamefont {M.~J.}\ \bibnamefont
  {Mitchell}}, \bibinfo {author} {\bibfnamefont {M.~M.}\ \bibnamefont
  {Billingsley}}, \bibinfo {author} {\bibfnamefont {R.~M.}\ \bibnamefont
  {Haley}}, \bibinfo {author} {\bibfnamefont {M.~E.}\ \bibnamefont {Wechsler}},
  \bibinfo {author} {\bibfnamefont {N.~A.}\ \bibnamefont {Peppas}}, \ and\
  \bibinfo {author} {\bibfnamefont {R.}~\bibnamefont {Langer}},\ }\href
  {\doibase 10.1038/s41573-020-0090-8} {\bibfield  {journal} {\bibinfo
  {journal} {Nature Reviews Drug Discovery}\ }\textbf {\bibinfo {volume}
  {20}},\ \bibinfo {pages} {101} (\bibinfo {year} {2021})}\BibitemShut
  {NoStop}%
\bibitem [{\citenamefont {Vijaykumar}\ \emph {et~al.}(2016)\citenamefont
  {Vijaykumar}, \citenamefont {Bolhuis},\ and\ \citenamefont {ten
  Wolde}}]{Vijaykumar2016}%
  \BibitemOpen
  \bibfield  {author} {\bibinfo {author} {\bibfnamefont {A.}~\bibnamefont
  {Vijaykumar}}, \bibinfo {author} {\bibfnamefont {P.~G.}\ \bibnamefont
  {Bolhuis}}, \ and\ \bibinfo {author} {\bibfnamefont {P.~R.}\ \bibnamefont
  {ten Wolde}},\ }\href {\doibase 10.1039/c6fd00104a} {\bibfield  {journal}
  {\bibinfo  {journal} {Faraday Discussions}\ }\textbf {\bibinfo {volume}
  {195}},\ \bibinfo {pages} {421} (\bibinfo {year} {2016})}\BibitemShut
  {NoStop}%
\bibitem [{\citenamefont {Vijaykumar}\ \emph {et~al.}(2017)\citenamefont
  {Vijaykumar}, \citenamefont {ten Wolde},\ and\ \citenamefont
  {Bolhuis}}]{Vijaykumar2017}%
  \BibitemOpen
  \bibfield  {author} {\bibinfo {author} {\bibfnamefont {A.}~\bibnamefont
  {Vijaykumar}}, \bibinfo {author} {\bibfnamefont {P.~R.}\ \bibnamefont {ten
  Wolde}}, \ and\ \bibinfo {author} {\bibfnamefont {P.~G.}\ \bibnamefont
  {Bolhuis}},\ }\href {\doibase 10.1063/1.5009547} {\bibfield  {journal}
  {\bibinfo  {journal} {The Journal of Chemical Physics}\ }\textbf {\bibinfo
  {volume} {147}},\ \bibinfo {pages} {184108} (\bibinfo {year}
  {2017})}\BibitemShut {NoStop}%
\bibitem [{\citenamefont {Winter}\ \emph {et~al.}(2020)\citenamefont {Winter},
  \citenamefont {Iranmanesh}, \citenamefont {Clark},\ and\ \citenamefont
  {Glover}}]{Winter2020}%
  \BibitemOpen
  \bibfield  {author} {\bibinfo {author} {\bibfnamefont {D.~L.}\ \bibnamefont
  {Winter}}, \bibinfo {author} {\bibfnamefont {H.}~\bibnamefont {Iranmanesh}},
  \bibinfo {author} {\bibfnamefont {D.~S.}\ \bibnamefont {Clark}}, \ and\
  \bibinfo {author} {\bibfnamefont {D.~J.}\ \bibnamefont {Glover}},\ }\href
  {\doibase 10.1021/acssynbio.0c00208} {\bibfield  {journal} {\bibinfo
  {journal} {ACS Synthetic Biology}\ }\textbf {\bibinfo {volume} {9}},\
  \bibinfo {pages} {2132} (\bibinfo {year} {2020})}\BibitemShut {NoStop}%
\bibitem [{\citenamefont {Dignon}\ \emph {et~al.}(2018)\citenamefont {Dignon},
  \citenamefont {Zheng}, \citenamefont {Kim}, \citenamefont {Best},\ and\
  \citenamefont {Mittal}}]{Dignon2018}%
  \BibitemOpen
  \bibfield  {author} {\bibinfo {author} {\bibfnamefont {G.~L.}\ \bibnamefont
  {Dignon}}, \bibinfo {author} {\bibfnamefont {W.}~\bibnamefont {Zheng}},
  \bibinfo {author} {\bibfnamefont {Y.~C.}\ \bibnamefont {Kim}}, \bibinfo
  {author} {\bibfnamefont {R.~B.}\ \bibnamefont {Best}}, \ and\ \bibinfo
  {author} {\bibfnamefont {J.}~\bibnamefont {Mittal}},\ }\href@noop {}
  {\bibfield  {journal} {\bibinfo  {journal} {PLoS Computational Biology}\
  }\textbf {\bibinfo {volume} {14}},\ \bibinfo {pages} {1} (\bibinfo {year}
  {2018})}\BibitemShut {NoStop}%
\bibitem [{\citenamefont {Best}\ and\ \citenamefont {Hummer}(2016)}]{Best2016}%
  \BibitemOpen
  \bibfield  {author} {\bibinfo {author} {\bibfnamefont {R.~B.}\ \bibnamefont
  {Best}}\ and\ \bibinfo {author} {\bibfnamefont {G.}~\bibnamefont {Hummer}},\
  }\href {\doibase 10.1073/pnas.1520864113} {\bibfield  {journal} {\bibinfo
  {journal} {Proceedings of the National Academy of Sciences of the United
  States of America}\ }\textbf {\bibinfo {volume} {113}},\ \bibinfo {pages}
  {3263} (\bibinfo {year} {2016})}\BibitemShut {NoStop}%
\bibitem [{\citenamefont {Senftle}\ \emph {et~al.}(2016)\citenamefont
  {Senftle}, \citenamefont {Hong}, \citenamefont {Islam}, \citenamefont
  {Kylasa}, \citenamefont {Zheng}, \citenamefont {Shin}, \citenamefont
  {Junkermeier}, \citenamefont {Engel-Herbert}, \citenamefont {Janik},
  \citenamefont {Aktulga}, \citenamefont {Verstraelen}, \citenamefont {Grama},\
  and\ \citenamefont {{Van Duin}}}]{Senftle2016}%
  \BibitemOpen
  \bibfield  {author} {\bibinfo {author} {\bibfnamefont {T.~P.}\ \bibnamefont
  {Senftle}}, \bibinfo {author} {\bibfnamefont {S.}~\bibnamefont {Hong}},
  \bibinfo {author} {\bibfnamefont {M.~M.}\ \bibnamefont {Islam}}, \bibinfo
  {author} {\bibfnamefont {S.~B.}\ \bibnamefont {Kylasa}}, \bibinfo {author}
  {\bibfnamefont {Y.}~\bibnamefont {Zheng}}, \bibinfo {author} {\bibfnamefont
  {Y.~K.}\ \bibnamefont {Shin}}, \bibinfo {author} {\bibfnamefont
  {C.}~\bibnamefont {Junkermeier}}, \bibinfo {author} {\bibfnamefont
  {R.}~\bibnamefont {Engel-Herbert}}, \bibinfo {author} {\bibfnamefont {M.~J.}\
  \bibnamefont {Janik}}, \bibinfo {author} {\bibfnamefont {H.~M.}\ \bibnamefont
  {Aktulga}}, \bibinfo {author} {\bibfnamefont {T.}~\bibnamefont
  {Verstraelen}}, \bibinfo {author} {\bibfnamefont {A.}~\bibnamefont {Grama}},
  \ and\ \bibinfo {author} {\bibfnamefont {A.~C.}\ \bibnamefont {{Van Duin}}},\
  }\href {\doibase 10.1038/npjcompumats.2015.11} {\bibfield  {journal}
  {\bibinfo  {journal} {npj Computational Materials}\ }\textbf {\bibinfo
  {volume} {2}} (\bibinfo {year} {2016}),\
  10.1038/npjcompumats.2015.11}\BibitemShut {NoStop}%
\bibitem [{\citenamefont {Rogal}\ \emph {et~al.}(2019)\citenamefont {Rogal},
  \citenamefont {Schneider},\ and\ \citenamefont {Tuckerman}}]{rogal2019}%
  \BibitemOpen
  \bibfield  {author} {\bibinfo {author} {\bibfnamefont {J.}~\bibnamefont
  {Rogal}}, \bibinfo {author} {\bibfnamefont {E.}~\bibnamefont {Schneider}}, \
  and\ \bibinfo {author} {\bibfnamefont {M.~E.}\ \bibnamefont {Tuckerman}},\
  }\href {\doibase 10.1103/PhysRevLett.123.245701} {\bibfield  {journal}
  {\bibinfo  {journal} {Physical Review Letters}\ }\textbf {\bibinfo {volume}
  {123}},\ \bibinfo {pages} {245701} (\bibinfo {year} {2019})},\ \Eprint
  {http://arxiv.org/abs/1905.01536} {arXiv:1905.01536} \BibitemShut {NoStop}%
\bibitem [{\citenamefont {Bouchaud}\ and\ \citenamefont
  {Cont}(1998)}]{Bouchaud2000}%
  \BibitemOpen
  \bibfield  {author} {\bibinfo {author} {\bibfnamefont {J.-p.}\ \bibnamefont
  {Bouchaud}}\ and\ \bibinfo {author} {\bibfnamefont {R.}~\bibnamefont
  {Cont}},\ }\href@noop {} {\bibfield  {journal} {\bibinfo  {journal} {Eur.
  Phys. J. B.}\ }\textbf {\bibinfo {volume} {6}},\ \bibinfo {pages} {543}
  (\bibinfo {year} {1998})},\ \Eprint {http://arxiv.org/abs/9801279}
  {arXiv:9801279 [arXiv:cond-mat]} \BibitemShut {NoStop}%
\bibitem [{\citenamefont {Haworth}\ and\ \citenamefont
  {Pope}(1986)}]{Haworth1986}%
  \BibitemOpen
  \bibfield  {author} {\bibinfo {author} {\bibfnamefont {D.~C.}\ \bibnamefont
  {Haworth}}\ and\ \bibinfo {author} {\bibfnamefont {S.~B.}\ \bibnamefont
  {Pope}},\ }\href {\doibase 10.1063/1.865723} {\bibfield  {journal} {\bibinfo
  {journal} {Phys. Fluids}\ }\textbf {\bibinfo {volume} {29}},\ \bibinfo
  {pages} {387} (\bibinfo {year} {1986})}\BibitemShut {NoStop}%
\bibitem [{\citenamefont {Jones}\ \emph {et~al.}(2020)\citenamefont {Jones},
  \citenamefont {Magdon-Ismail}, \citenamefont {Mersini-Houghton},\ and\
  \citenamefont {Meshnick}}]{Jones2020}%
  \BibitemOpen
  \bibfield  {author} {\bibinfo {author} {\bibfnamefont {L.~D.}\ \bibnamefont
  {Jones}}, \bibinfo {author} {\bibfnamefont {M.}~\bibnamefont
  {Magdon-Ismail}}, \bibinfo {author} {\bibfnamefont {L.}~\bibnamefont
  {Mersini-Houghton}}, \ and\ \bibinfo {author} {\bibfnamefont
  {S.}~\bibnamefont {Meshnick}},\ }\href {http://arxiv.org/abs/2008.10530}
  {\bibfield  {journal} {\bibinfo  {journal} {arXiv:2008.10530}\ ,\ \bibinfo
  {pages} {1}} (\bibinfo {year} {2020})},\ \Eprint
  {http://arxiv.org/abs/2008.10530} {arXiv:2008.10530} \BibitemShut {NoStop}%
\end{thebibliography}%


%

\newpage
\appendix

\end{document}